\documentclass[aps]{revtex4}

\def \vect#1{\hbox{\bf #1}}

\def \and{\& }

\def\RefAIP#1{\noindent\hangafter=1\hangindent=0.25in[#1]}
\def\by#1{#1,}
\def\and{and }
\def\yr#1{(#1)}
\def\paper#1{#1}
\def\jour#1{#1}

\def\vol#1{{#1}}
\def\issue#1{}
\def\pages#1{\hbox{#1.}}

\def\JCIS     {J.~Colloid Interface Sci.\ }
\def\JFM      {J.~Fluid Mech.\ }

\usepackage{amsmath}
\usepackage{amsfonts}
\usepackage{amssymb}
\usepackage{graphicx}
\usepackage{appendix}

\begin{document}

\title{\Large Relaxation of surface tension in the free-surface boundary layer of simple Lennard-Jones liquids.}
\bigskip
\bigskip

\author{\large A.V. Lukyanov, A.E. Likhtman}
\bigskip
\bigskip

\affiliation{ School of Mathematical and Physical Sciences, University of Reading, Reading RG6 6AX, UK}
\bigskip
\bigskip

\begin{abstract}
{ In this paper we use molecular dynamics (MD) to answer a classical question:
how does the surface tension on a liquid/gas interface appear? After defining
surface tension from the first principles and performing several consistency
checks,  we perform a dynamic experiment with a single simple liquid
nanodroplet. At time zero, we remove all molecules of the interfacial layer, creating a
fresh bare interface with the bulk arrangement of molecules. After that the
system evolves towards equilibrium, and the expected surface tension is
re-established. We found that the system relaxation consists of three distinct
stages. First, the mechanical balance is quickly re-established. During this
process the notion of surface tension is meaningless. In the second stage,
the surface tension equilibrates, and the density profile broadens to a value
which we call "intrinsic" interfacial width. During the third  stage, the density
profile continues to broaden due to capillary wave excitations, which does not
however affect the surface tension. We have observed this scenario for
monatomic Lennard-Jones (LJ) liquid as well as for binary LJ mixtures at different
temperatures, monitoring a wide range of physical observables.}
\end{abstract}
\medskip

\maketitle

\section{Introduction}
The phenomenon of capillarity and capillary flows is at the heart of numerous natural processes and technological
applications. They are ranging from coating devices, polymer films and emerging technologies, such as micro and nano-fluidics, to biological and medical
applications of fluid dynamics. There is a large class of flows involving liquid-gas interfaces which are controlled by surface tension. In many of them the free surface area undergoes significant changes over a period of time comparable to the characteristic diffusion time across the interfacial layer. The phenomena associated with this class of free-surface flows can be found in many applications and  include such processes as capillary pinch-off, coalescence and generation of drops, formation of cusp regions and collapse of bubbles [1--7]. For example, the characteristic time scale of a pinch-off process is of the order of $t_{cp} \sim\frac{\mu^3}{\gamma_0^2 \rho}$,  where $\rho$, $\mu$ and $\gamma_0$, are liquid density, viscosity and equilibrium surface tension respectively [1--3]. As an example, this characteristic time for water is $t_{cp}\sim  10^{-10}\,\mbox{s}$ at room temperature. On the other hand, typical relaxation time $\tau_{\Delta}$ associated with a diffusion process in the water-vapour interfacial layer is $\tau_{\Delta}\sim \frac{\Delta^2}{D} \sim 10^{-10}\, \mbox{s}$, taking characteristic width of the water-vapour interface at $\Delta\sim 3\times 10^{-10}\, \mbox{m}$ and the coefficient of self-diffusion of water molecules at room temperature and bulk conditions at $ D\sim  10^{-9}\,\mbox{m}^2/\mbox{s}$ [8, 9]. At this rate of change of the surface area, the interface will be at non-equilibrium conditions even for simple liquids. Apparently, this will result in transient non-equilibrium density profiles and as a consequence to {\em dynamic surface tension}. 

The effect of dynamic surface tension at a liquid-gas interface has been well studied for situations involving interfacially active molecules, or surfactants, when the interfacial tension is directly connected with the surfactant concentration at the interface [10, 11].  But as a matter of fact  little is known from the first microscopic principles about dynamic surface tension effects at the interfaces of simple liquids in the absence of surface active molecules and the topic is the subject of strong debates [12--24]. 

The difference between two kinds of interfacial layers, with and without surface active molecules, is fundamental. When surfactant molecules are present, the dynamic surface tension is controlled by diffusion from the bulk area in case of solvable surfactants or by surface diffusion in case of unsolvable surfactants. In both cases relevant characteristic length scale of the diffusion process is of the order of the whole system size, $L$, which is usually much larger than the interfacial thickness $L>>\Delta$. As a consequence characteristic relaxation time of surface tension, $\tau_{\gamma}$, is found to be much larger than the diffusion time on the length scale of the interfacial layer $\tau_{\Delta}$, and could be well of the order of seconds [10, 11]. Note that interfacially active molecules reduce the value of equilibrium surface tension, therefore surface tension always relaxes from higher to lower values. 

The situation is different in the case of liquid-gas interfaces of simple liquids in the absence of surfactants. First of all, the relaxation time can be only associated with a diffusion process on the length scale of the interfacial layer, $\Delta$. This length scale suggests a very short relaxation time, of the order of $\tau_{\gamma}\sim \tau_{\Delta}$.  Secondly,  the value of surface tension at a non-equilibrium interface of simple fluids could be in principle smaller than the equilibrium value. As a result, the relaxation process would have the character of an increasing  function of time, in contrast to the case when surfactant molecules are present. Apart from simple estimates, however, at present we have no direct reliable information about the properties of fresh bare non-equilibrium liquid-gas interfacial layers and the mechanism of relaxation process in that kind of systems.  At the same time, experiments and macroscopic analysis of liquid flows with forming interfaces have already demonstrated that even relaxation process on this short time scale, $\tau_{\Delta}$, should have substantial impact on the character of the whole flow [7, 15--17]. 

For example, theoretical analysis of the free-surface flow breakup has shown that asymptotic behaviour and scaling of solutions to the Navier-Stokes equation close to the point of pinch-off depends on the assumptions made about the relaxation time $\tau_{\gamma}$ of the surface phase [2, 3, 16, 17]. The first similarity solution to the pinch-off problem has been proposed by Eggers (1993, 1997) in the framework of a standard hydrodynamic model [2, 3]. In this approach, the surface tension relaxation time is simply $\tau_{\gamma}=0$ and the surface tension is always equal to its equilibrium value independent of the surface area rate of change. A comparison of the free-surface profiles calculated on the basis of the similarity solution with experiments has shown very good agreement, see recent results [18]. However, the integrity of the standard approach to the pinch-off problem has been questioned by Shikhmurzaev (2005, 2007) [16, 17] on the basis of apparent inconsistencies between parametric dependencies predicted by the standard theory and those found in experiments. This concerns, for example, the minimal diameter of the micro-thread, at which the pinch-off process is triggered. In the standard approach ($\tau_{\gamma}=0$), the diameter is a function of liquid viscosity. At the same time, it has been found in experiments on simple fluids and binary mixtures by Kowalewski (1996) [1] that the micro-thread diameter is independent of viscosity of the liquid. Similar conclusions have been drawn later by Shikhmurzaev (2005, 2007) [16, 17] on the basis of an interface formation theory, where the surface tension relaxation time is intrinsically finite $\tau_{\gamma}\neq 0$. 

Another specific feature that follows from the assumption $\tau_{\gamma}=0$ is a singularity of the axial velocity $v_z$ of the liquid jet, $v_z\sim (t_0-t)^{-1/2}$,  at the point of pinch-off $t_0$.  At the same time molecular dynamics simulations of liquid nano-jets rupture have shown rather smooth transition up to the moment of the break-up [25, 26]. One way to hinder the singularity of the axial velocity, as is shown in [16, 17], is to reduce the surface tension at the neck of the liquid thread, where the rate of change of the surface area is maximal. 

In summary, for the rigorous macroscopic description of free surface flows of simple liquids with substantial rate of change of the surface area, it is imperative to determine the relaxation time of the surface phase and the fundamental mechanism of the relaxation process. A similar conclusion has been drawn on the basis of experimental studies of fluid necks rupture in the presence of surfactants [11]. The main effect of surfactants on the pinch-off dynamics is an increase of the local interfacial tension at the location of the minimum neck radius when the point of pinch-off is approached. This trend,  as one can see, is opposite to what is expected during the pinch-off of simple fluids without surfactants [16, 17].

Despite the fact that the experimental evidence points out that the results of macroscopic analysis obtained in the assumption of finite relaxation time seems to be in a better agreement with the observations and molecular dynamics simulations, the issue of surface tension relaxation time is far from being resolved. There is still no consensus and the topic of surface tension relaxation time in simple fluids is the subject of disputes in the scientific community. 

In the current study, we are going to directly establish properties of non-equilibrium liquid-gas interfacial layers and focus on the relaxation phenomena at liquid-gas interfaces of simple fluids without surface active molecules. To achieve this, we turn to MD simulations. We will set up clear-cut model systems with fresh and sharp interfacial layers having bulk local structure. This way, we can observe for the first time the process of recreation of a liquid-gas interface from an initially sharp surface and measure the corresponding time-dependent surface tension. Our main objective is to address the following questions:

\begin{itemize}

\item What is the value of surface tension at the bare fresh non-equlibrium liquid-gas interfacial layer of simple fluids?

\item What is the characteristic relaxation time of an infinitely sharp non-equilibrium density profile and the surface tension associated with it? 

\item What are the mechanisms associated  with this relaxation process? 

\item  When can we assume that the surface tension of an evolving liquid-gas interfacial layer of simple liquids (or miscible mixtures) is at equilibrium? 

\end{itemize}

\section{Mathematical model and methodology}
We focus on a liquid-gas interface of simple monatomic and binary liquids consisting of particles interacting by means of 12-6 Lennard-Jones (LJ) potential $U(r_{ij})$ with the cut-off $r_c=2.5\,\sigma$
$$
U(r_{ij})=\left \{   
\begin{array}{cc}
  4 \varepsilon \left (  \left( \frac{\sigma}{r_{ij}}\right)^{12}- \left( \frac{\sigma}{r_{ij}}\right)^6  \right)  & r\le r_c \\
 0 & r>r_c
\end{array}\right.
$$ 
where $r_{ij}$ is the distance between particles $i$ and $j$, $\varepsilon$ and $\sigma$ are the energy and the length scale parameters of LJ potential. In what follows all length scales are normalised by $\sigma$, energy and temperature by $\varepsilon$, pressure or stress by $\varepsilon/\sigma^3$, surface tension by $\varepsilon/\sigma^2$, viscosity by $\sqrt{\varepsilon m}/\sigma^2$ and time by $t_0 = \sigma\sqrt{m/\varepsilon}$, where $m$ is the particle mass. To understand the actual parameter range, we note that in the case of liquid argon the best choice of LJ parameters with $r_c=2.5\,\sigma$ is $\sigma=3.345\times 10^{-10} \,\mbox{m}$, $\varepsilon/k_B = 125.7 \,\mbox{K} $ and $m=6.64\times 10^{-26}\,\mbox{kg}$ (where $k_B=1.38\times 10^{-23} \, \mbox{m}^2 \, \mbox{kg} \, \mbox{s}^{-2} \, \mbox{K}^{-1}$ denotes the Boltzmann constant) [27]. This choice of parameters yields exact agreement between the calculated by MD critical point density and temperature and the values determined from the measurements, such as $T_c=150.86\, \mbox{K}$. This results in the following estimates $t_0=2.07\times 10^{-12}\,\mbox{s}$, $\varepsilon/\sigma^2=15.5\, \mbox{mN/m}$, $\sqrt{\varepsilon m}/\sigma^2\approx 10^{-4}\, \mbox{Pa}\cdot \mbox{s}$ and $\varepsilon/\sigma^3=46.3\,\mbox{MPa}$. 

The system we study is a sufficiently large liquid drop consisting of up to 75000 particles giving the maximum droplet radius up to $R_0\approx 28$. The droplets have been prepared by means of an equilibration process at constant temperature $T$ during $\tau_{eq}=1000$ with the time integration step $\Delta t=0.01$, which is used in MD simulations throughout this study. The sizes of the droplets $22 \le R_0 \le 28$ in our study have been chosen to be sufficiently large in comparison to both interfacial thickness $\Delta\sim 1$ and the Tolman length $\delta_{Tl}\sim 0.1\div0.8$ so that one can work in the planar limit $\Delta/R_0\to 0$ and neglect the curvature dependence of surface tension [28--33]. The curvature dependence is proportional to $\sim 2\delta_{Tl}/R_0$, that is, neglecting higher order terms, $\gamma(R_0)=\gamma(\infty)(1-2\delta_{Tl}(R_0)/R_0+ ...)$ with Tolman length found in between $0.1\le \delta_{Tl}\le 0.8$ depending on the droplet size [32, 33]. The temperature has been controlled by means of a DPD thermostat with friction $\varsigma_{dpd}=0.5$ to preserve liquid motion. Note that the friction due to collisions is $\varsigma\sim 25$, so that DPD friction introduces only negligible corrections. At low temperatures, the drop had been first equilibrated at a higher temperature value during $\tau_{eq}$ and then brought to the lower temperature, by the thermostat, and equilibrated again during $\tau_{eq}$.  The computational box had reflecting boundary conditions and the size, which was usually larger by $\Delta r=5-10$ than the characteristic drop size to allow for the vapour phase to settle in without total evaporation of the droplet. 

\begin{figure}
\begin{center}
\includegraphics[trim=1cm 1cm 1cm 1cm,width=0.4\columnwidth]{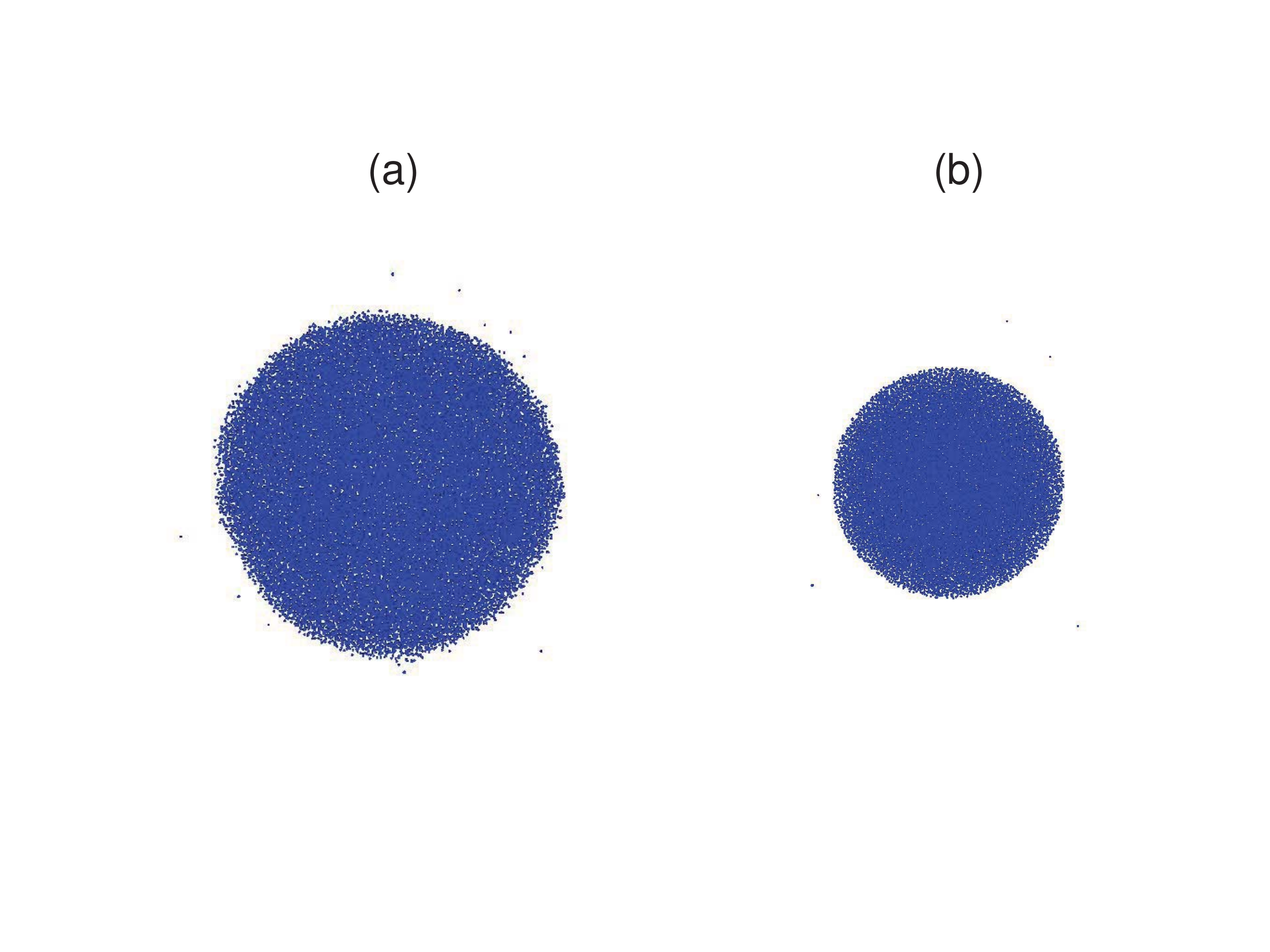}
\end{center}
\caption{A snapshot of a monatomic LJ liquid droplet consisting of $75000$ particles at $T=0.7$ (a) before the cut off (b) after the cut off, $\approx 40000$ particles left.} 
\label{Fig6}
\end{figure}

In the framework of this numerical approach, static properties of liquid-gas/liquid interfacial layers have been intensively studied by many groups over the last three decades, see for example [8, 34--42]. In the next two sections, we will discuss main procedures used in this study to analyse static and transient density profiles and to calculate the surface tension in liquid drops by MD simulations in equilibrium and non-equilibrium conditions. We will briefly summarize the results obtained in this area so far, while at the same time making a comparison with our own simulations to establish a robust connection between previous observations of interfaces in static conditions and our new results obtained in non-equilibrium conditions. In section 3, we will consider evolution of non-equilibrium interfacial density profiles and surface tension in monatomic LJ liquid drops. In section 4, we will extend our analysis of relaxation to binary LJ liquid drops. All non-equilibrium sharp density profiles in this study have been created by a cut off procedure, that is by removing all particles of the surface phase, see Fig. \ref{Fig6} for illustration. We will return to this cut off procedure for detailed consideration later and now consider density profile in a drop.  

\subsection{The density profile in monatomic LJ liquid drops}
\begin{figure}
\begin{center}
\includegraphics[trim=1cm 1cm 1cm 1cm,width=0.4\columnwidth]{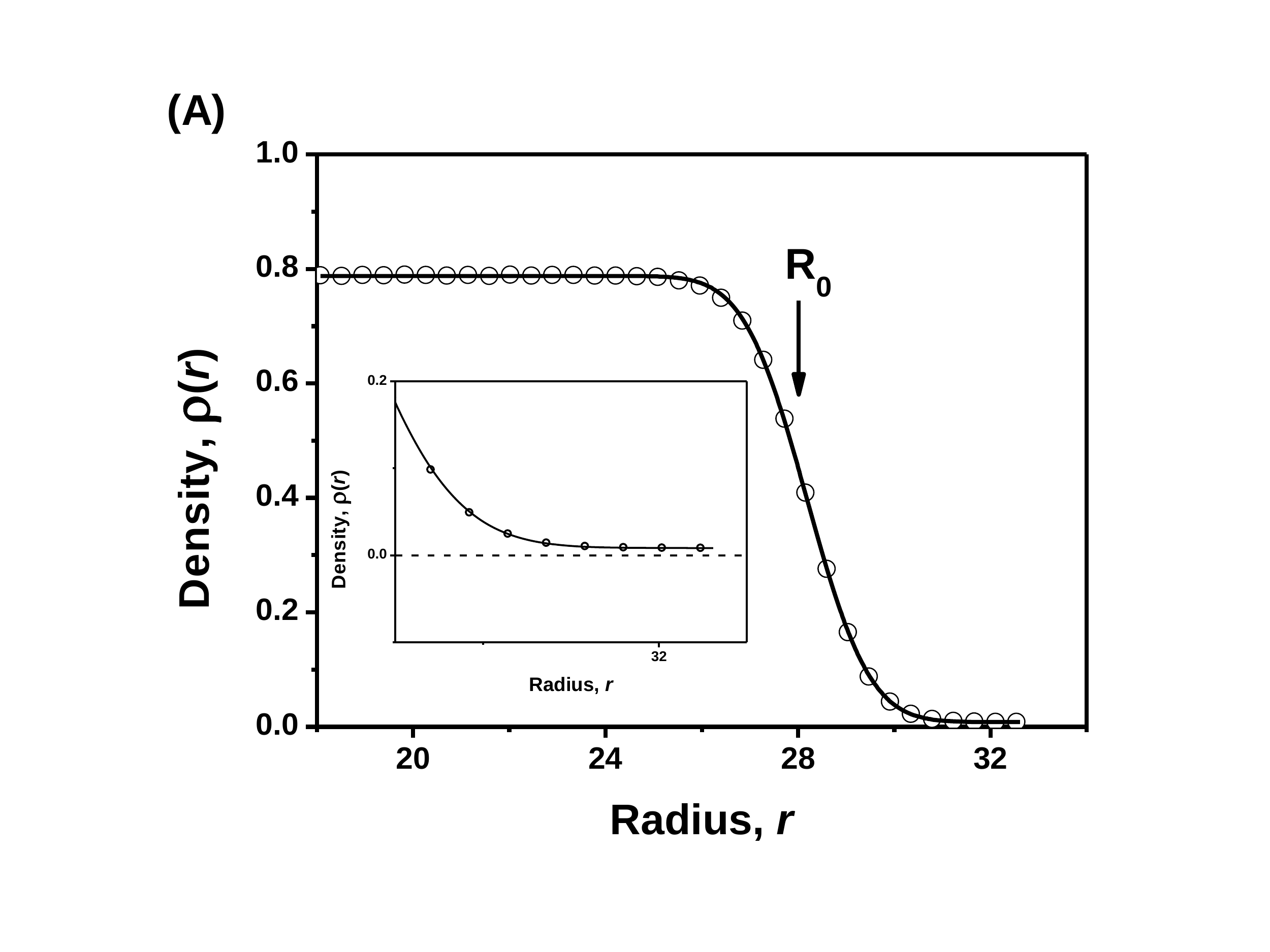}
\includegraphics[trim=1cm 1cm 1cm 1cm,width=0.4\columnwidth]{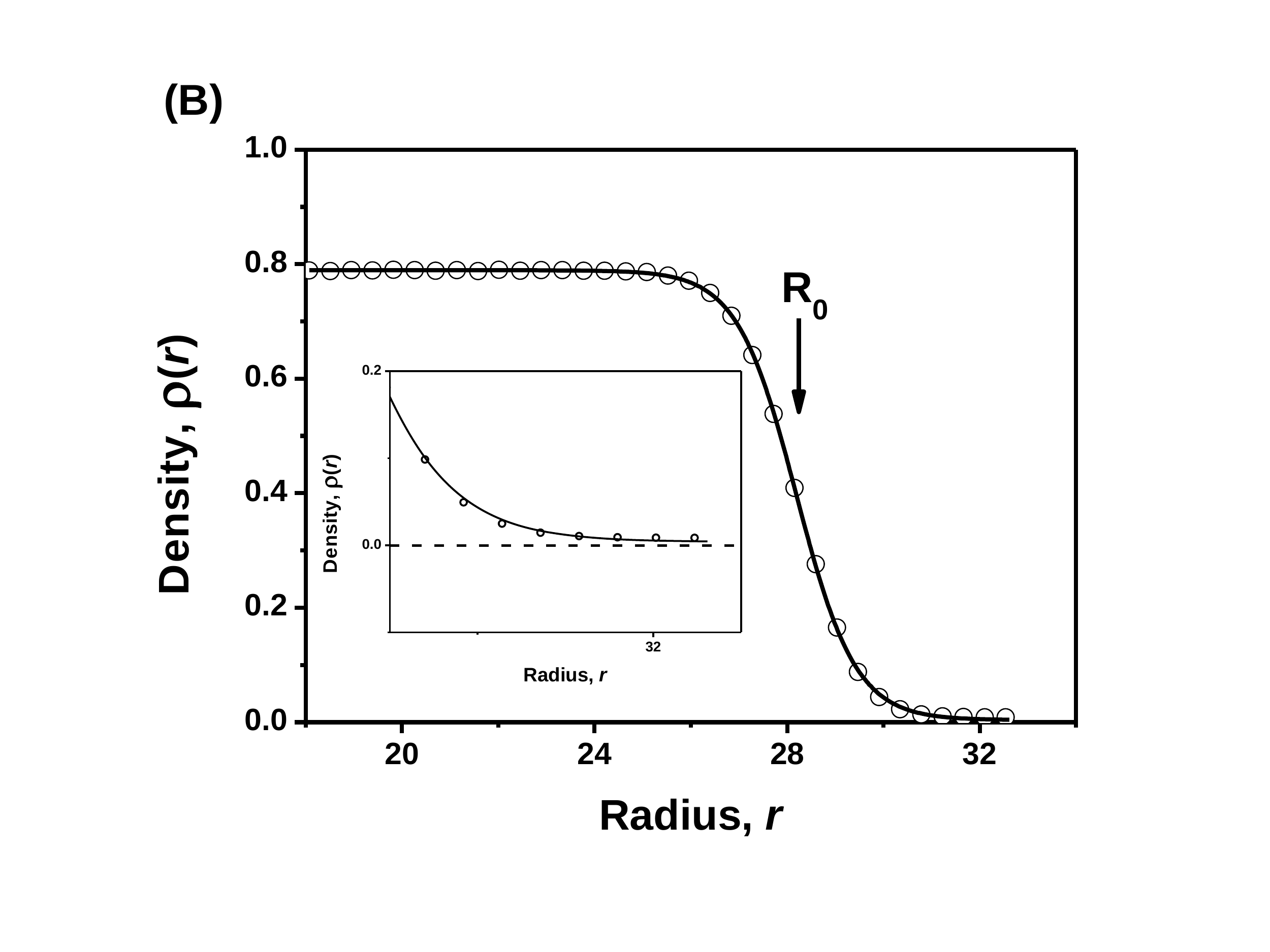}
\end{center}
\caption{Illustration of the density profile in a monatomic liquid drop consisting of 75000 particles at $T=0.7$ as a function of the reduced radius $r$ measured from the centre of mass. The result of MD simulations is shown by symbols (circles, only every 6th point is shown), while the solid lines designate (a) the four-parameter fit (\ref{fit-den}) at $\rho_L=0.7878\pm0.0002$, $\rho_G=0.0083\pm0.0004$, $R_0=28.186\pm 0.001$, $\Delta=1.032\pm 0.002$, and (b) the four-parameter fit (\ref{tanh}) at $\rho_L=0.7892\pm 0.0002$, $\rho_G=0.0039\pm 0.0004$, $R_0=28.194\pm 0.002$, $\Delta=1.116\pm 0.002$. The insets show a magnified view of the tail region.} \label{Fig1}
\end{figure}

\begin{figure}
\begin{center}
\includegraphics[trim=1cm 1cm 1cm 1cm,width=0.35\columnwidth]{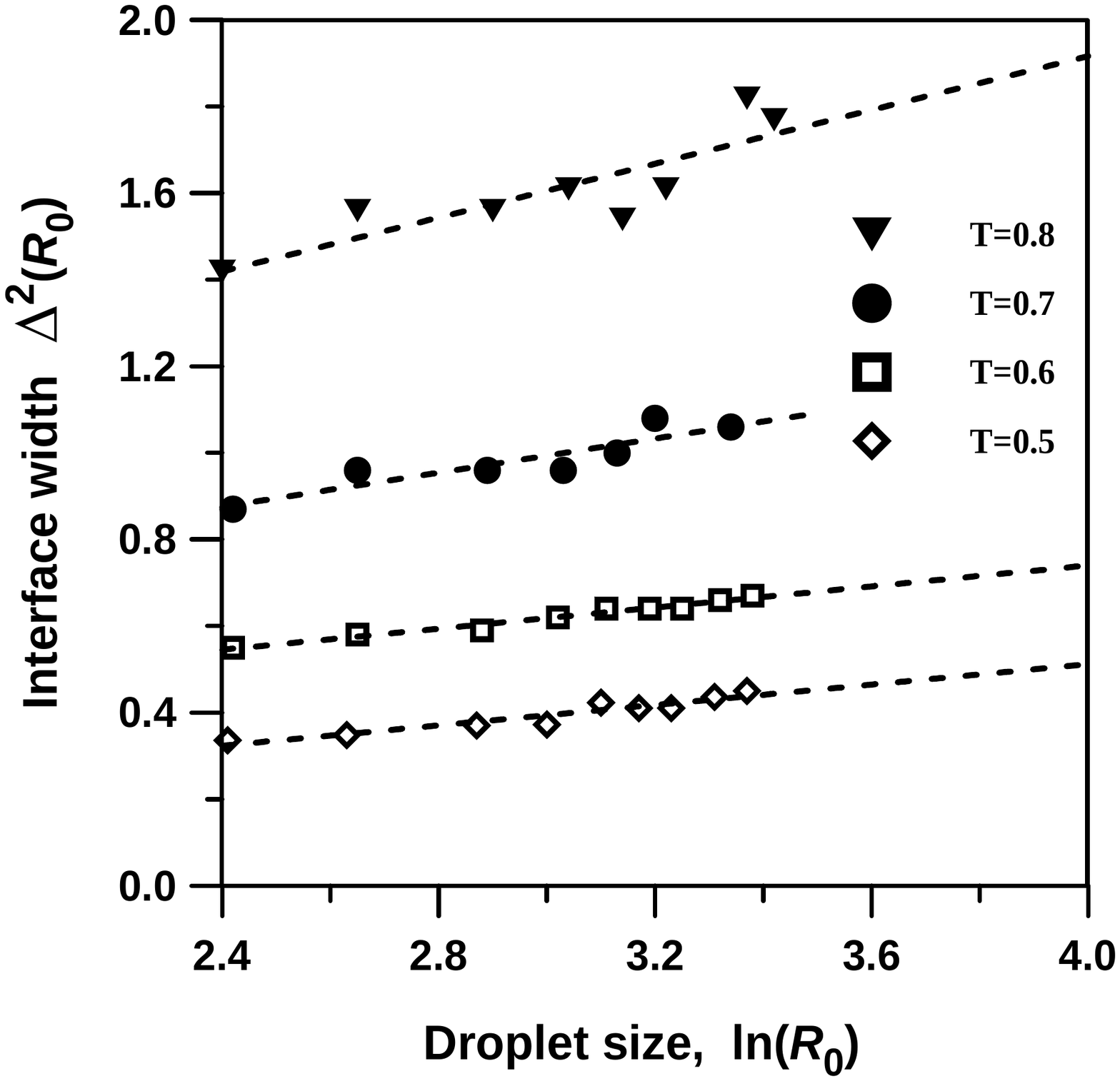}
\end{center}
\caption{Interfacial width, $\Delta^2$, as a function of the droplet size $\ln(R_0)$ at different temperatures. The results of MD simulations are shown by symbols and the dashed line is the two-parameter fit (\ref{LogDep}).} \label{Fig2}
\end{figure}

Typical equilibrium density profile $\rho(r)$ in a monatomic LJ drop consisting of 75000 identical particles, obtained as a result of averaging over $\Delta t_a =1000$ at $T=0.7$, is shown in Fig. \ref{Fig1}. The temperature $T=0.7$ is slightly above the value at the triple point of LJ liquids $T_t=0.68$ [43]. The observed values of the bulk $\rho_L=0.788$ and the vapour $\rho_G=0.008$ densities are in a good agreement with the values observed in similar conditions [37, 41].  As is seen in Fig. \ref{Fig1}, the calculated density profiles can be accurately approximated by means of an error function 
\begin{equation}
\rho(r)=\frac{1}{2}(\rho_L+\rho_G)-\frac{1}{2}(\rho_L-\rho_G)\, \mbox{erf}  \left( \frac{r-R_0}{\sqrt{2}\Delta} \right), 
\label{fit-den}
\end{equation}
where parameter $\Delta$ represents the interfacial thickness (half of the actual "visible" size of the interface) and $R_0$ is the average position of the interface (equimolar surface in terms of excess density). The origin of this approximation derives from the capillary wave theory, [44--50], and is used in this study for parametrisation of the density profiles. Alternatively, the density profile can be also approximated by a hyperbolic tangent fit
\begin{equation}
\rho(r)=\frac{1}{2}(\rho_L+\rho_G)-\frac{1}{2}(\rho_L-\rho_G)\, \mbox{tanh}  \left( \frac{\pi}{\sqrt{12}} \frac{r-R_0}{\Delta} \right), 
\label{tanh}
\end{equation}
which is based on a mean-field approach, [30].  Although it has been noted in [8] that the fit given by eq. (\ref{tanh}) may produce a less accurate approximation than that given by eq. (\ref{fit-den}), in our case we have not observed essential differences, Fig. \ref{Fig1}. That is, in a liquid drop consisting of $75000$ particles at $T=0.7$, the four-parameter fit (\ref{fit-den}) produces $\rho_L=0.7878\pm0.0002$, $\rho_G=0.0083\pm0.0004$, $R_0=28.186\pm 0.001$, $\Delta=1.032\pm 0.002$ and the four-parameter fit (\ref{tanh}) gives $\rho_L=0.7892\pm 0.0002$, $\rho_G=0.0039\pm 0.0004$, $R_0=28.194\pm 0.002$, $\Delta=1.116\pm 0.002$. The only difference observed is the value of the gas density $\rho_G$.  But, as one can see from the magnified view of the tail region (insets in Fig. \ref{Fig1}), it is the error function fit which produces much better approximation of this region. On the other hand, if parameter $\rho_G$ is fixed in the fitting procedure (\ref{tanh}), the other three parameters have shown only very weak variations with $\rho_G$ within $0.003\le \rho_G\le 0.008$. For example, at $\rho_G=0.008$, the four-parameter fit (\ref{tanh}) gives $\rho_L=0.7890\pm 0.0003$,  $R_0=28.184\pm 0.003$, $\Delta=1.102\pm 0.003$. So the observed difference can only indicate that the accuracy of (\ref{tanh}) in the tail region is lower than that of (\ref{fit-den}). We believe that the observed difference in approximation is not important, but the preference in the current study to use the error function fit (\ref{fit-den}) in the data profiling comes actually from the observed dependence of the interfacial width on the droplet size $R_0$, Fig. \ref{Fig2}, which is the characteristic feature according to the capillary wave theory, see further discussion. We will briefly return to this issue of the fitting procedure later in the manuscript, in the part dedicated to non-equilibrium situations for comparison. We will see, that the evolution curves of the density profile width are identical within the approximation error and only characteristic time scales are weakly dependant of the choice of the fitting function. 

According to the capillary wave theory, the observed interfacial density profile $\rho(r)$ is apparent and conditioned by the widening of an "intrinsic" density profile $\psi(r)$ by a spectrum of thermally excited capillary waves, that is by the fluctuations of density on the length scale larger than the width of the interface. Under the assumption that the density fluctuations on the large length scale are decoupled from the short length scale density fluctuations and neglecting effects of curvature (assuming  $R_0>>\Delta$) the density profile $\rho(z)$ can be represented as the convolution of $\psi(z)$ and the probability to find the interface at a position $z$, $P(z)$, centred at $z=0$,   
$$
\rho(z)=\int_{-\infty}^{\infty}\, \psi(z-z_0) P(z_0)\, dz_0 .
$$
The particular form of the density distribution (\ref{fit-den}) can be obtained exactly in the case of the Gaussian probability distribution function 
$$
P(z_0)=\frac{1}{\sqrt{2\pi \Delta^2}}\exp\left(-\frac{z_0^2}{2\Delta^2}\right)
$$ 
and infinitely sharp intrinsic profile
$$
\psi(z)=\rho_L \Theta(R_0-z) + \rho_G \Theta(z-R_0),
$$ 
where $\Theta(z)$ is the Heaviside step function. 

In general, though, when $\psi$ is not just a step function, the apparent interfacial width defined as $$
\Delta^2=\frac{\int_{-\infty}^{\infty} (z-R_0)^2 \rho^{'}(z)\, dz}{\int_{-\infty}^{\infty} \rho^{'}(z) \, dz}
$$
consists of two components 
$$ 
\Delta^2=\frac{\int_{-\infty}^{\infty} (z-R_0)^2 \rho^{'}(z)\, dz}{\int_{-\infty}^{\infty} \rho^{'}(z) \, dz}=\frac{\int_{-\infty}^{\infty} (z-R_0)^2 \psi^{'}(z)\, dz}{\int_{-\infty}^{\infty} \psi^{'}(z) \, dz}+\frac{\int_{-\infty}^{\infty} z_0^2 P(z_0)\, dz_0}{\int_{-\infty}^{\infty} P(z_0) \, dz_0}=\Delta_0^2+\Delta_c^2,
$$  
provided that $\displaystyle \int_{-\infty}^{\infty} (z-R_0)\psi^{'}(z)=0$, that is $\psi^{'}(z)$ is a symmetric function around $z=R_0$, $\psi^{'}(R_0+\xi)=\psi^{'}(R_0-\xi)$. 
Here $\rho^{\prime}$ and $\psi^{\prime}$ are used to designate the first derivatives $\frac{d\rho}{dz}$ and $\frac{d\psi}{dz}$.
The first contribution, 
$$
\displaystyle \Delta_0^2=\frac{\int_{-\infty}^{\infty} (z-R_0)^2 \psi^{'}(z)\, dz}{\int_{-\infty}^{\infty} \psi^{'} (z) \, dz},
$$ can be interpreted as the "intrinsic width" of the interfacial layer, which is a manifestation of the thermal fluctuations on the length scale shorter or comparable with $\Delta_0$, and the second contribution, 
$$
\displaystyle \Delta_c^2=\frac{\int_{-\infty}^{\infty} z_0^2 P(z_0)\, dz_0}{\int_{-\infty}^{\infty} P(z_0) \, dz_0},
$$ is due to the capillary waves [36, 45, 49]. 

At equilibrium, the interfacial width $\Delta^{(eq)}$ has a weak logarithmic dependence on the size of the droplet $R_0$ with the coefficient depending on the equilibrium value of the surface tension $\gamma_0$ [36]
\begin{equation}
\Delta^{(eq)2}=\Delta_0^{(eq)2}+\sum_{l=2}^{l_{max}}\bar{a^2_c}\simeq \Delta_0^{(eq)2}+\frac{T}{2\pi \gamma_0} \ln(R_0).
\label{LogDep}
\end{equation}
In eq. (\ref{LogDep}), the summation is over all spherical harmonics with the mode number $l$ running in between $2\le l\le l_{max}\approx \pi R_0$, with the frequency
\begin{equation}
\omega_c^2=\gamma_0 \frac{l(l-1)(l+2)}{\rho R_0^3}
\label{CapillaryWaveW}
\end{equation}
and the mean square amplitude
\begin{equation}
\bar{a^2_c}=\frac{T}{4\pi \gamma_0}\frac{2l+1}{(l-1)(l+2)}.
\label{CapillaryWaveA}
\end{equation}
This effect of the capillary wave motion has been clearly observed and studied in MD simulations and in the experiments on simple fluids [8, 36, 48, 49, 51-55]. In particular, MD simulations of nanoscale capillary waves (with amplitude $\approx 0.3$ particle diameters) in liquid argon drops have decisively demonstrated that capillary wave motion closely follows predictions of continuum hydrodynamics (frequency and damping coefficient), the observed temperature dependence of the density profile width is in a very good agreement with experiments on liquid argon [51, 52]. The width of the density profile was found to be within $1.15\le \Delta^{(eq)}\le 1.75$ for $0.81\le T\le 0.97$  for a drop of $R_0\approx 46$. At $T\approx0.8$, this width $\Delta=1.15$ is slightly smaller than the one $\Delta=1.32$ observed in our MD simulations ($R_0\approx 28$) due to larger cut-off distance $r_c=4\,\sigma$ of LJ potential introduced in [51] to reproduce real argon. This effect of the LJ potential cut-off  has been clearly observed in [41], due to, in particular, larger surface tension values leading to smaller contribution from the capillary wave widening. The effects of capillary waves at a water-vapour interface have been studied in detail in [8], confidently establishing that at $T=300\,\mbox{K}$ the interface width is $\Delta^{(eq)}\approx0.125\,\mbox{nm}$ and the intrinsic width $\Delta_0^{(eq)}\approx0.08\,\mbox{nm}$, that is about $0.46$ and $0.29$ of the water molecule diameter ($0.275\,\mbox{nm}$) respectively. A similar study of capillary waves at interfaces between coexisting phases in
polymer blends has been conducted in [49].

\begin{table}[htbp]
\begin{tabular}{ | c | c | c | c |  c | c | c | c | c |}
\hline
  $T$ &  $\gamma^{\dag}_0$  & $\gamma_0$ & $\Delta_0^{\dag(eq)}$  & $\Delta^{(eq)}$  & $\Delta_0^{(eq)}$ \\  
\hline
  0.5 &   $0.88\pm 0.09$ & $0.93\pm 0.03$  & $0.32\pm 0.06$  &  $0.63\pm0.003$ &  $0.35\pm0.01$\\ 
\hline
  0.6 & $0.78\pm 0.05$ & $0.73\pm 0.02$  & $0.50\pm 0.03$  &  $0.80\pm0.003$   &  $0.50\pm0.01$ \\ 
\hline
  0.7 &  $0.56\pm 0.05$ & $0.55\pm 0.02$  & $0.63\pm 0.05$  &  $1.01\pm0.003$  &  $0.63\pm0.02$ \\ 
\hline
  0.8 &   $0.44\pm 0.06$ & $0.36\pm 0.02$  & $0.84\pm 0.06$  &  $1.32\pm0.003$ &  $0.82\pm0.04$ \\ 
\hline                   
\end{tabular}
\caption{Surface tension $\gamma_0$, intrinsic width $\Delta_0^{(eq)}$ and total width $\Delta^{(eq)}$ at equilibrium in monatomic LJ liquid drops at different temperatures. 
The total interfacial width $\Delta^{(eq)}$  has been obtained using a four-parameter fit, eq. (\ref{fit-den}), for a drop consisting of $\approx 40000$ particles, that is at $R_0\approx 22$.
Note, the parameters marked by $\dag$ have been calculated by means of a two-parameter fit, eq. (\ref{LogDep}), while $\gamma_0$ has been obtained by direct MD simulations in this study and 
$\Delta_0^{(eq)}$ has been calculated from $\Delta^{(eq)}$ at $R_0=22$ using $\gamma_0$ and (\ref{LogDep}).}
\vspace{12pt}
\label{Table1}
\end{table} 

The effect of the capillary waves is also clearly observed in our MD simulations, as is illustrated in Fig. \ref{Fig2}, where the interface width is shown as a function of drop size $R_0$. A logarithmic two-parameter fit to that dependence (Fig. \ref{Fig2}) given by eq. (\ref{LogDep}) results in, for example at $T=0.7$,  $\Delta_0^{\dag(eq)}=0.63\pm 0.05$ and  $\gamma^{\dag}_0=0.56\pm 0.05$. This value of the surface tension $\gamma^{\dag}_0$ obtained from the fitting procedure, eq. (\ref{LogDep}), is very close to $\gamma_0=0.55$, directly obtained in [40] and in this study using MD simulations in similar conditions at $r_c=2.5$ and $T=0.7$. The accuracy of $\Delta_0^{\dag(eq)}$ calculation appears to be lower than that of directly obtained values of $\Delta_0^{(eq)}$, though the obtained values are essentially the same. But, one needs to note that the latter does not probably include all approximation errors coming from calculation of the density profiles from MD simulations, while the former probably does account for that. For consistency, we will use directly obtained values of $\Delta_0^{(eq)}$ in what follows. Note, that in our simulations, we have been able to go below the temperature at the triple point of monatomic LJ liquids, $T_t$, without crystallisation to expand the range of parameters, such as viscosity of the liquid. To ensure that crystallisation does not occur in the simulations, the state of the liquid has been monitored by means of the mean square displacement of particles both in the whole drop and in the equivalent conditions in the bulk phase with periodic boundary conditions 
\begin{equation}
R^2_{msd}(t)=<\frac{1}{N_p} \sum_{i} ({\bf r}_i(t)-{\bf r}_i(0))^2>,
\label{RMSD}
\end{equation}
where $N_p$ is the total number of particles. 

The obtained values of "intrinsic width" $\Delta_0^{(eq)}$ and  surface tension $\gamma_0$ at different temperatures are presented in Table \ref{Table1}. One can see that in general the equilibrium intrinsic width $\Delta_0^{(eq)}$ is of the order of less than one particle diameter and is monotonically increasing function of temperature. The value of the surface tension $\gamma_0^{\dag}$ calculated from the logarithmic fit deviates in general by less than 10\% from the values obtained independently and in this study by direct methods. In what follows, we will use this approach derived from the capillary wave theory in the analysis of relaxation of the interfacial phase, although with caution, since the separation of the density fluctuations  into two parts is artificial and the density profile and the state of the surface phase are the result of the full spectrum of density fluctuations [47].

\subsection{Calculation of surface tension in the drops}
Instead of using empirical fit eq. (\ref{LogDep}), the value of the surface tension generated in the interfacial layer of a drop is calculated in our MD experiments in the planar interface limit  $\Delta/R_0<<1$ directly from the microscopic stress tensor $T_{\alpha\beta}$ by means of  
\begin{equation}
\gamma=\int_0^{\infty} \frac{R_s}{r}(T_T-T_N) \, dr \approx \int_0^{\infty} (T_T-T_N) \, dr,
\label{SurfTens}
\end{equation}
where $T_T$ and $T_N$ are the tangential and normal components of the stress tensor respectively, $R_s$ is the surface of tension $R_s\approx R_0$. For example, if the interface is flat and is perpendicular to $z$-axis, $T_N=T_{zz}$ and $T_T=\frac{T_{xx}+T_{yy}}{2}$. 
The microscopic stress tensor components $T_{\alpha\beta}$ have been calculated through the procedure developed in [29, 37, 56], the details of the method can be found in the Appendix. 

\begin{figure}
\begin{center}
\includegraphics[trim=1cm 1cm 1cm 1cm,width=0.5\columnwidth]{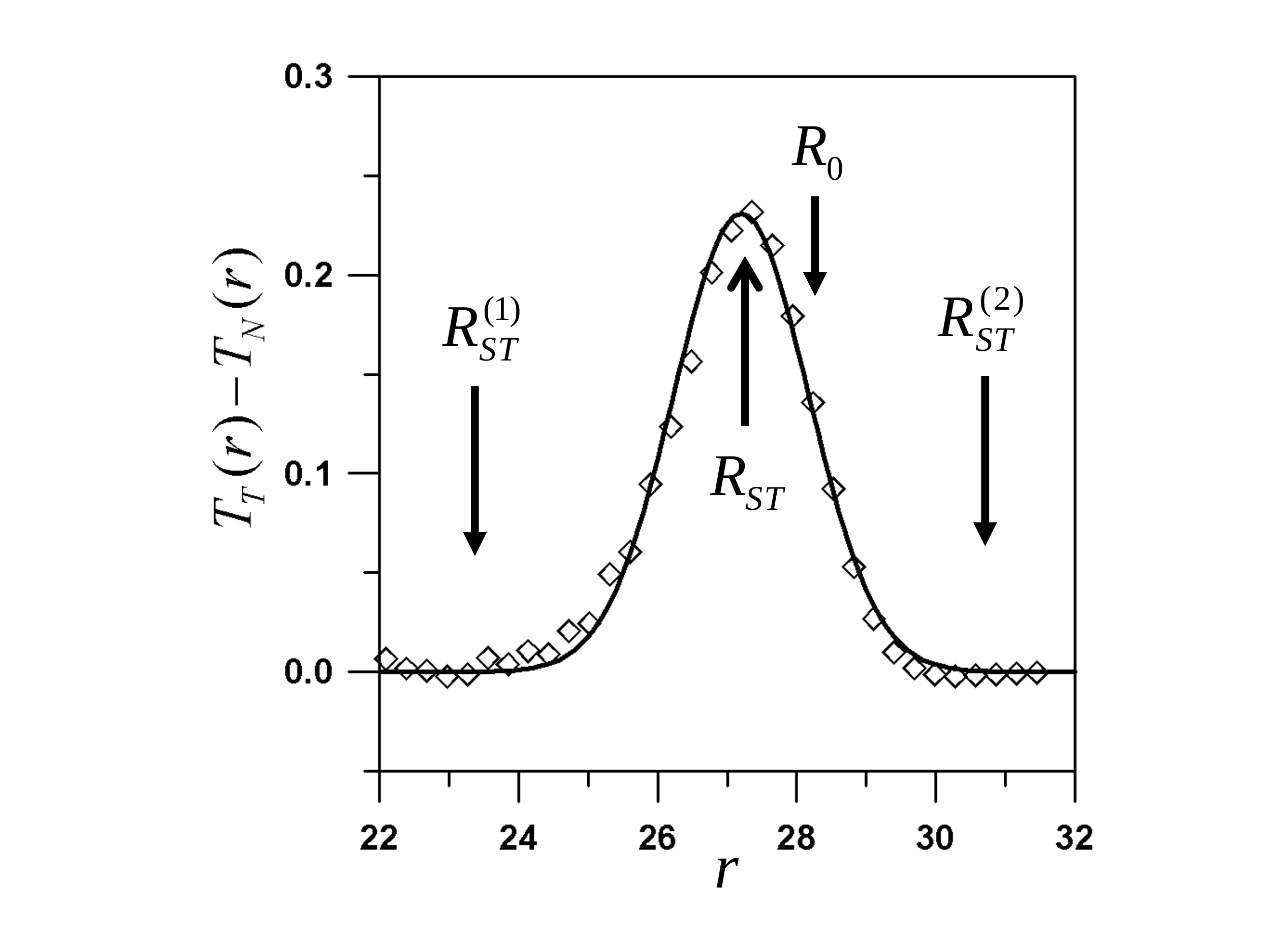}
\end{center}
\caption{Distribution of $T_T({\bf r})-T_N({\bf r})$ in a monatomic liquid drop consisting of 75000 particles at $T=0.7$. The result of MD simulations is shown by symbols and the solid line is the Gaussian fit (\ref{GaussFit}) at $A_{ST}=0.56\pm 0.01$, $\Delta_{ST}=0.97\pm 0.02$ and $R_{ST}=27.20\pm 0.02$. The average positions of the stress distribution $R_{ST}$ and the density profile $R_0$ are indicated by the arrows.} \label{Fig4}
\end{figure}

In practice, the actual integration in eq. (\ref{SurfTens}) takes place in the interval $r\in [R_{ST}^{(1)}, R_{ST}^{(2)},]$, that is between the points where the integrand $T_T({\bf r})-T_N({\bf r})$ vanishes, see Fig. \ref{Fig4}. 

The definition of the surface tension through the integral (\ref{SurfTens}) results from a mechanical argument, which is thermodynamically well-defined in the macroscopic limit ($\Delta/R_0\to0$) and valid in case of sufficiently slow motion, when, should we neglect the inertia of the surface layer, the general condition of mechanical equilibrium is fulfilled 
$$
\frac{\partial T_{\alpha\beta}}{\partial x_{\alpha}}=0.
$$
In case of a spherically symmetric geometry, this condition is reduced to
\begin{equation}
T_T(r)-T_N(r)=\frac{r}{2}\frac{dT_N}{dr}.
\label{BalanceSph}
\end{equation}
Equation (\ref{BalanceSph}) is the necessary condition for the correct evaluation of surface tension in drops, which should be always monitored under non-equilibrium conditions.

The distribution of stresses $T_T({\bf r})-T_N({\bf r})$ in the radial direction can be accurately approximated by a Gaussian function 
\begin{equation}
T_T({ r})-T_N({ r})=\frac{ A_{ST} }{  \sqrt{2\pi \Delta_{ST}^2} }\exp\left( -\frac{(r-R_{ST})^2}{2\Delta_{ST}^2}  \right),
\label{GaussFit}
\end{equation}
where $R_{st}$ is the point of maximum of the distribution $T_T({r})-T_N({r})$, Fig. \ref{Fig4}.

The fit, eq. (\ref{GaussFit}), gives for a drop consisting of 75000 particles at equilibrium at $T=0.7$ (this case is identical to the one shown in Fig. \ref{Fig1}) $A_{ST}=\gamma_0=0.55\pm 0.02$, $\Delta_{ST}^{(eq)}=0.97\pm 0.02$ and $R_{ST}=27.20\pm 0.02$. As one can see from Fig. \ref{Fig4}, the maximum of the distribution of $T_T({r})-T_N({r})$ is shifted into the drop interior by $\delta r\approx 1$, which is a typical value in our simulations. That distance is close to the values of Tolman length $\delta_{Tl}\sim 0.8$ reported in [32, 33] and could be interpreted as the distance between equimolar surface and the surface of tension, though, as is known, the later is not well-defined thermodynamically and path dependant [29, 30, 56]. The obtained width of the profile of $T_T({r})-T_N({r})$, $\Delta_{ST}^{(eq)}$, if we were to assume that it is also a result of capillary wave widening, allows to calculate intrinsic contribution $\Delta_{ST}^{0(eq)}$. At $T=0.7$, this intrinsic width is $\Delta_{ST}^{0(eq)}=\sqrt{\Delta_{ST}^{(e)2}-\frac{T}{2\pi\gamma_0}\ln(R_{ST})}=0.52$  obtained from eq. (\ref{LogDep}). 
So, in general, the intrinsic width of the stress difference distribution is on the same length scale as the density profile intrinsic width $\Delta_0^{(eq)}$, Tables \ref{Table1} and \ref{Table3}. Note however that unlike the distribution of density, distribution of stresses  $T_T({r})-T_N({r})$ has no direct physical meaning, since it depends on the choice of the contour connecting interacting particles [29, 30, 56], see Appendix for details. So those values of $\Delta_{ST}^{0(eq)}$ can only be used to check the consistency of the results. On the other hand, the integral (\ref{SurfTens}) of that distribution is invariant of the choice of contour in the leading order of the limit of a planar interface $\Delta/R_0\to0$, as expected, since in macroscopic limit it represents the measurable quantity  - surface tension. The equilibrium values of the distribution parameters, given by eq. (\ref{GaussFit}) at different temperatures are presented in Table \ref{Table3} for liquid drops consisting of  $ \approx 40000$ particles.

\begin{table*}[htbp]
\begin{tabular}{ | c | c | c | c | c | c |  c | c | c |}
\hline
  $T$ & $A_{ST}^{(eq)}$ & $\Delta_{ST}^{(eq)}$ & $\Delta_{ST}^{0(eq)}$ &  $A_{ST}(t=1)$  & $A_{ST}(t=2)$ & $A_{ST}(t=3)$ & $\Delta_{ST}(t=3)$ & $\tau_{\gamma}$ \\  
\hline
  0.5 &  $0.94\pm 0.01$  & $0.70\pm 0.01$ & $0.46\pm 0.01$ &  $0.59 \pm 0.02$ & $0.87\pm 0.02$ & $0.88\pm 0.02$  &  $0.44\pm 0.02$ & $1.1\le \tau_{\gamma}\le 2$\\ 
\hline
  0.6 & $0.74\pm 0.01$  & $0.80\pm 0.01$ & $0.46\pm 0.01$  &  $0.51\pm 0.02$ & $0.73\pm 0.02$ & $0.71\pm 0.02$  & $0.50\pm 0.02$  & $1.7\pm 0.8$\\ 
\hline
  0.7 & $0.56\pm 0.01$  & $0.97\pm 0.02$ & $0.52\pm 0.02$  &  $0.42\pm 0.01$& $0.56\pm 0.02$ & $0.55\pm 0.02$  &  $0.57\pm 0.02$ & $1.9\pm 0.3$\\ 
\hline
  0.8 &  $0.37\pm 0.01$ & $1.19\pm 0.03$  & $0.50\pm 0.03$ &  $0.47\pm 0.02$ & $0.41\pm 0.02$ & $0.40\pm 0.02$   & $0.64\pm 0.03$ & $6.6\pm 0.6$ \\ 
\hline                   
\end{tabular}
\caption{Coefficients $A_{ST}$ and $\Delta_{ST}$ of the fit (\ref{GaussFit}) at equilibrium and shortly after the cut off (at $t_{\mbox{\tiny cut} \mbox{\tiny off}}=0$), intrinsic width $\Delta_{ST}^{0(eq)}$ at equilibrium and characteristic time $\tau_{\gamma}$ in monatomic LJ liquid drops consisting of $\approx 40000$ particles at different temperatures.}
\vspace{12pt}
\label{Table3}
\end{table*} 

In the next part, we consider evolution and recreation of interfaces from an initially infinitely sharp density profile and apply described density profile fitting and surface tension calculation techniques at non-equilibrium conditions. 

\section{Interfacial dynamics of monatomic liquids}

To study dynamic properties of liquid-gas interfacial layers, we create a fresh sharp interfacial surface by removing all particles at $r>R_{ST}^{(1)}$ (Fig. \ref{Fig4}) in an initially equilibrated drop consisting of 75000 particles, so that after the cut off the drop size is reduced to approximately 40000 particles, Fig. \ref{Fig6}a,b. We can then observe the recreation process of a new interfacial layer and measure its properties directly after the cut off as functions of time to reveal relevant relaxation times. 

In this study, we are primarily interested in the relaxation process in isothermal conditions.  But, even that the DPD thermostat friction introduces only negligible corrections, we have done control runs for the calculations presented in the paper at non-equilibrium conditions with the thermostat switched off to understand if the results would be qualitatively different. We have found no substantial evidence of any influence from the thermostat on the data so far, which can give rise to qualitatively different results. Although, a detailed study of the relaxation process in non-isothermal conditions wil be done later.

There are several characteristic time scales that one can anticipate here. First of all, it is the stress relaxation time $\tau_{stress}$, which is usually associated with Non-Newtonian behaviour of liquids in classical hydrodynamics. This characteristic time can be calculated in the bulk from the off-diagonal ($\alpha\neq \beta$) stress-stress correlation function, see for example [57],
$$
G(t)=\frac{V}{T}<T_{\alpha\beta}(t)T_{\alpha\beta}(0)>.
$$ 
The integral of this correlation function also provides the value of the zero shear rate viscosity in the liquid at the bulk conditions.
The values of the longest stress relaxation time in monatomic LJ liquids and corresponding viscosities at different temperatures are summarised in Table. \ref{Table2}. We note that the stress relaxation time is smaller than $1$ in all cases.

If we were to assume that the density fluctuations on the large and small length scales at the interface are decoupled, then one can expect another two separate time scales: one associated with the intrinsic width of the interfacial density profile $\Delta_0^{(eq)}$ and another one associated with the excitation of capillary waves. In the later case we may expect a full spectrum of capillary waves, which depends on the size of the droplet $R_0$ and is quantified by the mode number $l$, eq. (\ref{CapillaryWaveW}). The largest contribution to $\Delta$ has to come from the mode with $l=2$, which has the maximum amplitude at equilibrium, eq. (\ref{CapillaryWaveA}).  

The excitation time of the capillary waves $t_{c}$
is defined for each mode by the wave damping coefficient $\eta_c$ [58],
$$
\eta_c=\frac{2\mu \omega_c^{4/3}}{\rho^{1/3}\gamma_0^{2/3}}.
$$
In the underdamped regime, $\eta_c^2-4\omega_c^2<0$,
\begin{equation}
t_{cu}=\eta_c^{-1},
\label{underD}
\end{equation}
and in the overdamped regime $\eta_c^2-4\omega_c^2>0$,
\begin{equation}
t_{co}=\frac{2}{\eta_c-\sqrt{\eta_c^2-4\omega_c^2}}.
\label{overD}
\end{equation}
That is, from (\ref{underD}) and (\ref{overD}), for the mode $l=2$ with the dominant contribution into $\Delta$ 
\begin{equation}
t_{cu}^{max}=\frac{ \rho R_0^2}{8\mu},
\label{underD2}
\end{equation}
and 
\begin{equation}
t_{co}^{max}=\frac{\rho R_0^2}{4\mu}\frac{1}{1-\sqrt{1-\frac{\rho \gamma_0 R_0}{2\mu^2}}},
\label{overD2}
\end{equation}
which is in the limit of strong damping $\frac{\rho \gamma_0 R_0}{2\mu^2}<<1$
$$
t_{co}^{max}\approx \frac{R_0 \mu}{\gamma_0}.
$$
The later implies that $t_{c}$ has different scaling with viscosity in the underdamped and overdamped regimes.

For monatomic liquid drops, the mode $l=2$ is either in the underdamped or overdamped regime. 
For example at $T=0.7$ for a liquid drop  consisting of $\approx 40000$ particles ($R_0\approx 22$), the characteristic frequency is  $\omega_c\approx 0.023$, $\eta_c\approx 0.036$ and thus $\eta_c^2-4\omega_c^2\approx -0.0008<0$, but at $T=0.5$, $\eta_c^2-4\omega_c^2\approx 0.003 >0$. The surface waves relaxation time then will be found in the range $21 \le t_{cu,o}^{max} \le 41 $ depending on the temperature  of the liquid, see Table \ref{Table2}. The estimated value of the damping coefficient $\eta_c=0.027$ at $T=0.8$ is close to $\eta_c=0.022$ actually observed in MD experiments on capillary waves in monatomic LJ nano droplets [51]. Note that in the case of binary liquids studied later in the second part of our work, the capillary wave motion will be in the overdamped regime.

\begin{figure}
\begin{center}
\includegraphics[trim=1cm 1cm 1cm 1cm,width=0.5\columnwidth]{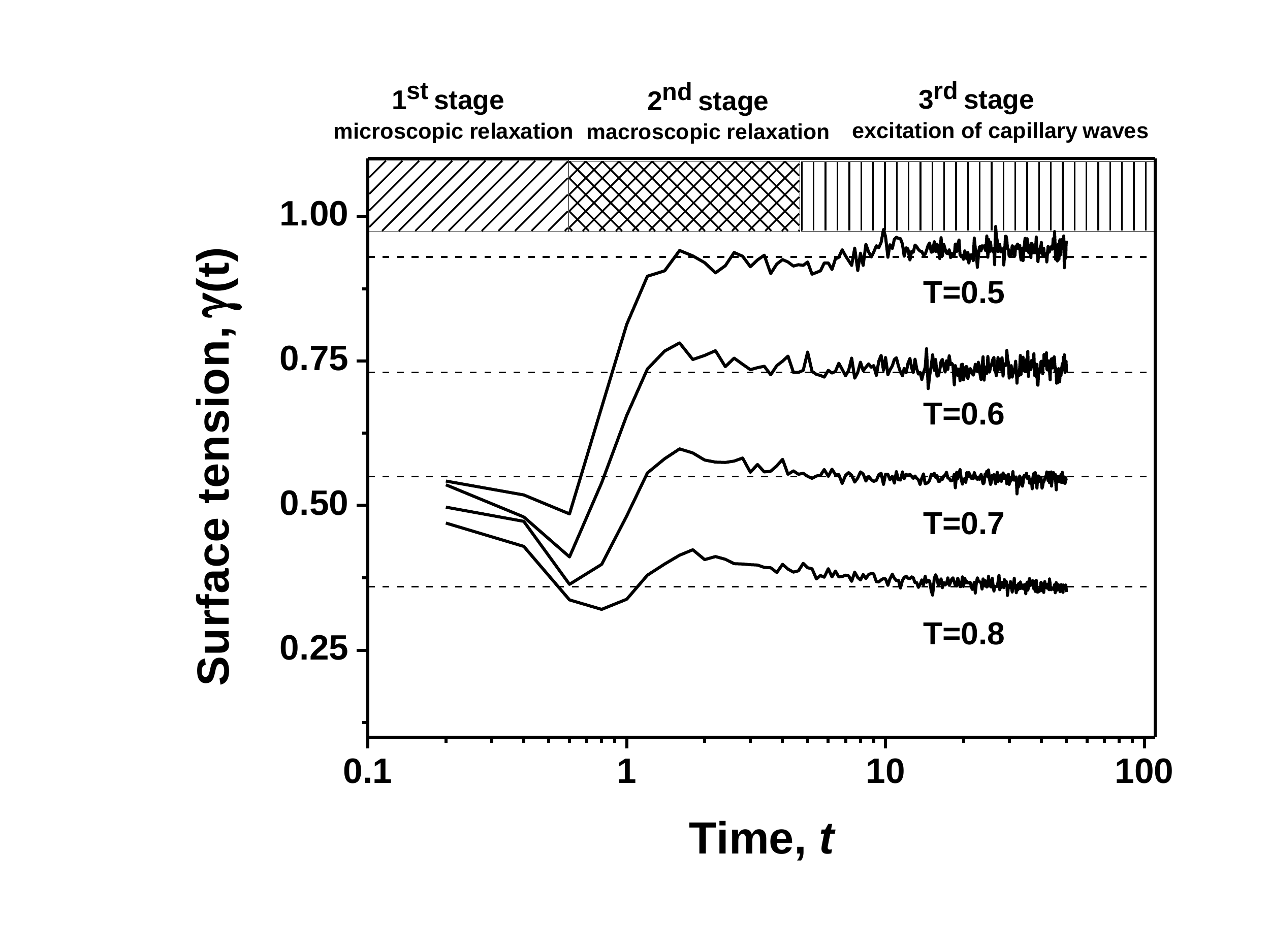}
\end{center}
\caption{Evolution of the surface tension $\gamma(t)$ as a function of time in monatomic LJ liquid drops consisting of approximately 40000 particles after the cut off at $t=0$ at different temperatures. The dashed lines correspond to equilibrium values of the surface tension.} \label{Fig7}
\end{figure}

\begin{figure}
\begin{center}
\includegraphics[trim=1cm 1cm 1cm 1cm,width=0.5\columnwidth]{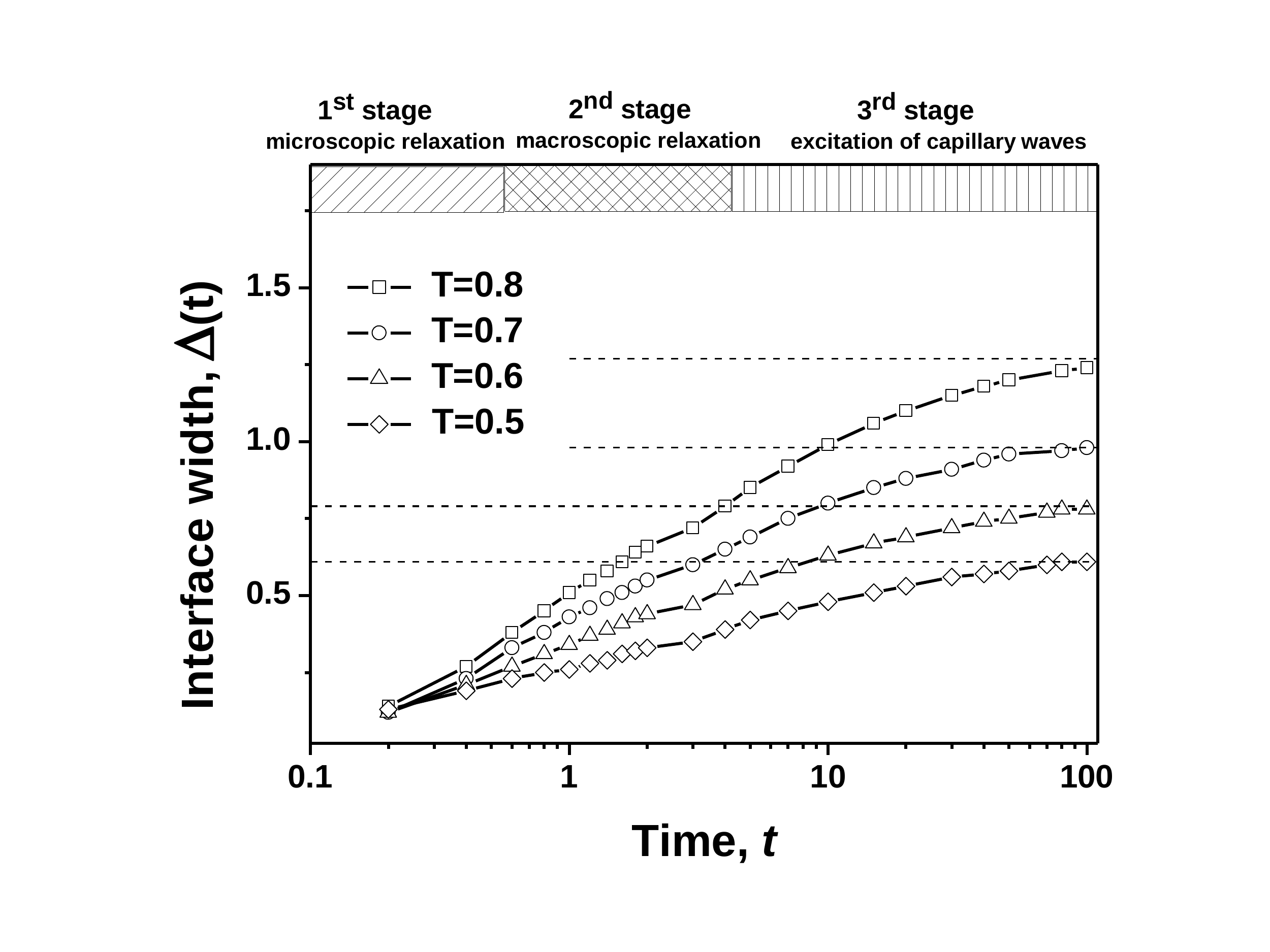}
\end{center}
\caption{Evolution of the width of the density profile $\Delta(t)$ as a function of time in monatomic LJ liquid drops consisting of approximately 40000 particles after the cut off at different temperatures $T$. The dashed lines correspond to equilibrium width of the density profiles, $\Delta^{(eq)}$. The relative error in obtaining $\Delta$ was about $3\%$.} \label{Fig8}
\end{figure}

Consider now results of MD simulations at non-equilibrium conditions. Evolution of  surface tension and  the density profiles after the cut off (at $t_{\mbox{\tiny cut} \mbox{\tiny off}}=0$, which is always the case unless otherwise stated) is shown in  Figs. \ref{Fig7}--\ref{Fig8} at different temperatures. 
The observed values of surface tension have been averaged over 20 time steps ($\delta t=0.2$) and over approximately 150--200 statistically independent experiments to reduce the noise, which is inherently present in calculations of stresses in molecular dynamics simulations. As is seen in Figs. \ref{Fig7}--\ref{Fig8}, the evolution of the surface tension $\gamma(t)$ and the width of the density profile $\Delta(t)$ has three distinguishable and separated stages, $ t\le 0.5$, $0.5 \le t \le 3-5$ and $ t >3-5$. 

\begin{figure}
\begin{center}
\includegraphics[trim=1cm 1cm 1cm 1cm,width=0.4\columnwidth]{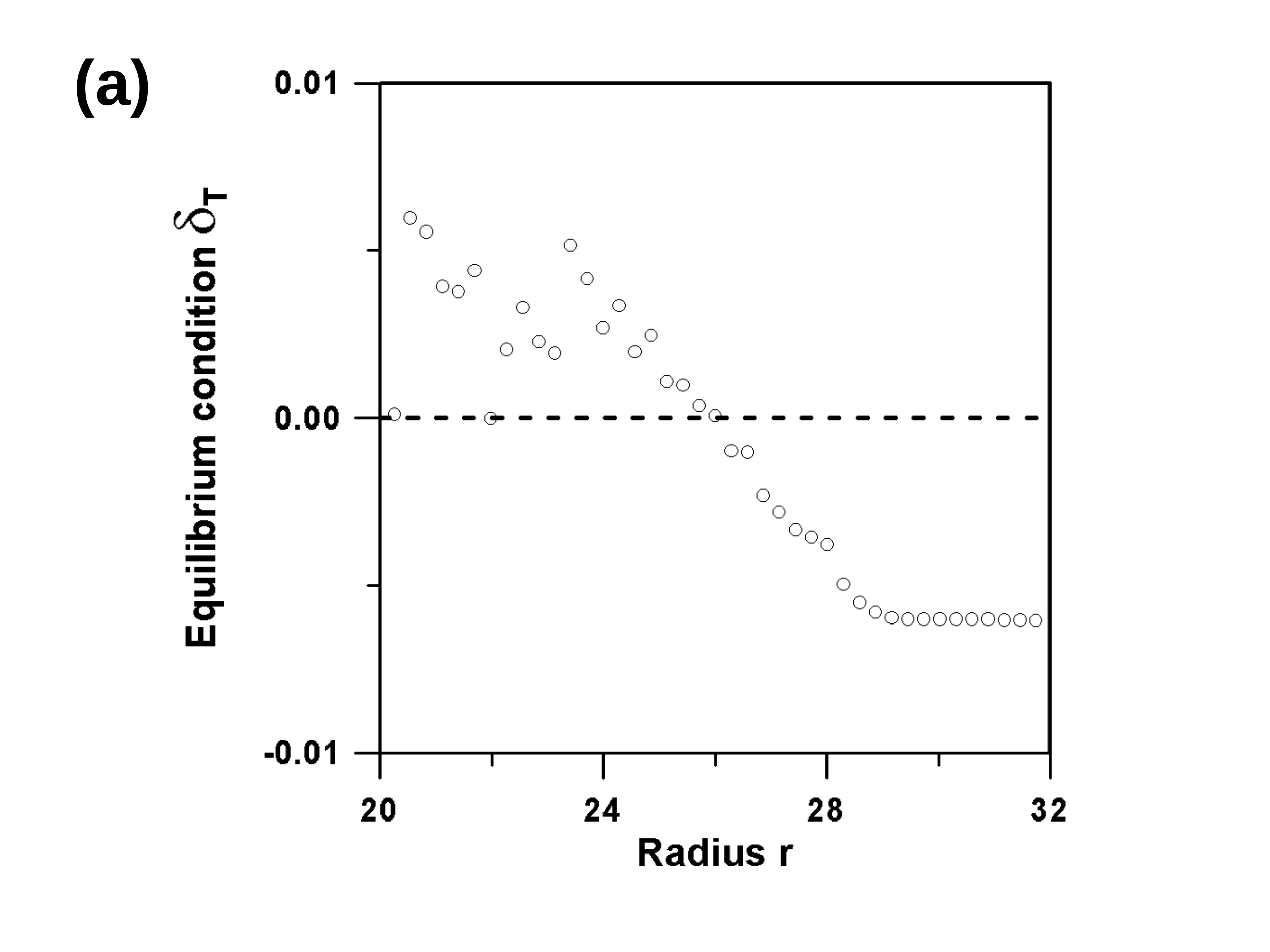}
\includegraphics[trim=1cm 1cm 1cm 1cm,width=0.4\columnwidth]{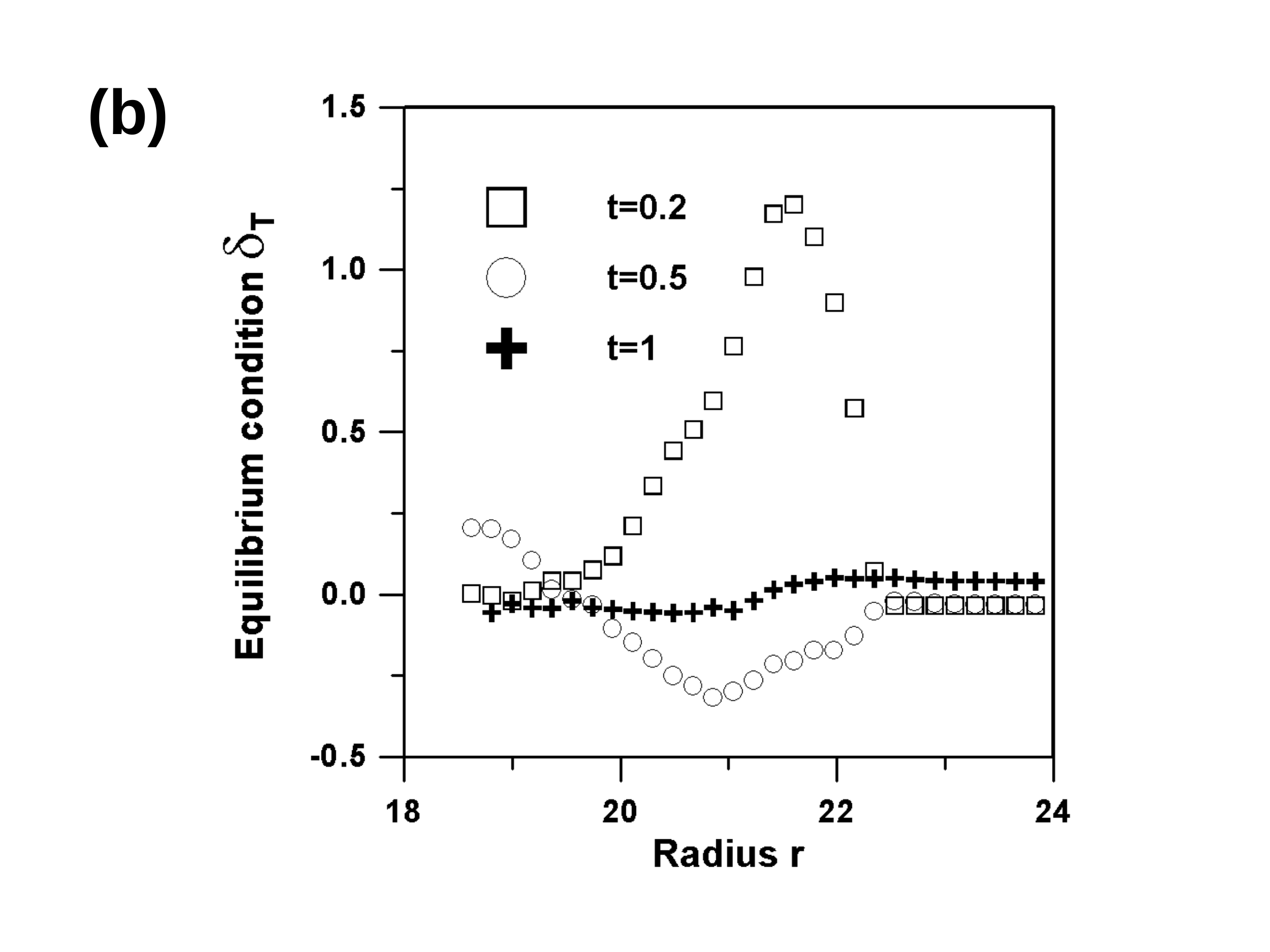}
\end{center}
\caption{Normalised condition of mechanical equlibrium, eq. (\ref{EqCondSph}),  at $T=0.5$ (a) in an equilibrated monatomic LJ drop with 75000 particles before the cut off (b) after the cut off at $t=0.2, t =0.5$ and $t=1$ in a monatomic LJ liquid drop with $\approx 40000$ particles. } \label{Fig9}
\end{figure}

The first stage takes place immediately after the cut off during $\Delta t \simeq 0.5$, which is on the brink between macroscopic and molecular time scales ($t\sim 0.3$). The appearance of this characteristic time scale seems to be solely due to the artificial nature of our experiment. Once the particles of the equilibrium interfacial layer have been removed, the particles in the newly formed interfacial layer immediately experience strong uncompensated force acting from the bulk particles. Strictly speaking, the notion of macroscopic stress tensor is not well-defined during this stage and a definition of the stress tensor could only be given in terms of the ensemble average. If we were to calculate surface tension using eq. (\ref{SurfTens}) and check the condition of mechanical equilibrium (\ref{BalanceSph}), we would see that the condition is not fulfilled.  Indeed, if we present eq. (\ref{BalanceSph}) in an integral normalised form
\begin{equation}
\delta_T(r)=\frac{2\int_0^r(T_T-T_N)r^{-1}\, dr- T_N}{\max(|T_T|, |T_N|)}=0,
\label{EqCondSph}
\end{equation}
we will see that condition (\ref{EqCondSph}) is fulfilled at equilibrium within the accuracy of $1\%$, Fig. \ref{Fig9}a. But after the cut off at $t=0.2$, not surpsingly, condition (\ref{EqCondSph}) is violated and the system is out of mechanical balance at the interfacial layer, Fig \ref{Fig9}b. This simply implies that the definition and the measurement of the surface tension during this initial phase have no direct macroscopic physical meaning and may be described as a microscopic relaxation phase. The mechanical balance should be of course restituted either on the time scale of the shear stress relaxation time $\tau_{stress}$ or on the time scale of the pressure wave generated by the uncompensated force in the normal to the interface direction. 

This relaxation process of the restitution of the mechanical equilibrium is indeed observed. For example, if we consider the evolution of condition (\ref{EqCondSph}) in a drop at $T=0.5$, then one can see that at $t = 0.5$ the mechanical balance has been almost restored, Fig. \ref{Fig9}b, and at $t= 1$ the balance is observed with almost the same accuracy as at equilibrium conditions. The separation between the first two stages is slightly smeared at higher temperatures, for example at $T=0.8$, Fig. \ref{Fig7}, as the relaxation process seems to take longer due to lower viscosity.

We can conclude that the observation of the surface tension that would have any macroscopic meaning should start after the first transitional phase of relaxation. The observed minimum value of surface tension (Fig. \ref{Fig7}) at $t=0.6$, when the interface is still relatively sharp ($\Delta<\Delta_0^{(eq)}$ for all temperatures, Fig. \ref{Fig8}) is about $60\%$ of the equilibrium value. 

\begin{figure}
\begin{center}
\includegraphics[trim=1cm 1cm 1cm 1cm,width=0.4\columnwidth]{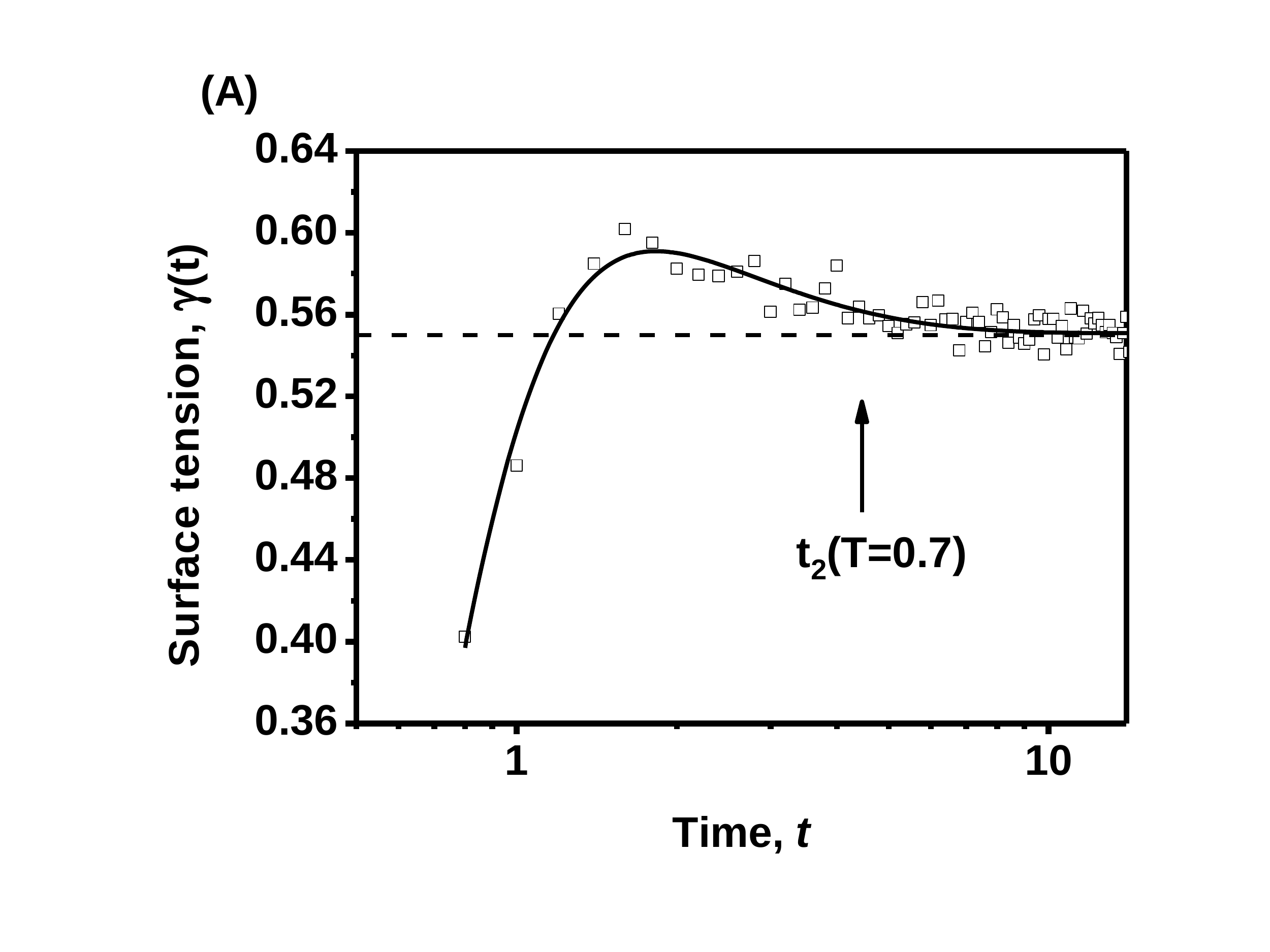}
\includegraphics[trim=1cm 1cm 1cm 1cm,width=0.4\columnwidth]{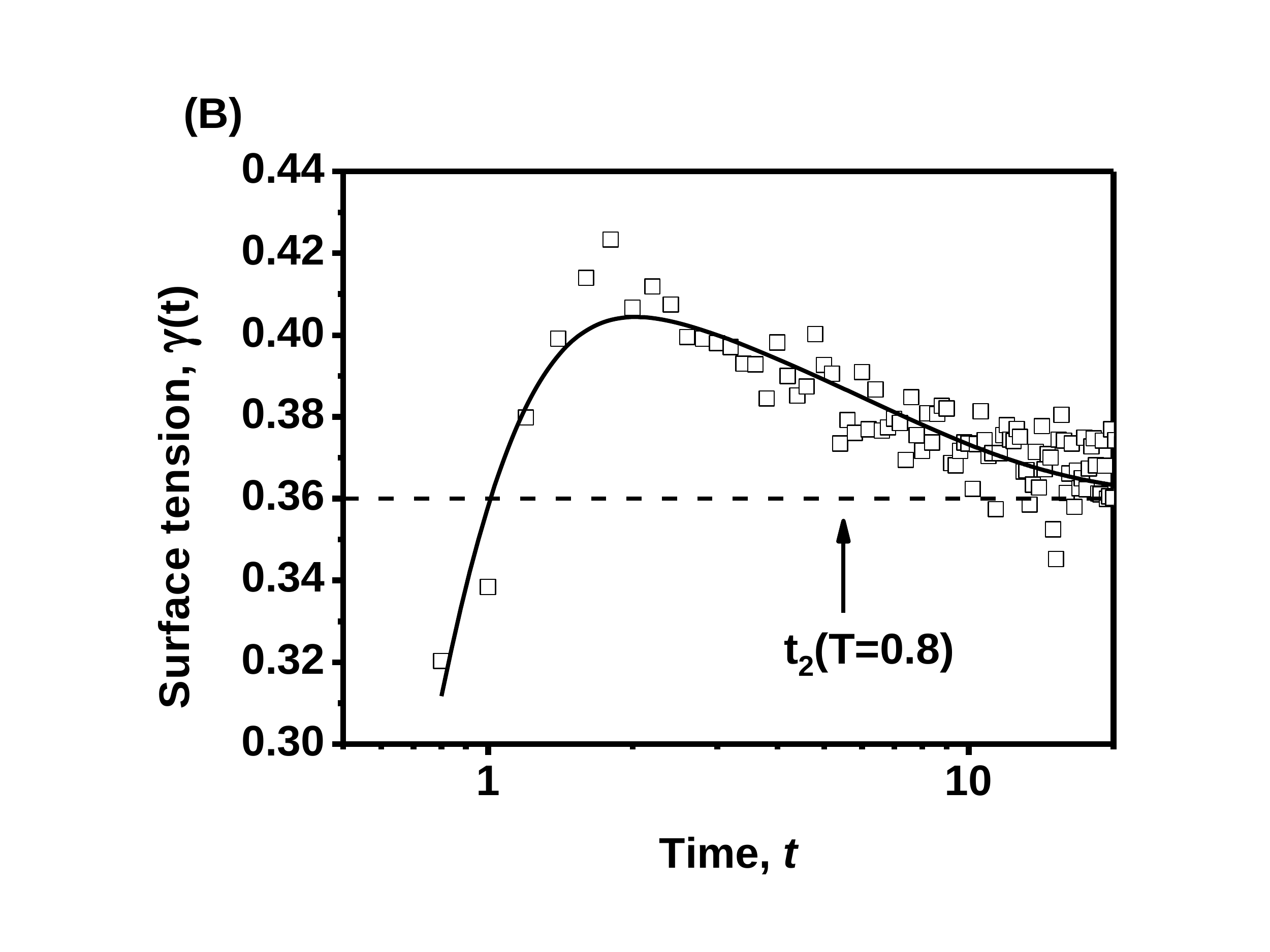}
\end{center}
\caption{Evolution of the surface tension $\gamma(t)$ as a function of time in a monatomic LJ liquid drop consisting of  $\approx 40000$ particles after the cut off. The results of MD simulations are shown by symbols and the fit (\ref{FitST}) is shown by the solid lines, (a) at $T=0.7$ and $A_1=-3.4$, $A_2=0.1$, $\tau_{\gamma}=1.9\pm 0.3$ and $\tau_{\gamma}^{1}=0.3\pm0.03$, and (b) at $T=0.8$ and $A_1=-1.3$, $A_2=0.06$, $\tau_{\gamma}=6.6\pm 0.6$ and $\tau_{\gamma}^{1}=0.32\pm0.05$. The dashed lines correspond to equilibrium values of the surface tension.} \label{ST-relaxation-time}
\end{figure}

In the second stage, the mechanical equilibrium is observed. During this stage, the surface tension relaxes to its equilibrium value, while, at the same time, the density profile is still in transition to equilibrium, Fig. \ref{Fig8}. This indicates that we observe two separate time scales; the first time scale is associated with the relaxation of the surface tension, while the second time scale is associated with the relaxation of the density profile to its equilibrium shape.  
The surface tension relaxation time $\tau_{\gamma}$ can be determined by fitting the time dependence $\gamma(t)$ with
\begin{equation}
\gamma(t)=\gamma_0+A_1 e^{-t/\tau_{\gamma}^{1}}+A_2 e^{-t/\tau_{\gamma}},
\label{FitST}
\end{equation}
where $\tau_{\gamma}^{1}<<\tau_{\gamma}$ is introduced to account for the transition from the first, fast relaxation stage. This particular choice of the fitting function is motivated by simplicity and serves to roughly reveal temperature dependence of the characteristic time. It may not reflect the nature of the surface tension as a derivative of the density distribution of particles, for example. The fit is applied after the first stage is completed ($t>0.5$) as illustrated in Fig. \ref{ST-relaxation-time}. Note that the accuracy of that definition due to non-monotonic and noisy character of function $\gamma(t)$ is low and leaves some room for uncertainty. While the errors bars for $\tau_{\gamma}$ in Table. \ref{Table3} may look good, we think they are too optimistic, since the overall deviation of the fit from the fitting data points in this case was in between $2\%$ and $5\%$. As a function of temperature surface tension relaxation time is found to be in the range $1.1\le \tau_{\gamma}\le 6.6$, Table. \ref{Table3}. As is seen, this characteristic time is always larger (much larger at high temperatures) than the stress relaxation time $\tau_{stress}$, $\tau_{\gamma}>>\tau_{stress}$ and has a clear temperature dependence, which is in anti-correlation with that of $\tau_{stress}$, so that one can completely rule out their possible connection.

\begin{table*}[htbp]
\begin{tabular}{ | c | c | c | c | c | c | c | c | c | c | c | c | c | c |}
\hline
  $T$ & $\rho$  & $D^{\dag} $ & $\mu^{\dag}$  & $\tau_{stress}^{\dag}$ & $ \tau_{\Delta_0}$ & $\tau_{\Delta_0}^{\star}$ & $\tau_{\Delta_0}^{I}$ & $\tau_{\gamma}$ & $t_1^{\rho}$ & $t_2^{\rho}$ & $t_3^{\rho}$ & $\Delta(t_2^{\rho})$ &  $t_{cu}^{max}$ \\  
\hline
  0.5 & $0.885$ &   $0.017$ & $4.1$ & $0.97$ & $3.5$ & $2.8$  & $1.8$ & $1.1\le \tau_{\gamma} \le 2$     & $0.30\pm 0.06$  & $3.4\pm 0.5$  & $26\pm 3$ & $0.36$ &   $41$ \\ 
\hline
  0.6 & $0.841$ &   $0.032$ & $2.5$ & $0.45$ & $3.9$ & $3.8$  & $1.94$ & $1.7\pm0.8$    & $0.48\pm 0.05$  & $4.3\pm 0.4$  & $31\pm 3$  & $0.51$  &   $21$ \\ 
\hline
  0.7 & $0.788$ &   $0.052$ & $1.7$ & $0.22$ & $3.8$  & $3.8$ & $1.76$ & $1.9\pm0.3$    & $0.47\pm 0.03$  & $4.2\pm 0.5$  & $25\pm 3$  & $0.64$  &   $28$ \\ 
\hline
  0.8 & $0.732$ &   $0.078$ & $1.2$ & $0.16$ & $4.3$  & $4.3$ & $1.8$ & $6.6\pm0.6$  & $0.50\pm 0.03$  & $5.4\pm 0.4$  & $42\pm 4$    & $0.82$  &   $37$ \\ 
\hline                  
\end{tabular}
\caption{Parameters of monatomic LJ liquids in the drops and of interfacial profiles: density $\rho$, coefficients of self-diffusion $D$, dynamic shear viscosity $\mu$ and stress relaxation time $\tau_{stress}$, characteristic times $\tau_{\Delta_0}$, $\tau_{\Delta_0}^{\star}$ and $\tau_{\Delta_0}^{I}$ of particle motion on the length scale of $\Delta_0^{(eq)}$, surface tension relaxation time $\tau_{\gamma}$,  characteristic times of the evolution of the density profile, $t_l^{\rho}$, obtained from the fit (\ref{ExpF}) after the cut off and the width of the density profile at $t_2^{\rho}$. The maximum relaxation time of the capillary waves $t_{cu}^{max}$ is given in the last column and is calculated by means of eq. (\ref{underD2}) for a drop consisting of $40000$ particles, that is at $R_0\approx 22$.  Note that parameters marked by $\dag$ have been calculated by MD simulations with periodic boundary conditions at the densities equivalent to those in the drops.}
\vspace{12pt}
\label{Table2}
\end{table*} 

\begin{figure}
\begin{center}
\includegraphics[trim=1cm 1cm 1cm 1cm,width=0.5\columnwidth]{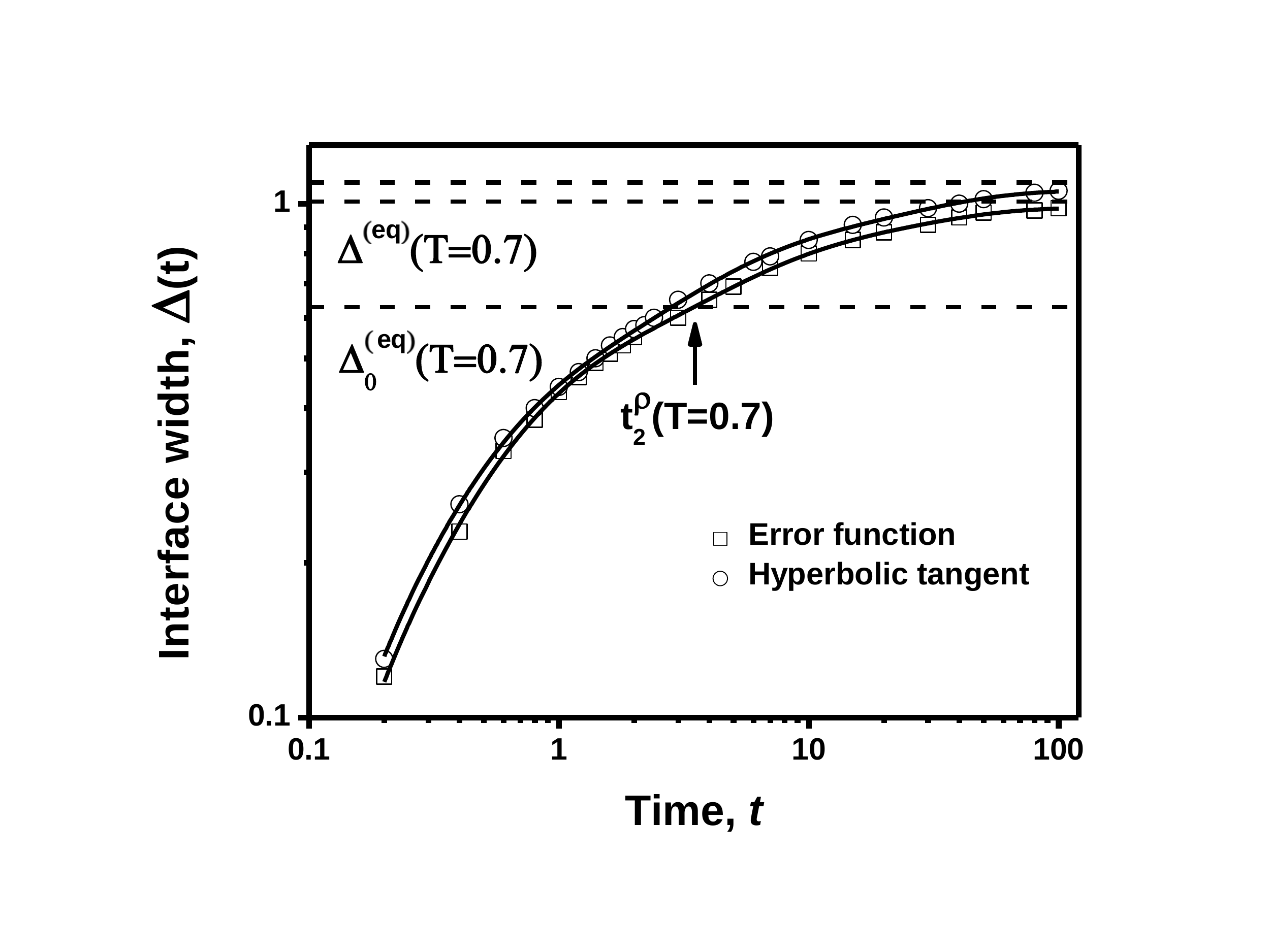}
\end{center}
\caption{Evolution of the width of the density profile in a monatomic LJ drop consisting of $\approx40000$ particles at $T=0.7$ obtained from the fitting functions (\ref{fit-den}) and (\ref{tanh}). For comparison, the characteristic time scales obtained using fit (\ref{ExpF}) are, for the error function fit, $t_{1}^{\rho}=0.47\pm0.03, t_{2}^{\rho}=4.2\pm0.5, t_{3}^{\rho}=25\pm4$, and for the hyperbolic tangent  fit  $t_{1}^{\rho}=0.37\pm0.04, t_{2}^{\rho}=3.2\pm0.3, t_{3}^{\rho}=24\pm3$. The dashed lines are the equilibrium values of $\Delta$ (obtained from both fitting procedures) and intrinsic width $\Delta_0^{(eq)}$. The profile widths have been obtained with approximation of $\approx 3\%$.} \label{Fig11comp}
\end{figure}

To quantify the time scales relevant to the relaxation process at the interface, consider evolution of $\Delta(t)$, Fig. \ref{Fig8}, in more detail. These evolution curves have monotonic time dependence and have substantially lower level of noise and thus represent less formidable task for interpolation and time scale analysis than the evolution curves of the surface tension. To extract characteristic time scales, we fit function $\Delta(t)$ by    
\begin{equation}
\Delta(t)=\Delta^{(eq)}+\sum_{k} A_k e^{-t/t_k^{\rho}}
\label{ExpF}
\end{equation}
We have found that it was sufficient to use just three characteristic time scales, $k=1..3$, in eq. (\ref{ExpF}) for accurate approximation of $\Delta(t)$, see Fig. \ref{Fig11comp} and Table. \ref{Table2}, where the results of the fitting procedure are summarised for different temperatures. For comparison, the same evolution curve has been reproduced using hyperbolic tangent fit (\ref{tanh}), Fig. \ref{Fig11comp}. One can see that the results are essentially identical within the approximation errors. The characteristic time scales extracted using (\ref{ExpF}) are sufficiently close, considering sensitivity of the exponential fits and the errors of approximation.

The first relaxation time, $t_1^{\rho}$, as expected from the evolution of the surface tension, is in the range $0.3\le t_1^{\rho} \le 0.5$ and corresponds to the first, microscopic stage of surface tension relaxation.  This characteristic time is very short and just on the verge of the resolution $\delta t=0.2$. The second relaxation time is found to be in the range $3.4 \le t_2^{\rho} \le 5.4$, which is close to the range of $\tau_{\gamma}$, $1.1\le \tau_{\gamma}\le 6.6$, in the same temperature interval. The temperature dependence of $t_2^{\rho}$, Table \ref{Table2},  is in a good qualitative agreement with the temperature dependence of $\tau_{\gamma}$. While the characteristic time scales are not very much apart, considering the approximation errors involved, exact quantitative comparison is difficult due to the difficulty in defining $\tau_{\gamma}$ from non-monotonic and noisy $\gamma(t)$ curves. Anyway, the characteristic values and the trend with temperature of both $\tau_{\gamma}$ and $t_2^{\rho}$ suggest that these characteristic time scales correspond to the second, macroscopic stage of the surface tension and intrinsic density profile relaxation in the system. 

\begin{figure}
\begin{center}
\includegraphics[trim=1cm 1cm 1cm 0.1cm,width=0.4\columnwidth]{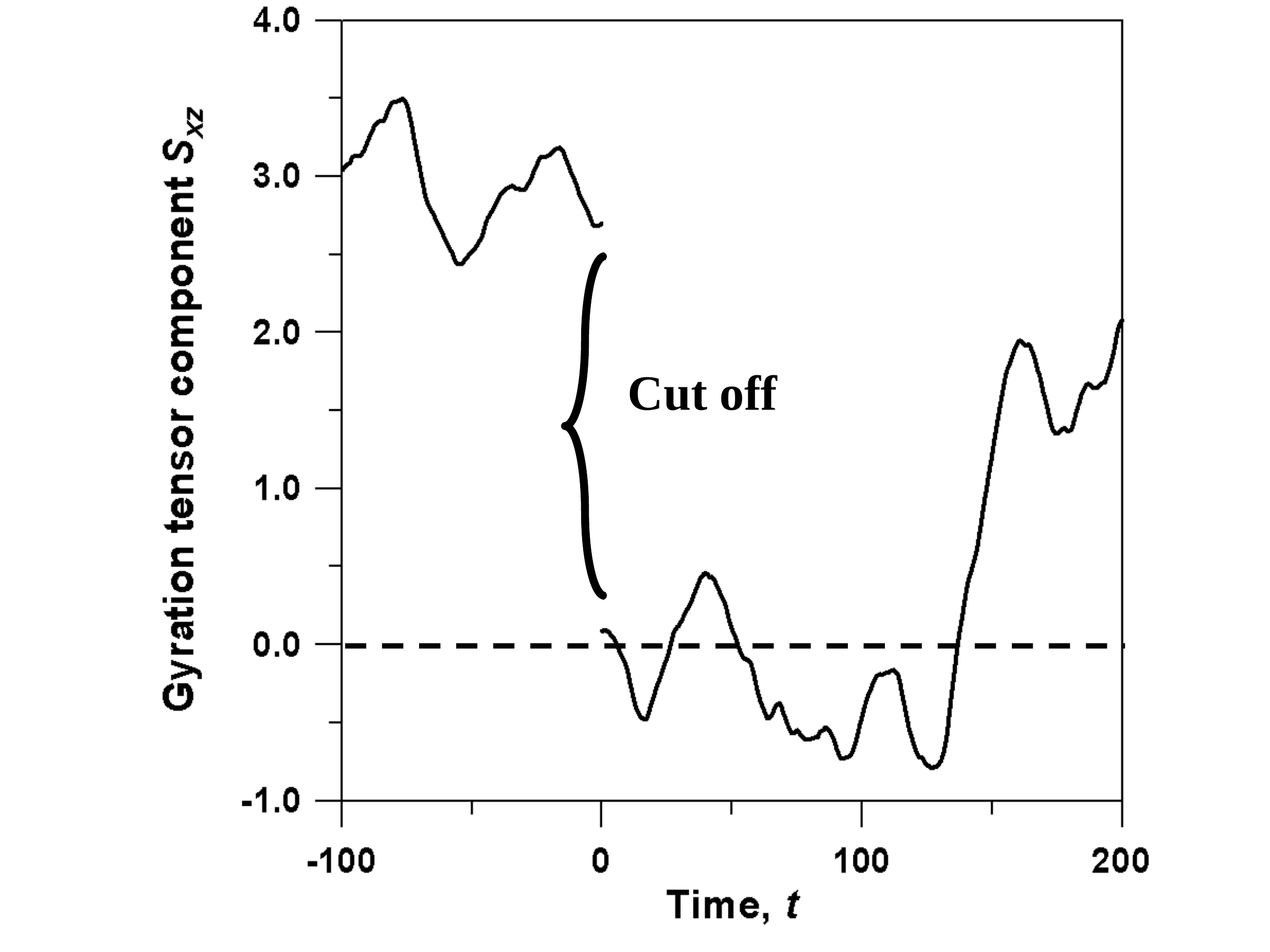}
\end{center}
\caption{Evolution of the gyration tensor component $S_{xz}$ as a function of time before and after the cut off in a monatomic LJ liquid drop consisting of 40000 particles at $T=0.5$.} \label{Fig10}
\end{figure}

The third fitting parameter $t_3^{\rho}$, on the other hand, is apparently related with the longest relaxation time associated with the excitation of capillary waves $t^{max}_{cu}$. This may be evidenced from the scaling of both $t_3^{\rho}$ and $t^{max}_{cu}$ with temperature, see Table \ref{Table2}. One can observe a sufficiently good quantitative correlation between those two characteristic times. The appearance of capillary waves can be also illustrated if we consider evolution of the off-diagonal components of the gyration tensor 
$$
S_{\alpha\beta}=\frac{1}{N}\sum_{i=1}^{N} r^i_{\alpha} r^{i}_{\beta}
$$
with time, for example $S_{xz}$, which is an indicator of deviations from sphericity. From this dependence, Fig. \ref{Fig10}, one can see that the shape of the drop is very close to a sphere just after the cut off, $S_{xz}\simeq 0$ (some small deviation is actually expected due to fluctuations in the bulk area) and then the asymmetry develops.

So far we have established that  the first, shortest relaxation time of the surface phase after the cut off  is related to microscopic processes, while the third, the longest relaxation time should be attributed to the excitation of capillary waves. To understand the nature of the second time scale, $t_2^{\rho}$, we consider transport properties of LJ liquids.

\subsection{Particle transport in the liquid and at the interface.}

\begin{figure}
\begin{center}
\includegraphics[trim=1cm 1cm 1cm 1cm,width=0.35\columnwidth]{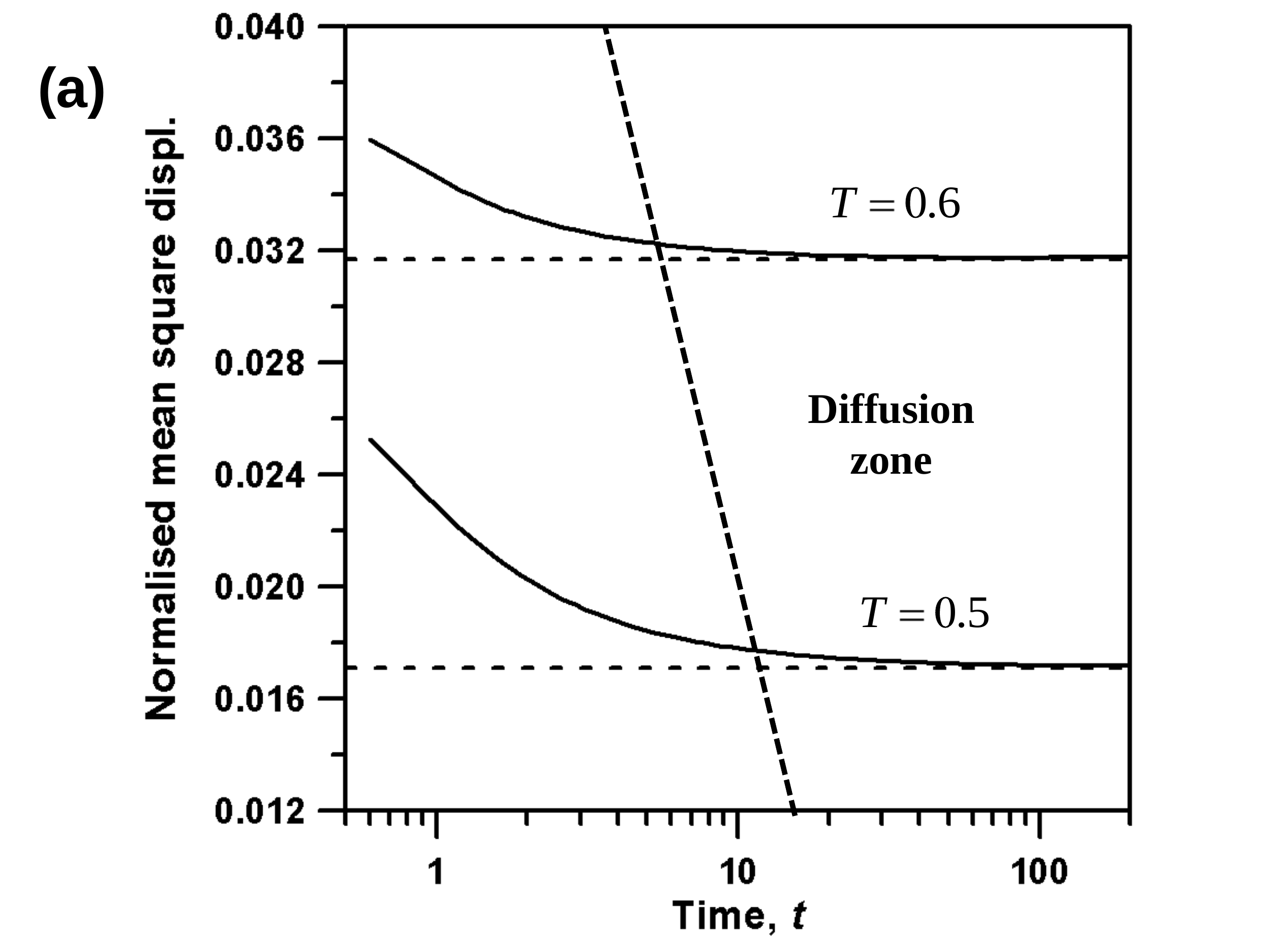}
\includegraphics[trim=1cm 1cm 1cm 1cm,width=0.35\columnwidth]{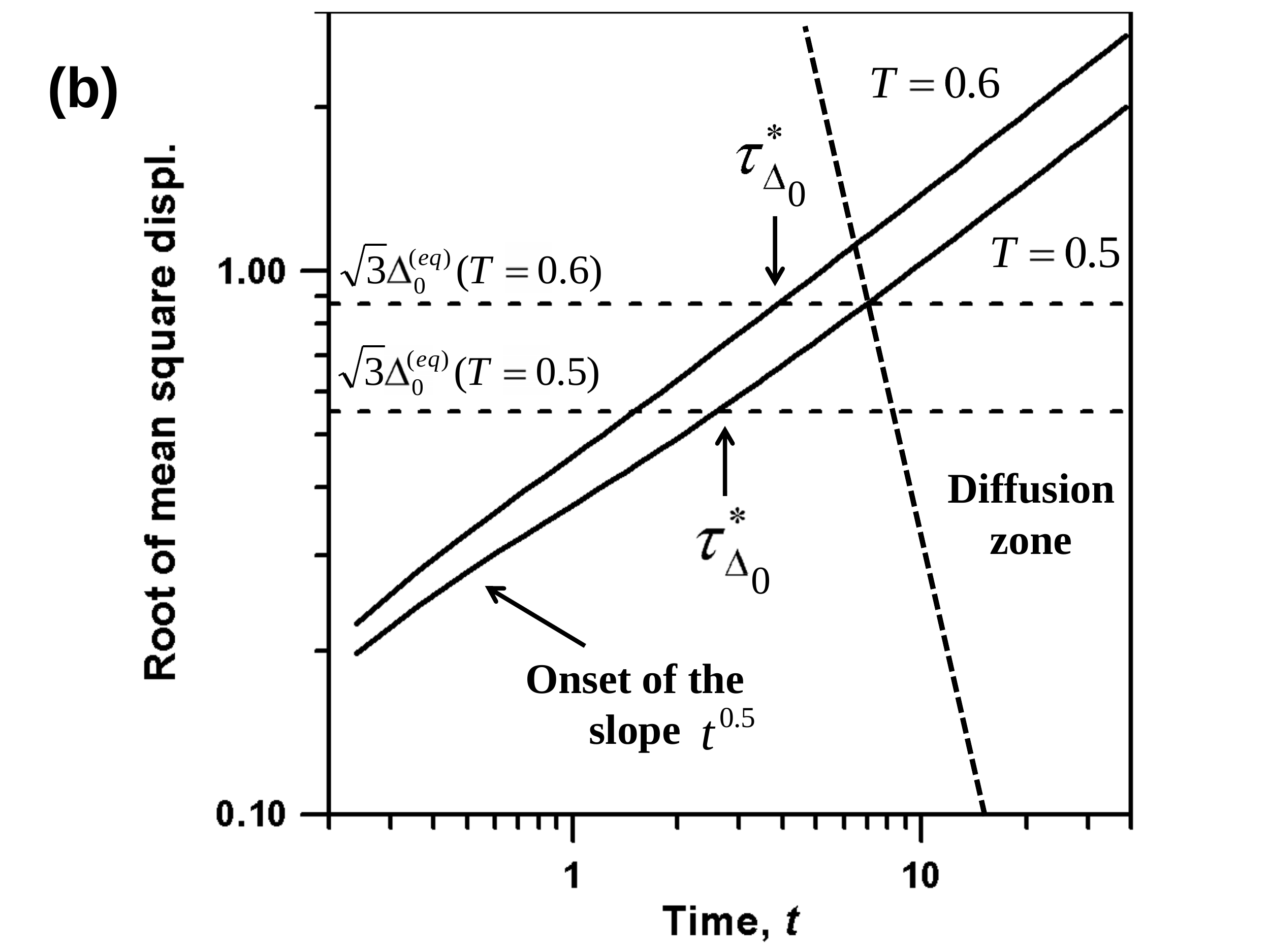}
\end{center}
\caption{(a) Normalised mean square displacement $\frac{R_{msd}^2}{6t}$ (b) Square root of the mean square displacement $\sqrt{R^2_{msd}}$ as functions of time in the conditions similar to the conditions in the bulk of monatomic LJ liquid drops.} \label{Fig3}
\end{figure}

To understand the characteristic time scales of particle transport, consider the mean square displacement $R^2_{msd}(t)$, eq. (\ref{RMSD}), in the bulk. The mean square displacement  is used to calculate coefficient of self-diffusion in the bulk 
$$
D=\lim_{t\to \infty}\frac{R^2_{msd}(t)}{6t},
$$ 
as illustrated in Fig. \ref{Fig3}a.  This, in turn, is used to estimate the characteristic diffusion time $\tau_{\Delta_0}$ on the length scale of intrinsic equilibrium width $\Delta_0^{(eq)}$ 
$$
\tau_{\Delta_0}=\frac{\Delta^{(eq)2}_0}{2D}.
$$ 
Alternatively, the effective time $\tau_{\Delta_0}^{\star}$, which a particle needs to travel the distance $\Delta_0^{(eq)}$ in the direction across the interface (in the bulk conditions) can be obtained directly via
\begin{equation}
R^2_{msd}(\tau_{\Delta_0}^{\star})=3\Delta_0^{(eq)2},
\label{DirectDiffTime}
\end{equation}
assuming isotropic motion. The values of the self-diffusion coefficient in the bulk, characteristic diffusion times $\tau_{\Delta_0}$ and $\tau_{\Delta_0}^{\star}$ at different temperatures are summarized in Table \ref{Table2}. One can see that the estimate $\tau_{\Delta_0}$ is close to  $\tau_{\Delta_0}^{\star}$, while $\tau_{\Delta_0}\ge \tau_{\Delta_0}^{\star}$. This is anticipated since if we were to calculate the characteristic time $\tau_{\Delta_0}^{\star}$ directly from (\ref{DirectDiffTime}), we would find that particle motion on this time scale is close to but not yet in the fully developed diffusive regime, which is reached at approximately $t\sim  10$ (though this value depends, of course, on temperature), Fig. \ref{Fig3}a.  At the same time,  for monatomic liquids in the range of temperatures $2.8\le \tau_{\Delta_0}^{\star}\le 4.3$, see Table \ref{Table2} and Fig. \ref{Fig3}b for illustration.  The temperature dependences of $\tau_{\Delta_0}$ and $\tau_{\Delta_0}^{\star}$ reflect the fact that both coefficient of self-diffusion and intrinsic width of the density profile are monotonically increasing functions of temperature. Two effects almost compensate each other and both $\tau_{\Delta_0}$ and  $\tau_{\Delta_0}^{\star}$ demonstrate rather weak  dependence on temperature, weaker than it is suggested by the dependence of $\Delta^{(eq)}_0$, for example. 

\begin{figure}
\begin{center}
\includegraphics[trim=1cm 0.5cm 1cm 0.5cm,width=0.4\columnwidth]{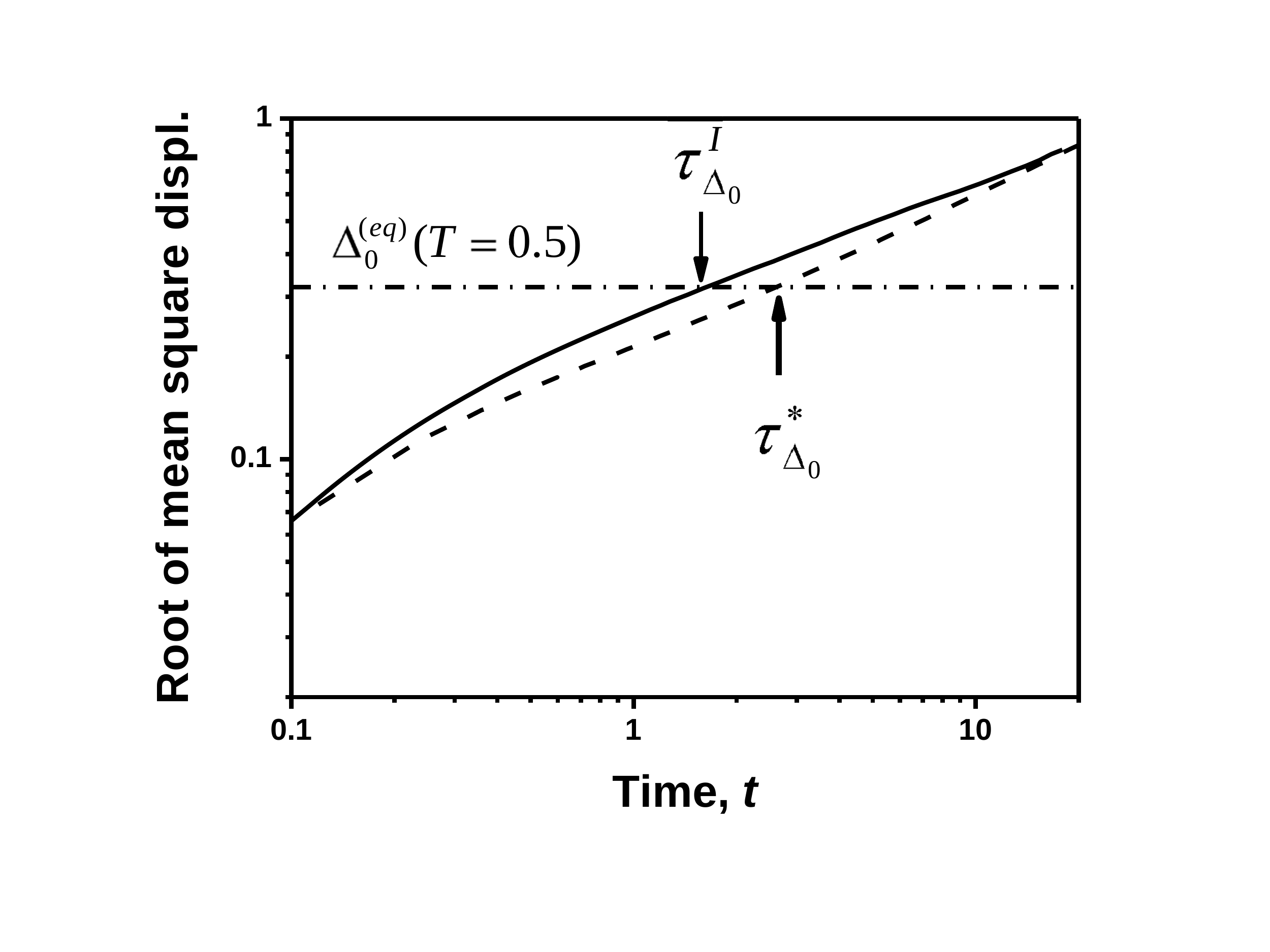}
\end{center}
\caption{Square root of the mean square displacement in the radial direction $\sqrt{R^2_{\Delta_0, r}}$ as a function of time in a monatomic liquid drop at $T=0.5$. The solid line is for particles $r_i$ initialy located at the interfacial layer $R_0 -\Delta_0^{(eq)}\le r_{i}(0) \le R_0  + \Delta_0^{(eq)}$ ($R_0$ is the mean position of the interface in the drop) and the dashed line is for the bulk conditions.} \label{Fig_MSD_SL}
\end{figure}

One needs  to note that while those estimations may be practical and informative, they do not tell, strictly speaking, the whole story. Firstly, the particle motion is not in a fully developed diffusive regime. Secondly, the particles in the first layer at the interface are more mobile than those in the bulk, though this can be moderated by the uncompensated force, which keeps the interface width finite. Thus, the characteristic times $\tau_{\Delta_0}$ and  $\tau_{\Delta_0}^{\star}$ may only provide an upper limit on the contribution of the diffusion into the relaxation at the interface. On the other hand, a direct calculation of the residence time in the interfacial layer may be even less accurate. The interface is widened by capillary wave motion. Given that on average $\Delta^2/\Delta_0^2\sim 2.5$, it would be difficult to exactly locate particles belonging to the "intrinsic" interface. To estimate characteristic residence time of particles at the interface, we calculate the mean square displacement in the radial direction of particles initially contained in a layer of thickness $2\Delta_0^{(eq)}$ at the interface located at $R_0$, that is $R_0 -\Delta_0^{(eq)}\le r_i(0)\le R_0 +\Delta_0^{(eq)}$,
\begin{equation}
R^2_{\Delta_0, r}(t)=<\frac{1}{N_{\Delta_0}} \sum_{i=1}^{N_{\Delta_0}} ({ r}_i(t)-{ r}_i(0))^2>,
\label{RMSD-r}
\end{equation}
where $N_{\Delta_0}$ is the number of particles in the layer. The result is shown in Fig. \ref{Fig_MSD_SL} in comparison with the mean square displacement obtained in the radial direction in the bulk. The corresponding time $\tau_{\Delta_0}^I$ calculated via
$$
R_{\Delta_0, r}^{2}(\tau_{\Delta_0}^I)=\Delta_0^{(eq)2}
$$
is listed in Table \ref{Table2}. One can see that in general $\tau_{\Delta_0}^I$ is close to $\tau_{\Delta_0}, \,\, \tau_{\Delta_0}^{\star}$ and as expected $\tau_{\Delta_0}^I<\tau_{\Delta_0}, \,\, \tau_{\Delta_0}^{\star}$. So on average particles at the interface are more mobile, though it is difficult to instrument an exact measure of their mobility due to the interfacial "roughness", that is uncertainty in their positions. 

\begin{figure}
\begin{center}
\includegraphics[trim=1cm 1cm 1cm 1cm,width=0.5\columnwidth]{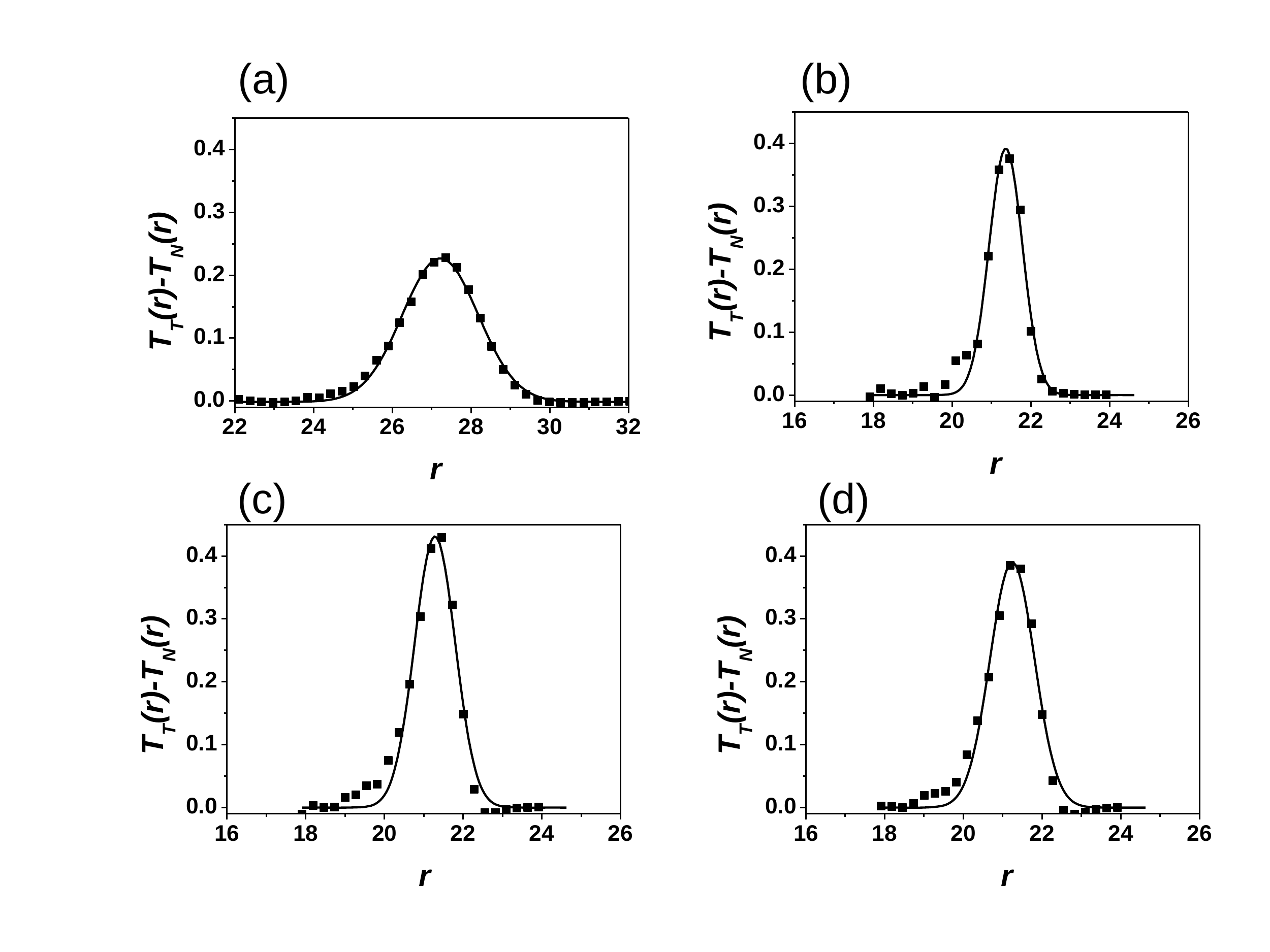}
\end{center}
\caption{Snapshots of distributions of $T_T({r})-T_N({r})$ averaged over $\delta t=1$ in a droplet consisting initially of $75000$ particles at $T=0.7$ (a) before (b), (c), (d) after the cut off at $t=1,2,3$ respectively. The numerical data are shown by symbols and the solid line is the Gaussian fit (\ref{GaussFit}) with parameters shown in Table \ref{Table3}.} \label{Fig11}
\end{figure}

\begin{figure}
\begin{center}
\includegraphics[trim=1cm 1cm 1cm 1cm,width=0.5\columnwidth]{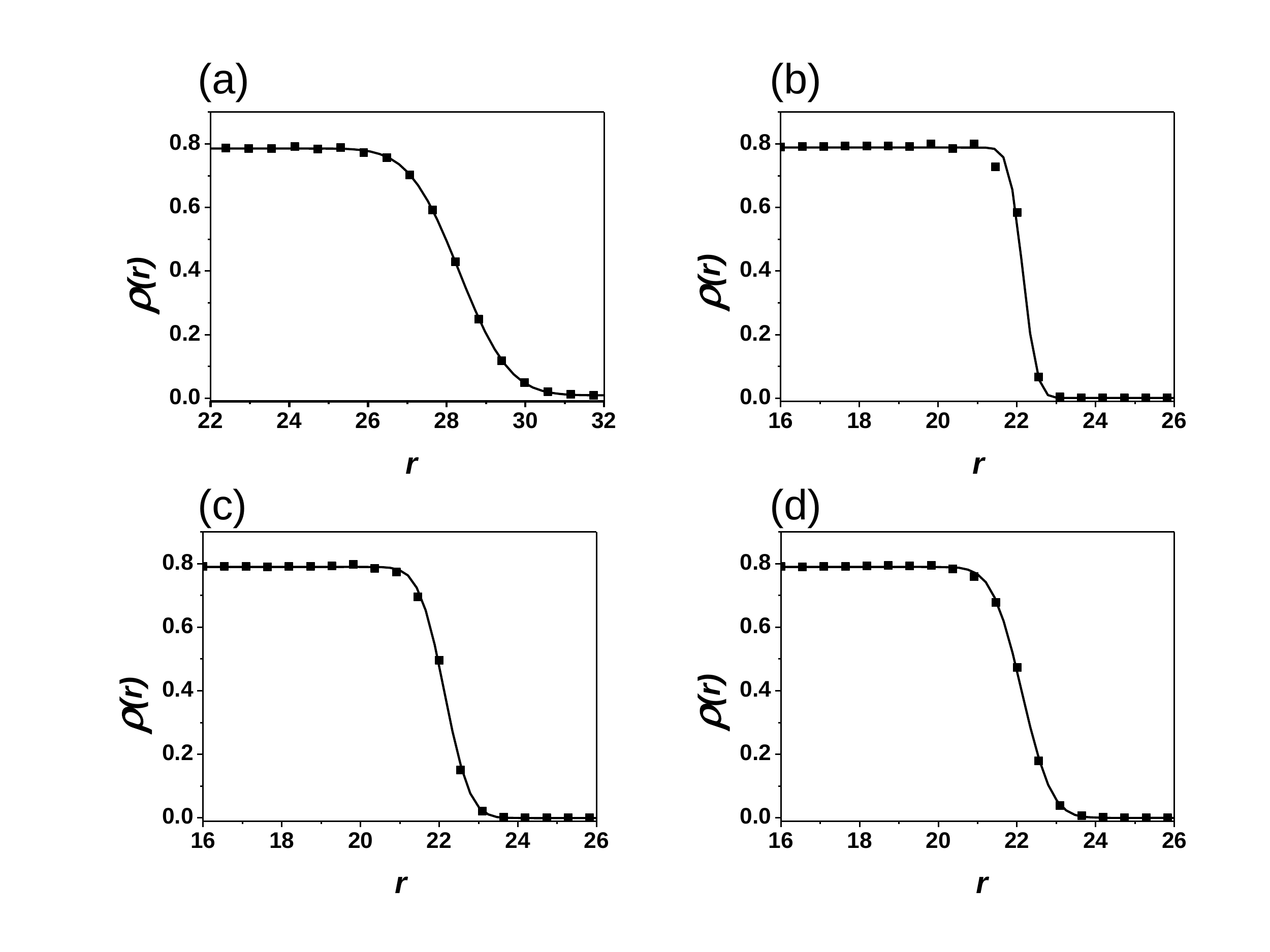}
\end{center}
\caption{Snapshots of distributions of density averaged over $\delta t=1$ in a droplet consisting initially of $75000$ particles initially at $T=0.7$ (a) before (b), (c), (d) after the cut off at $t=1,2,3$ respectively. The numerical data are shown by symbols and the solid line is the error function fit (\ref{fit-den}) with parameters shown in Table \ref{Table3}.} \label{Fig12}
\end{figure}

Compare now the obtained values of the second relaxation time $t_{2}^{\rho}$ and its temperature dependence with those of $\tau_{\Delta_0}^{\star}$ and $\tau_{\Delta_0}$. One can observe that all of them, $t_{2}^{\rho}$, $\tau_{\Delta_0}$ and $\tau_{\Delta_0}^{\star}$ are in a very good quantitative agreement in the temperature range used in our MD experiments. This correlation between $t_{2}^{\rho}$, $\tau_{\Delta_0}$, $\tau_{\Delta_0}^{\star}$ and $\tau_{\gamma}$ is a clear indication of their connection.  If we now compare the width of the interface at $t=t_2^{\rho}$, $\Delta(t_2^{\rho})$, then we will see that it is only slightly above $\Delta_0^{(eq)}$, that is  $\Delta_0^{(eq)}\simeq \Delta(t_2^{\rho})$. This is also illustrated in Fig. \ref{Fig11comp}. All this implies that during the second relaxation stage, the interfacial width $\Delta$ has reached equilibrium intrinsic width $\Delta_0^{(eq)}$, while the surface tension has relaxed to its equilibrium value $\gamma_0$. To elaborate on this statement, consider snapshots of the profiles of distribution of stresses, Fig. \ref{Fig11}, and density, Fig. \ref{Fig12}. The snapshots of the profiles are shown before the cut off and at the subsequent times $t=1,2,3$ after the cut off. Corresponding values of the fitting parameters (eq. (\ref{GaussFit}) ) are given in Table \ref{Table3}. One can see that at $t=3$ (the value at the beginning of the interval $3\le t_2^{\rho} \le 5$) the amplitude $A_{ST}$ of the profile $T_T({r})-T_N({r})$ (which is the value of the surface tension) has just returned to its equilibrium value or is close to it at high temperatures.  At the same time the width $\Delta_{ST}$ is only slightly larger than $\Delta_{ST}^{0(eq)}$, which is expected if we consider simultaneous relaxation to $\Delta_{ST}^{0(eq)}$ and excitation of the capillary waves. Thus, not only the density profile has relaxed to its intrinsic value, but also the distribution of $T_T({r})-T_N({r})$.  Therefore Fig. \ref{Fig11}d and Fig. \ref{Fig12}d show the snapshots of the virgin interface without apparent widening by the capillary waves. We note that we did not clearly observe the oscillating intrinsic interface profiles as in [59], though at $t=1$ after the cut-off, Fig. \ref{Fig12}b,  the density profile shows some signs of the fine structure similar to the one observed in simulations [59]. The reason for that "discrepancy" might be that the effect was weak and simply below our resolution in dynamic simulations. This will need further studies.  

One can then directly relate the relaxation of the surface tension and the relaxation of the density profile to its intrinsic value, and simply define the surface tension relaxation time in monatomic liquids as 
$$
\tau_{\gamma}= \tau_{\Delta_0}.
$$  

So from this one can conclude that, first of all, the liquid-gas interface in purely monatomic LJ liquids without apparent widening by the capillary waves is basically very thin and structureless. Secondly,  the relaxation time of this density profile, $t^{\rho}_2$, is directly related to a diffusion-like process on the length scale of intrinsic width of the density profile $\Delta_0^{(eq)}$. Thirdly, the surface tension relaxation time is directly connected with the relaxation time of the density profile and therefore is also defined by the diffusion-like process on the length scale of the intrinsic width of the density profile $\Delta_0^{(eq)}$. And finally, the process of surface tension relaxation is separated from the process of excitation of capillary waves, described by the third relaxation time $t^{\rho}_3$ of the density profile, which is much larger than $\tau_{\gamma}$.

We will further clarify these statements in the next section dedicated to similar dynamical studies but in binary LJ liquids, where we would be able to go into slightly different parameter range, especially with different viscosities, so that we can further separate the time scales discovered in monatomic liquids.

\section{Interfacial dynamics of a binary liquid} 

In this section, we turn our attention to a drop of a binary LJ liquid to test generality of our findings.  The binary liquid consists of two particles, types A and B, such that their number density ratio in the mixture is $N_{A}/N_{B}=4$ . The particles interact with each other by means of a LJ potential with parameters $\varepsilon_{AA}=1, \varepsilon_{BB}=0.5, \varepsilon_{AB}=1.5$, $\sigma_{AA}=1, \sigma_{BB}=0.88, \sigma_{AB}=0.8$. This choice of LJ parameters corresponds to the well-known Kob-Anderson model, which is  less susceptible to crystallisation and has different range of viscosities in the larger operational window of temperatures in comparison to the monatomic LJ liquids [60]. Macroscopic parameters of the binary liquid used in this study, namely density, viscosity, coefficients of self-diffusion for both components $A$ and $B$, surface tension and the stress relaxation time $\tau_{stress}$ calculated in a drop consisting of $75000$ particles (or by the bulk MD simulations with periodic boundary conditions at the equivalent density) are listed in Table \ref{Table4} at different temperatures. The minimum temperature of the binary liquid used in our simulations $T=0.45$ was above the critical temperature $T>T_c=0.435$, when the coefficient of self-diffusion for both particles $D^{A,B}\propto(T-T_c)^{\lambda}$ goes to zero with $\lambda=1.7\div2$ [60]. 

\begin{table*}[htbp]
\begin{tabular}{ | c | c | c | c | c | c | c |}
\hline
  $T$  & $\rho$ & $\mu^{\dag}$  & $D_A^{\dag}$        & $D_B^{\dag}$    		   & $\gamma_0$    & $\tau_{stress}^{\dag}$ \\      
\hline
  0.45 & $1.14$ & $16.2$  &   $0.0044$ 	& $0.0087$	     & $1.18\pm0.04$			 	 & $2.4$ \\
\hline
  0.50 & $1.12$ & $10.1$  &   $0.0074$   & $0.0143$       & $1.11\pm0.04$         & $1.5$ \\
\hline
  0.55 & $1.10$ & $7.2$  &   $0.0111$   & $0.0198$			 & $1.01\pm0.04$         & $1.0$ \\
\hline
  0.60 & $1.06$ & $4.6$  &   $0.0179$   &	$0.0314$       & $0.92\pm0.04$         & $0.65$ \\ 
\hline
  0.65 & $1.04$ & $3.8$  &   $0.0230$   & $0.0392$			 & $0.84\pm0.03$         & $0.53$ \\
\hline                   
\end{tabular}
\caption{Parameters of the LJ binary liquid in the drops: density $\rho$, dynamic shear viscosity $\mu$, coefficients of the self-diffusion of particles $A$ and $B$, $D_{A,B}$, equilibrium surface tension $\gamma_0$ and the stress relaxation time $\tau_{stress}$ at different temperatures. Parameters marked by $\dag$ have been calculated by MD bulk simulations with periodic boundary conditions at the densities equivalent to those in the drops.}
\vspace{12pt}
\label{Table4}
\end{table*}

\begin{table*}[htbp]
\begin{tabular}{ | c | c | c | c | c | c | c | c | c |}
\hline
  $T$     & $\Delta_A^{(eq)}$       & $\Delta_B^{(eq)}$ 			 & $\Delta_A^0$ & $\Delta_B^0$  & $\Delta_{ST}$	& $R_{ST}$ & $R_0^A$ & $R_0^B$  \\      
\hline
  0.45 	      & $0.53\pm 0.01$ 	& $0.62\pm 0.03$   & $0.28\pm0.01$  			& $0.46\pm0.02$  			&  $0.67$       & $24.5$   & $25.3$  & $24.3$	  \\
\hline
  0.50        & $0.59\pm 0.01$ 	& $0.65\pm 0.03$   & $0.34\pm0.02$  			& $0.44\pm0.02$  			&  $0.71$       & $24.7$   & $25.5$  & $24.5$   \\
\hline
  0.55        & $0.65\pm 0.01$ 	& $0.74\pm 0.03$   & $0.38\pm0.02$  			& $0.52\pm0.02$  			&  $0.72$       & $24.8$   & $25.7$  & $24.7$   \\
\hline
  0.60        & $0.72\pm 0.01$ 	& $0.81\pm 0.03$   & $0.43\pm0.02$  			& $0.57\pm0.02$  			&  $0.78$       & $24.9$   & $25.9$  & $24.9$   \\ 
\hline
  0.65        & $0.79\pm 0.02$ 	& $0.86\pm 0.04$   & $0.49\pm0.02$  			& $0.58\pm0.03$  			&  $0.83$       & $25.1$   & $26.0$  & $25.1$   \\
\hline                   
\end{tabular}
\caption{Parameters of interfacial profiles in the binary LJ liquid drops consisting of 75000 particles at different temperatures. Widths of the interfacial density profiles $\Delta^{(eq)}_{A,B}$, intrinsic widths $\Delta_{A,B}^{0}$, width of the distribution of $T_T({r})-T_N({r})$,  $\Delta_{ST}$, and the locations of interfacial profiles $R_0^{A,B}$ and $R_{ST}$. The relative error of determining $\Delta_{ST}$ was $\approx3\%$ and $R_0^{A,B}$, $R_{ST}$ less than $1\%$.}
\vspace{12pt}
\label{Table5}
\end{table*} 

\begin{table*}[htbp]
\begin{tabular}{ | c | c | c | c | c | c |  c | c | c | c | }
\hline
  $T$ & $A_{ST}^{(eq)}$ & $\Delta_{ST}^{(eq)}$ & $\Delta_{ST}^{0(eq)}$ & $A_{ST}(t=1)$ & $A_{ST}(t=2)$  & $A_{ST}(t=3)$ & $A_{ST}(t=5)$ & $\Delta_{ST}(t=5)$ & $\tau_{\gamma}$ \\  
\hline
  0.45 & $1.18\pm0.03$  & $0.67$ & $0.51$  & $0.66\pm0.02$ & $1.07\pm0.03$ & $1.11\pm0.03$ & $1.14\pm0.04$  & $0.50$  & $6.5\pm1.2$\\ 
\hline
  0.50 & $1.11\pm0.03$  & $0.71$ & $0.52$  & $0.65\pm0.02$ & $1.03\pm0.03$& $1.05\pm0.03$ & $1.07\pm0.03$  &  $0.53$ & $7.2\pm1.6$ \\ 
\hline
  0.65 &  $0.81\pm0.02$ & $0.83$  & $0.53$ & $0.72\pm0.03$ & $0.81\pm0.03$ & $0.81\pm0.04$ & $0.82\pm0.04$   & $0.62$ & $1\le \tau_{\gamma}\le 2$ \\ 
\hline                   
\end{tabular}
\caption{Coefficients of the fit (\ref{GaussFit})  at equilibrium and shortly after the proportional cut off in binary LJ liquids at different temperatures. Temperature dependence of  intrinsic width $\Delta_{ST}^{0(eq)}$ and surface tension relaxation time $\tau_{\gamma}$. Parameters $\Delta_{ST}^{(eq)}$, $\Delta_{ST}^{0(eq)}$ have been determined with the relative accuracy of $\approx3\%$, while $\Delta_{ST}(t=5)$ has been obtained with $\approx4\%$ relative accuracy.}
\vspace{12pt}
\label{Table7}
\end{table*}

\begin{table*}[htbp]
\begin{tabular}{ | c | c | c | c | c | c | c | c | c  | c | c |}
\hline
  $T$   & $\tau_{\Delta_A^0}$  & $\tau_{\Delta_B^0}$  &  $t_{A2}^{\rho}$   & $t_{A3}^{\rho}$ & $\tau_{\gamma}$ & $t_{B2}^{\rho}$  & $t_{B3}^{\rho}$ & $\Delta_A(t_{A2}^{\rho})$ & $\Delta_B(t_{B2}^{\rho})$ & $t_{co}^{max}$\\
\hline
  0.45       & $8.9$ 	  & $12.0$ 	      &  $6.5\pm 1$     & $127 \pm 20$ 	& $6.5\pm1.2$ & $9\pm 2$  	&  $240\pm 60$ & $0.30$   & $0.28$ & $149$ \\
\hline
  0.50       & $7.8$    & $6.8$ 	      &  $5.2\pm 0.4$   & $99\pm 11$  	& $7.2\pm1.6$ & $4.6\pm 0.9$ &  $116\pm 21$ & $0.34$ & $0.32$ & $97$\\
\hline
  0.55       & $6.5$    & $6.8$         &  N/A     	      & N/A  	          & N/A  	& N/A				&  N/A 	 & N/A &  N/A & $74$\\
\hline
  0.60       & $5.2$    & $5.2$         & N/A     	      & N/A  	          & N/A  		& N/A		  &  N/A 	 & N/A & N/A & $47$\\ 
\hline
  0.65       & $5.2$    & $4.3$         & $2.6\pm 0.2$     & $30\pm 2$  	    & $1\le \tau_{\gamma}\le 2$ & $3.6\pm 0.2$  &  $53\pm 5$ & $0.48$ & $0.48$ & $39$ \\
\hline                   
\end{tabular}
\caption{Characteristic diffusion times on the length scale of  $\Delta^0_{A,B}$,  characteristic times of the evolution of the density profiles of $A$ and $B$ components, $t_{A,B l}^{\rho}$, obtained from the fit (\ref{ExpF}) after the proportional cut off and the characteristic time scale of the capillary waves, calculated via eq. (\ref{overD}) at $l=2$,  in the binary LJ liquid drops at different temperatures.}
\vspace{12pt}
\label{Table6}
\end{table*}

\subsection{Structure of interfacial  profiles in LJ binary liquid drops}
\begin{figure}
\begin{center}
\includegraphics[trim=1cm 1cm 1cm 1cm,width=0.4\columnwidth]{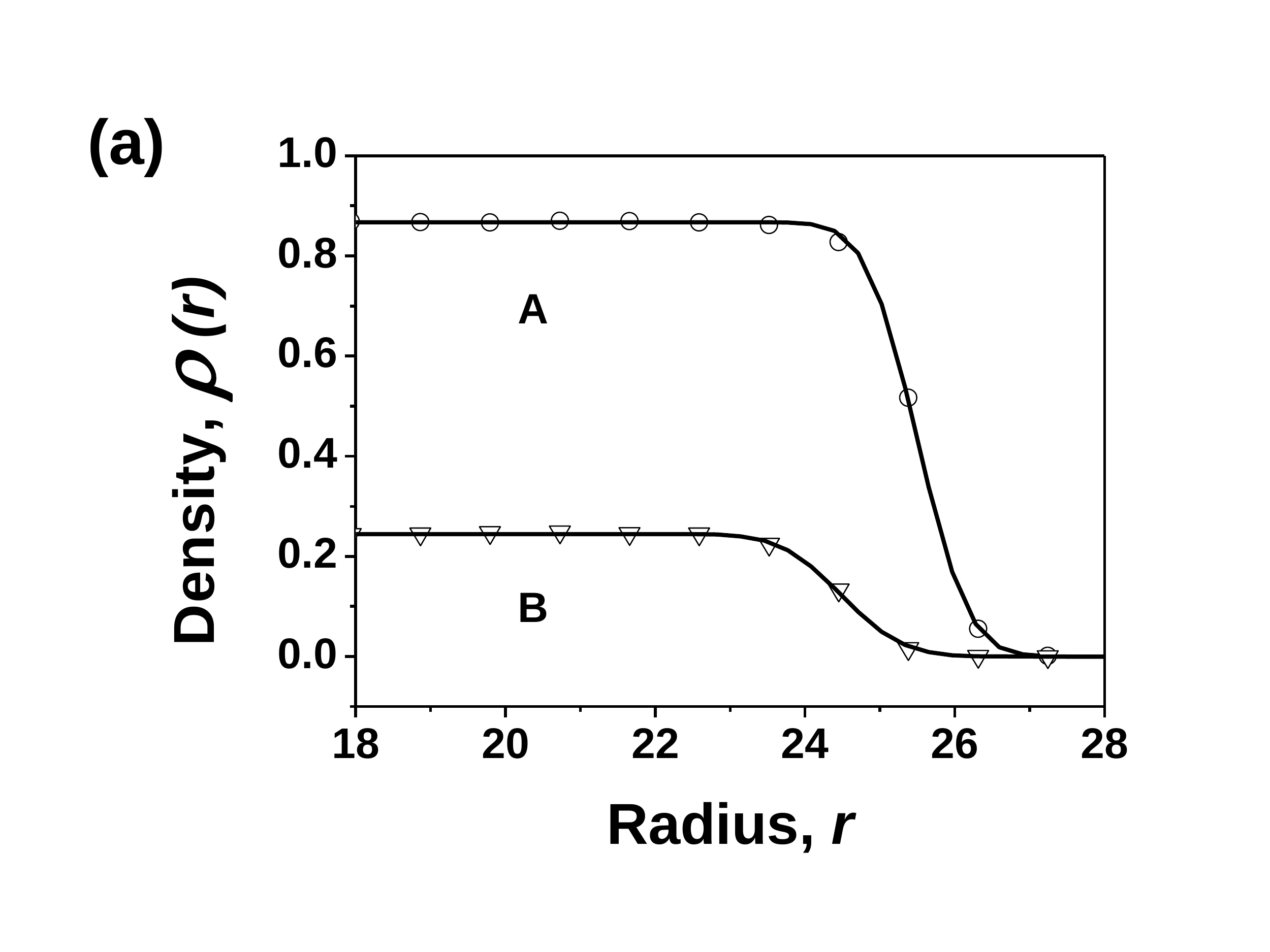}
\includegraphics[trim=1cm 1cm 1cm 1cm,width=0.4\columnwidth]{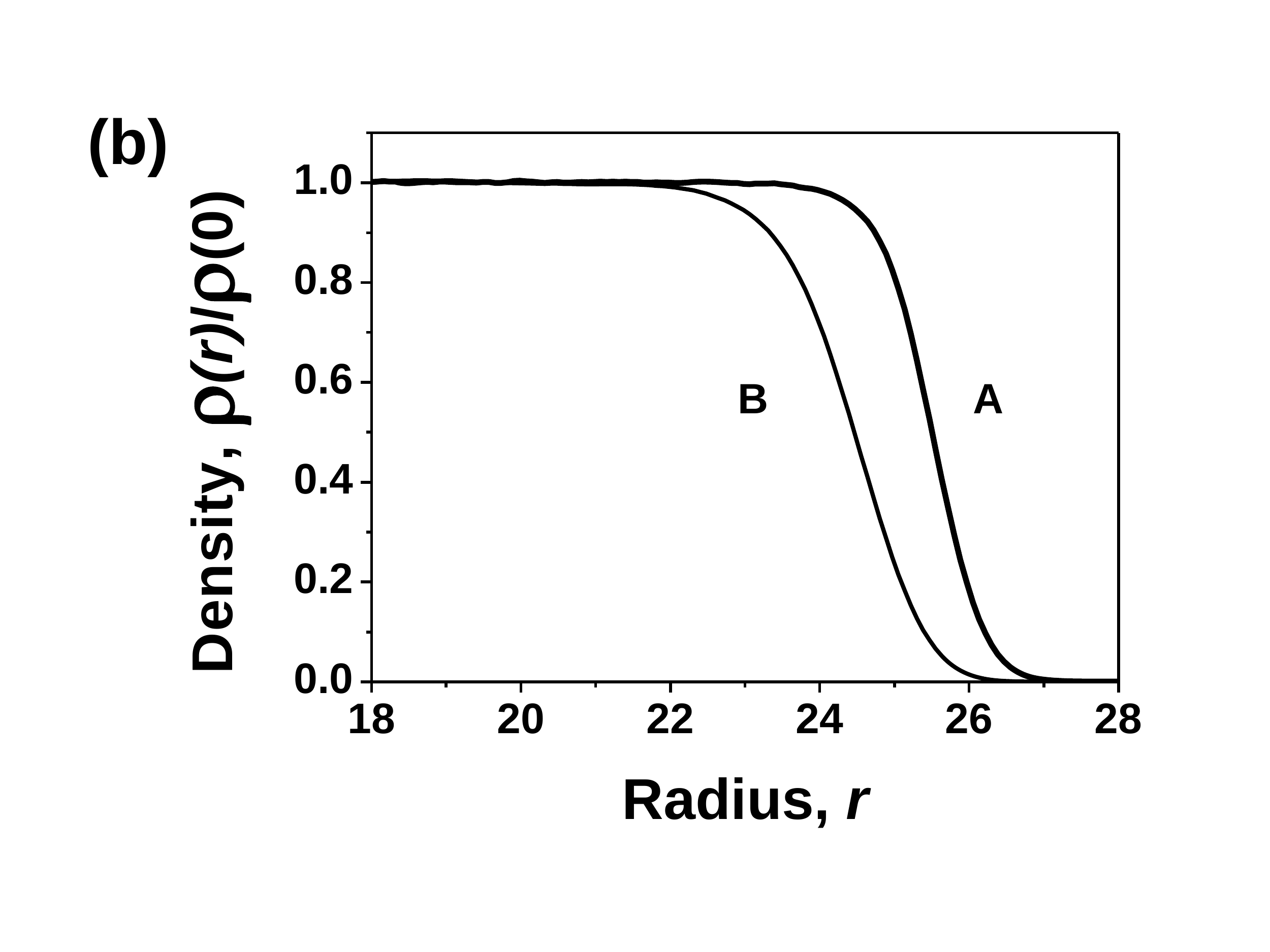}
\end{center}
\caption{Equilibrium density profile in a binary LJ liquid drop consisting of 60000 particles of type A and 15000 particles of type B at $T=0.5$ as a function of radius $r$ measured from the centre of mass, where (a) absolute density (b) normalised density. The results of MD simulations are shown by symbols (only every 10th point is shown), while the solid lines are the fit (\ref{fit-den}). } \label{Fig13}
\end{figure}

\begin{figure}
\begin{center}
\includegraphics[trim=1cm 1cm 1cm 1cm,width=0.5\columnwidth]{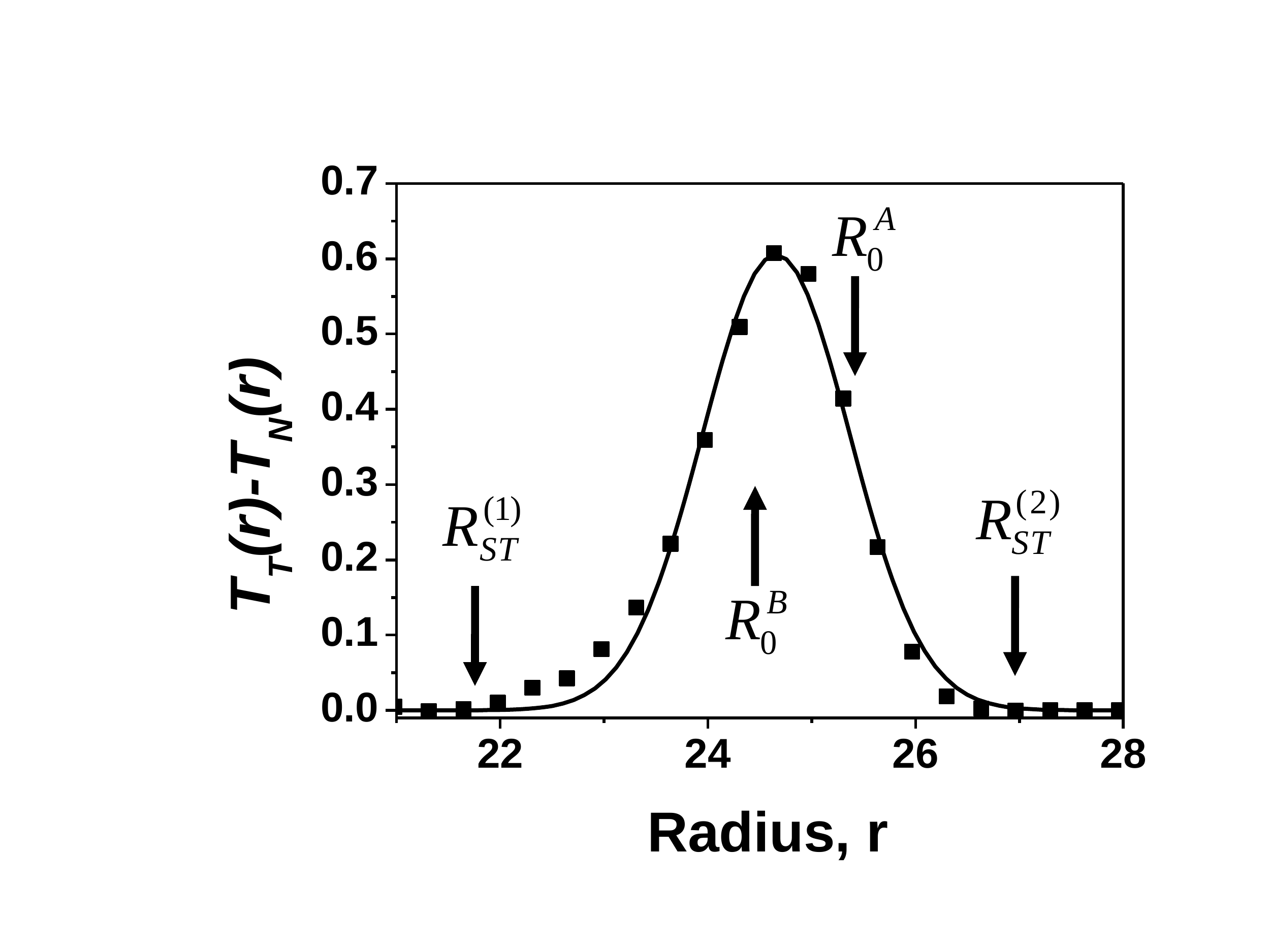}
\end{center}
\caption{Equilibrium distribution of $T_T({r})-T_N({r})$ in a binary LJ liquid drop consisting of $60000$ particles of type $A$ and $15000$ particles of type $B$ at $T=0.5$ as a function of radius $r$. The mean positions of the density profiles $R_0^{A,B}$ are shown by the arrows.} \label{Fig14}
\end{figure}

Typical density profile found in a drop consisting of $60000$ particles of type $A$ and $15000$ particles of type $B$ (similar to the total number of particles used in the study of monatomic liquid drops in the previous section) is shown in Fig. \ref{Fig13}a,b. One can see that at equilibrium the first layer of molecules at the interface only consists of particles of type $A$, which are slightly bigger than $B$. This interfacial structure is typical in the whole range of temperatures used in this study, as one can see from Table \ref{Table5}, where the position and the width of the equilibrium density profiles of both components are presented.  One can notice that intrinsic widths of the density profile for both components are very similar to the ones found in monatomic LJ liquid drop at equivalent temperatures, the density profile of $B$ particles sitting deeper inside the drop and being wider than that of particles $A$. The distribution of stresses $T_T-T_N$, Fig. \ref{Fig14},  is located inside the drop and is shifted by  $\delta r \simeq 1$ relatively to the density profile of particles $A$, as it is in the case of monatomic LJ liquids. 

\begin{figure}
\begin{center}
\includegraphics[trim=1cm 1cm 1cm 1cm,width=0.4\columnwidth]{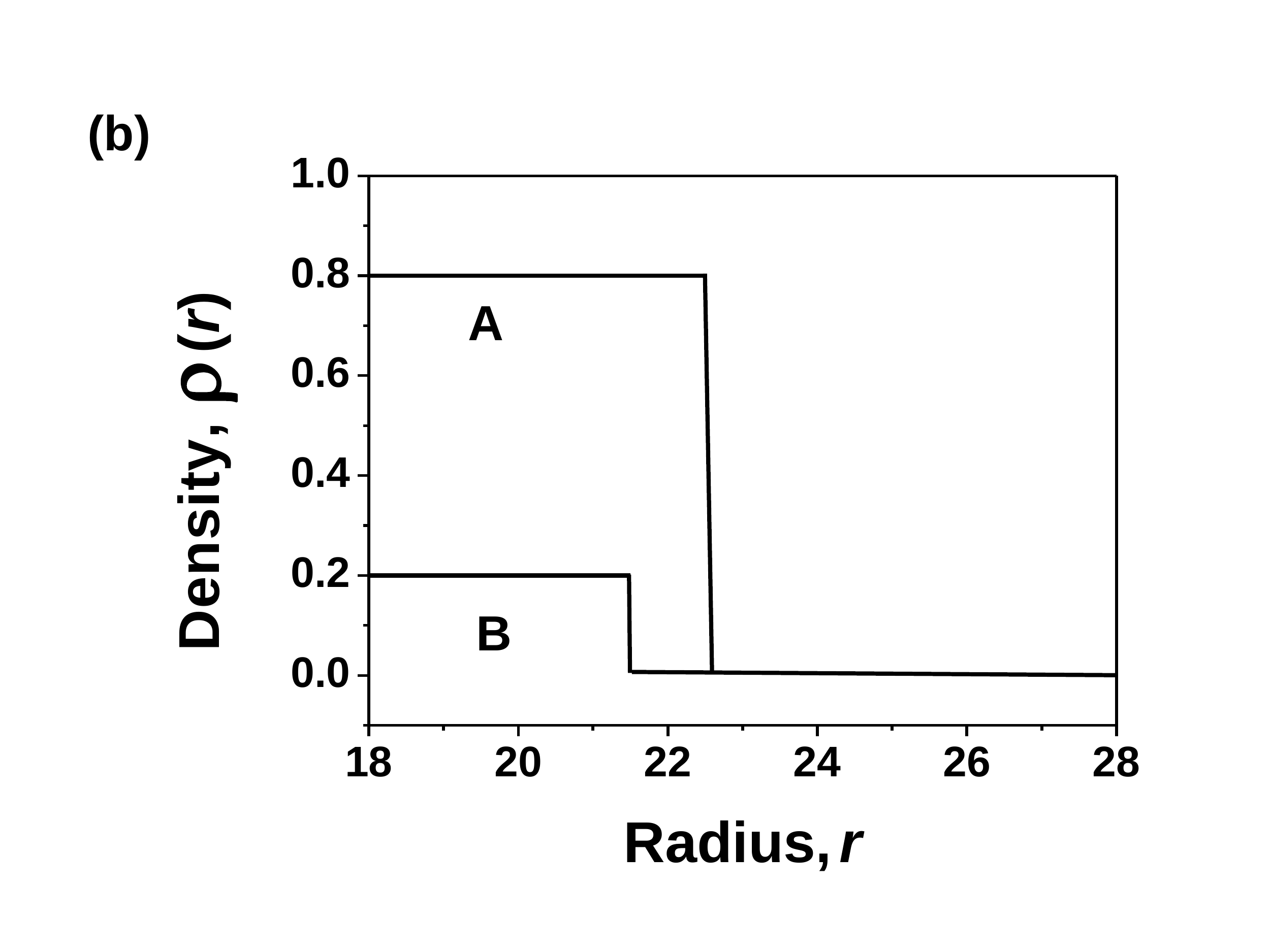}
\includegraphics[trim=1cm 1cm 1cm 1cm,width=0.4\columnwidth]{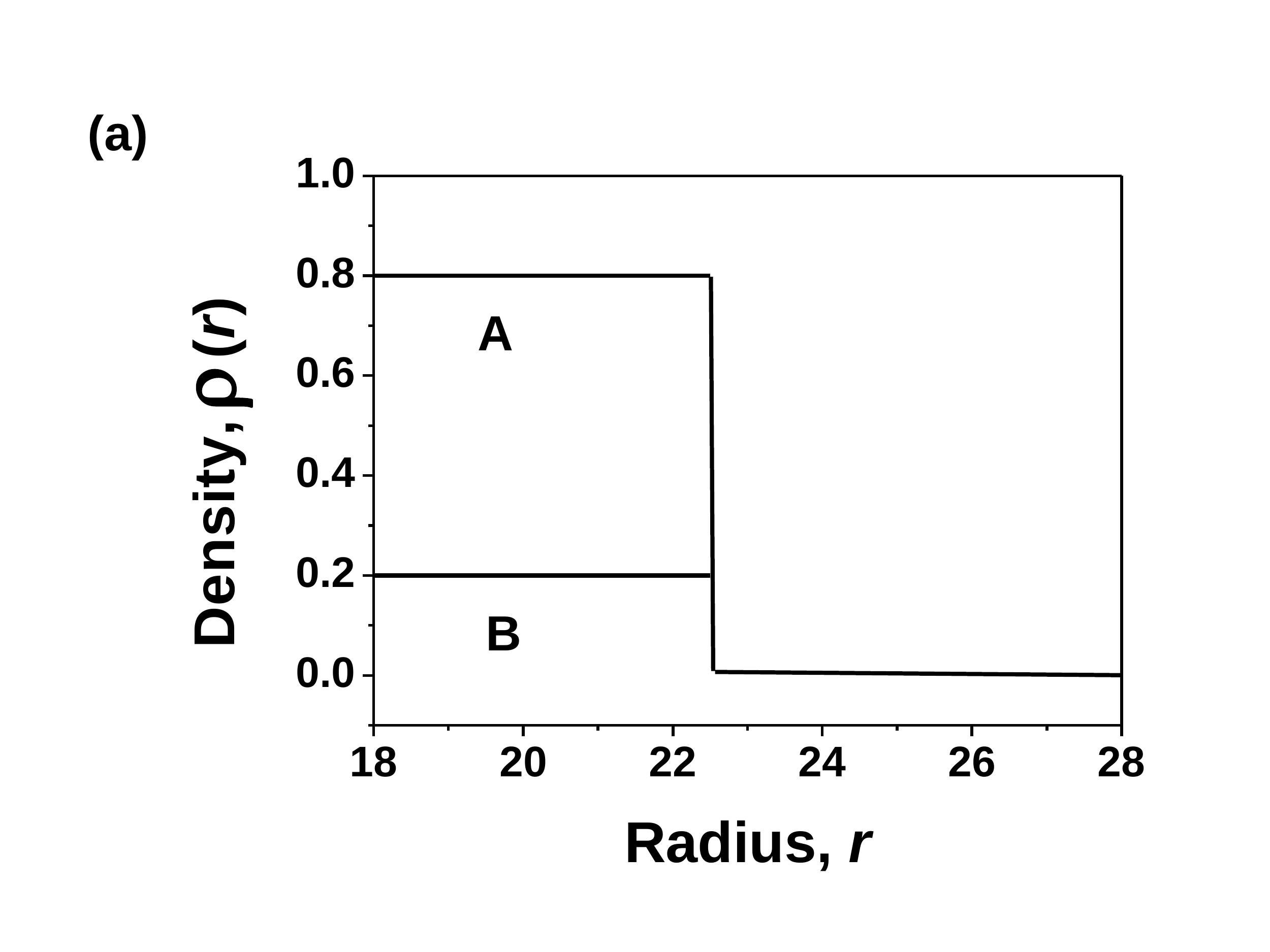}
\end{center}
\caption{Illustration of the bulk and proportional cut off in binary LJ liquid drops.} \label{Fig15}
\end{figure}

We now perform similar numerical experiments by creating fresh interfacial profile by means of removing particles of the equilibrium interfacial layer. The structure of the interfacial layer suggests that one can create two different systems with sharp interfaces. In the first case, one can just remove all particles at $r>R_{ST}^{(1)}$, Fig. \ref{Fig15}a,  which we shall call a bulk cut off. It is easy to see that in this case one can anticipate an additional time scale associated with restoration of the two-layer boundary system from the homogeneous bulk structure.  That is, the excess of particles $B$ at the interface with respect to the equilibrium values have to redistribute by diffusion into the bulk area before the system can reach an equilibrium state. In this case, one expects a much longer relaxation time in comparison to the relaxation on the length scale of $\Delta_{A,B}^0$. Indeed, this process is associated with diffusion of $B$ particles over the distance of at least two particle diameters (corresponding to the double-layer interfacial structure) $\Delta_{DL} \sim 2$. That is $\max(\tau_{\gamma})\sim \frac{\Delta_{DL}^2}{D_B}\sim 200$ at $D_B=0.02$, $T=0.55$. In the second case, which is designated as a proportional cut off, one can remove particles $A$ at $r>R_{ST}^{(1)}$ (as in the case of the bulk cut off), but keeping the same number ratio as in the original Kob-Anderson model $N_A/N_B=4$. The proportional cut off is shown in Fig. \ref{Fig15}b. Effectively, this procedure means that after the bulk cut off, we also remove some particles $B$ from approximately the first layer of the newly formed sharp interface to achieve exactly the ratio $N_A/N_B=4$.  In this case, one might expect almost identical behaviour to what was found in the case of one-component monatomic LJ liquid drops.

\subsection{Proportional cut off in binary liquid drops}

\begin{figure}
\begin{center}
\includegraphics[trim=1cm 1cm 1cm 1cm,width=0.5\columnwidth]{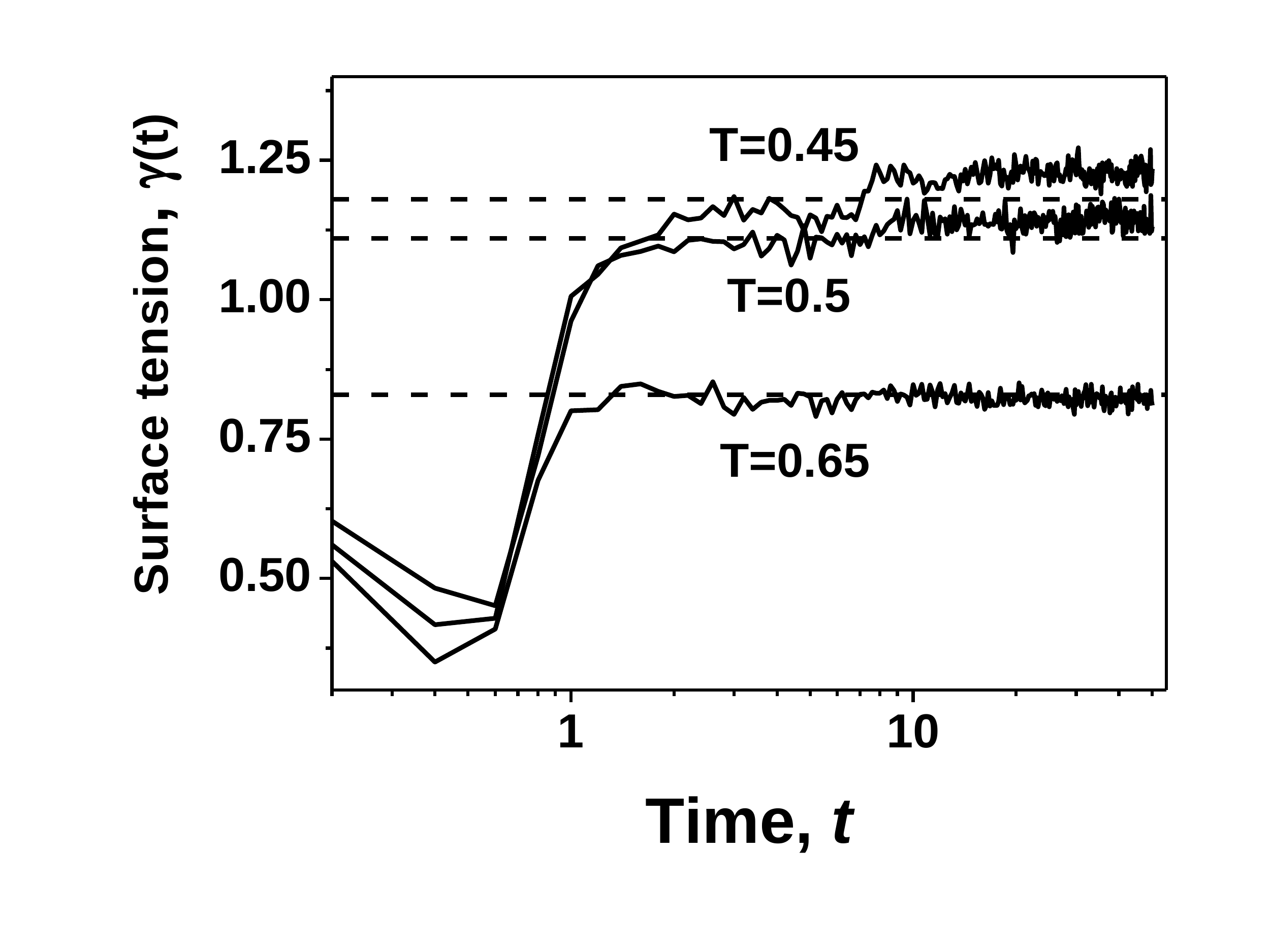}
\end{center}
\caption{Evolution of surface tension $\gamma(t)$ in binary LJ liquid drops at different temperatures after the proportional cut off. The dashed lines correspond to equilibrium values of surface tension.} \label{Fig16}
\end{figure}

\begin{figure}
\begin{center}
\includegraphics[trim=1cm 1cm 1cm 1cm,width=0.5\columnwidth]{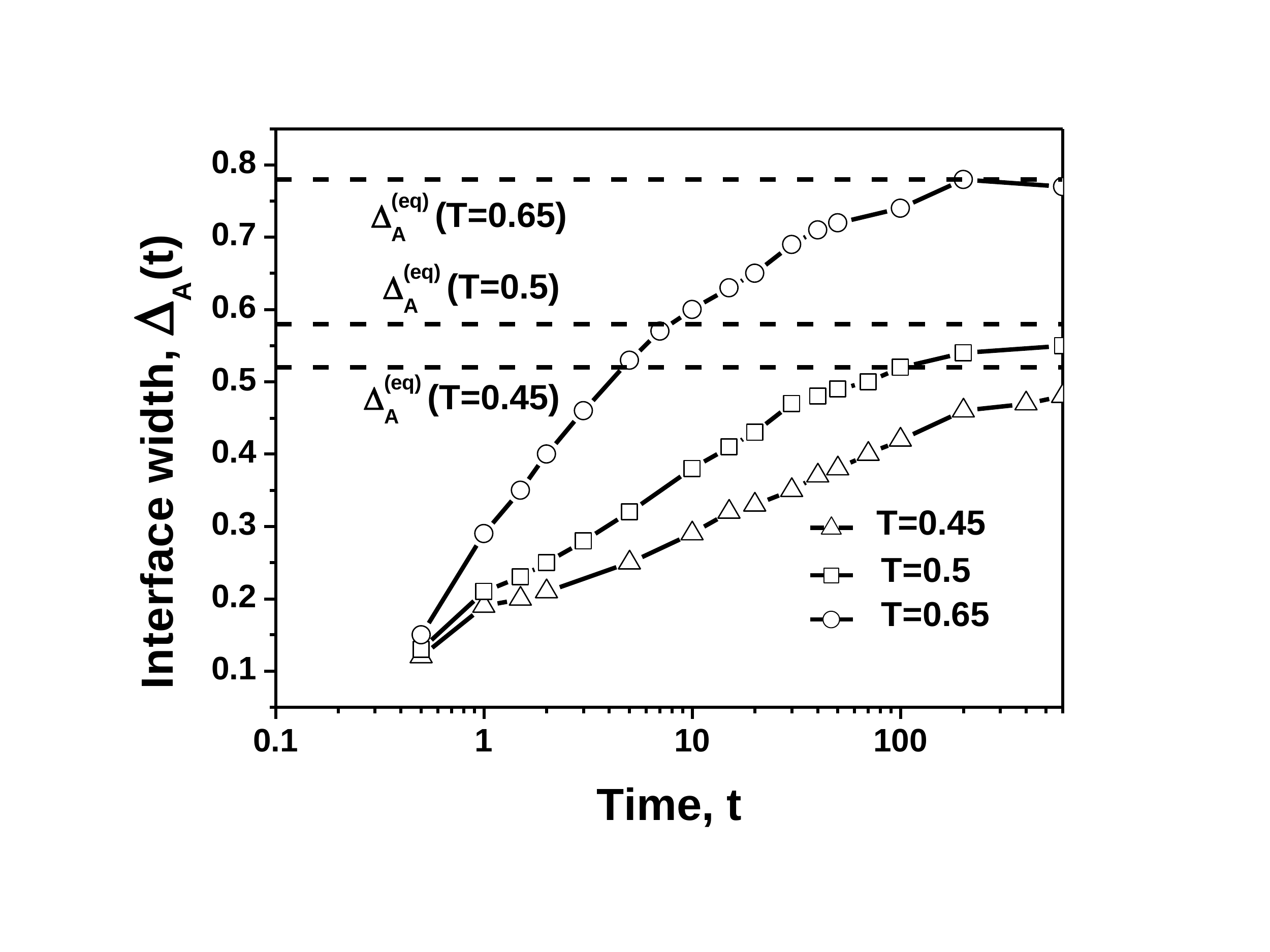}
\end{center}
\caption{Evolution of the width of the density profiles $\Delta_{A}(t)$ in binary LJ liquid drops at different temperatures after the proportional cut off. The dashed lines correspond to equilibrium values $\Delta_{A}^{(eq)}$. The profile width has been obtained with approximation $\approx 4\%$.} \label{Fig17}
\end{figure}

\begin{figure}
\begin{center}
\includegraphics[trim=1cm 1cm 1cm 1cm,width=0.5\columnwidth]{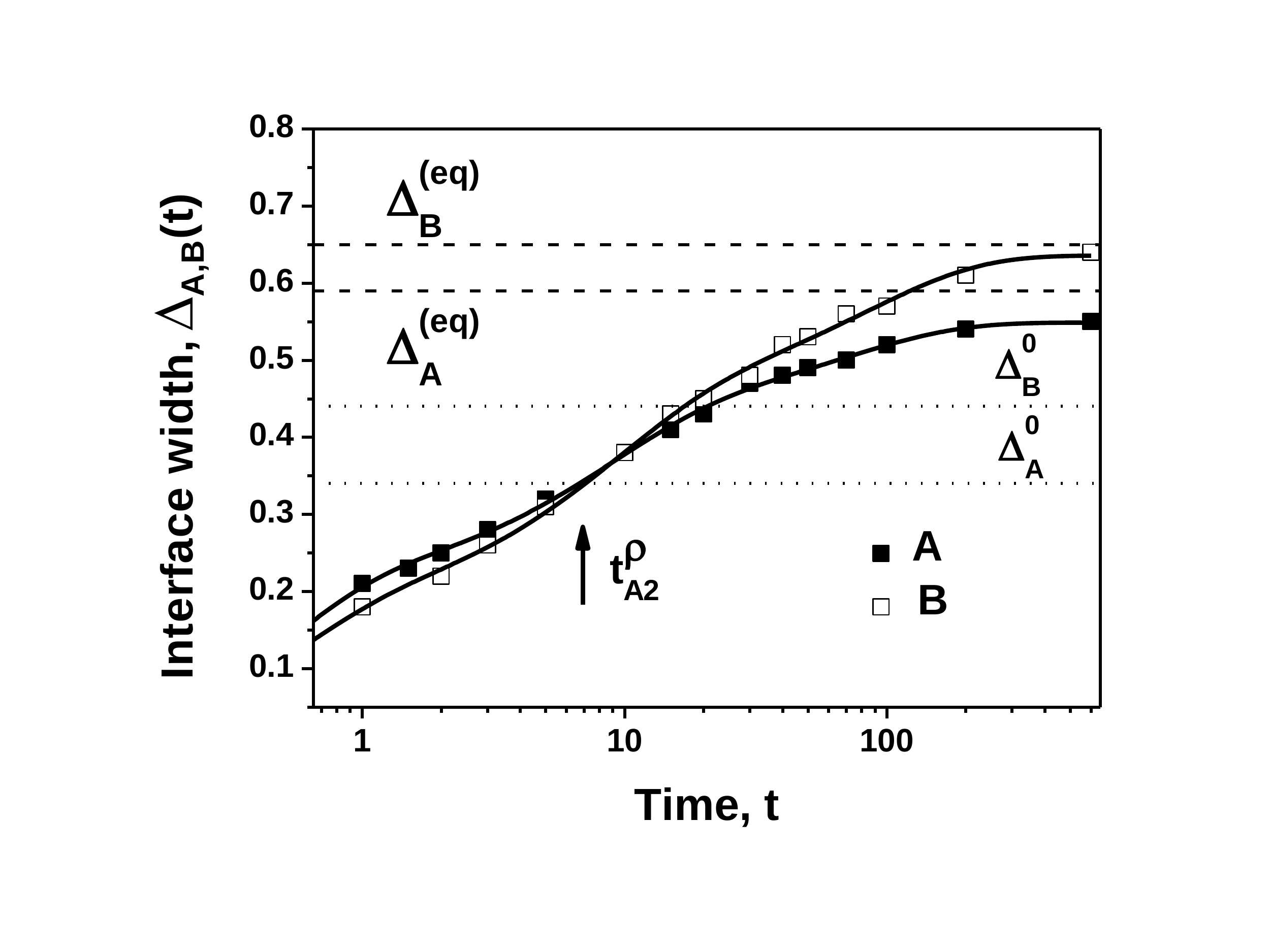}
\end{center}
\caption{Evolution of the width of the density profiles $\Delta_{A,B}(t)$ in binary LJ liquid drops at $T=0.5$ after the proportional cut off. The result of MD simulations is shown by symbols and the solid lines are the fit (\ref{ExpF}). The profile width has been obtained with approximation $\approx 4\%$.} \label{Fig18}
\end{figure}

\begin{figure}
\begin{center}
\includegraphics[trim=1cm 1cm 1cm 1cm,width=0.5\columnwidth]{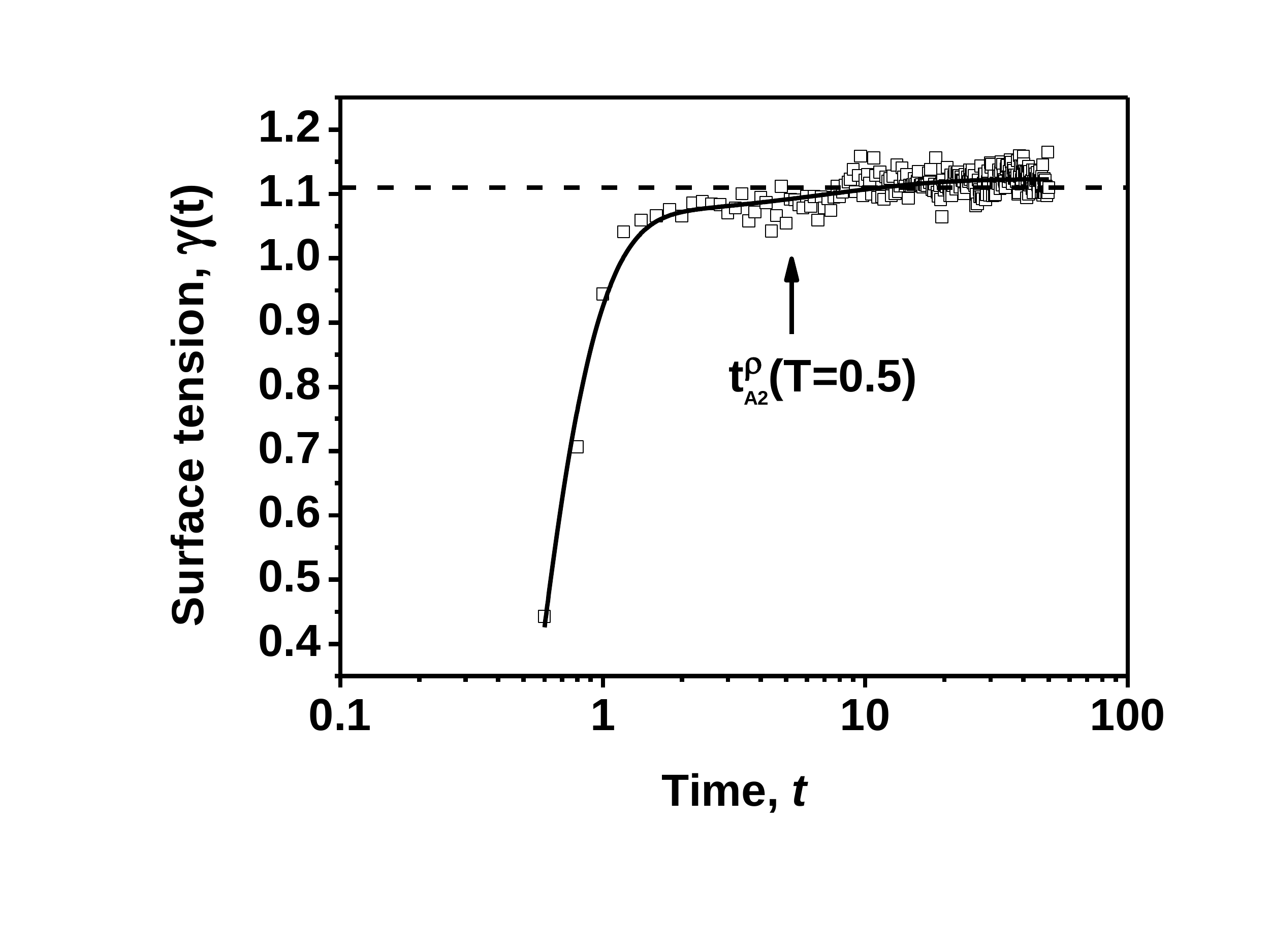}
\end{center}
\caption{Evolution of the surface tension $\gamma(t)$ as a function of time in a binary LJ liquid drop consisting of  $\approx 40000$ particles after the proportional cut off at $T=0.5$ (MD simulations, symbols) and the fit (\ref{FitST}) (solid line) at $A_1=-6.1$, $A_2=-0.06$, $\tau_{\gamma}=7.2\pm1.6$ and $\tau_{\gamma}^{1}=0.27\pm 0.02$. The dashed line corresponds to equilibrium value of surface tension.} \label{ST-relaxation-time-bin}
\end{figure}

\begin{figure}
\begin{center}
\includegraphics[trim=1cm 1cm 1cm 1cm,width=0.4\columnwidth]{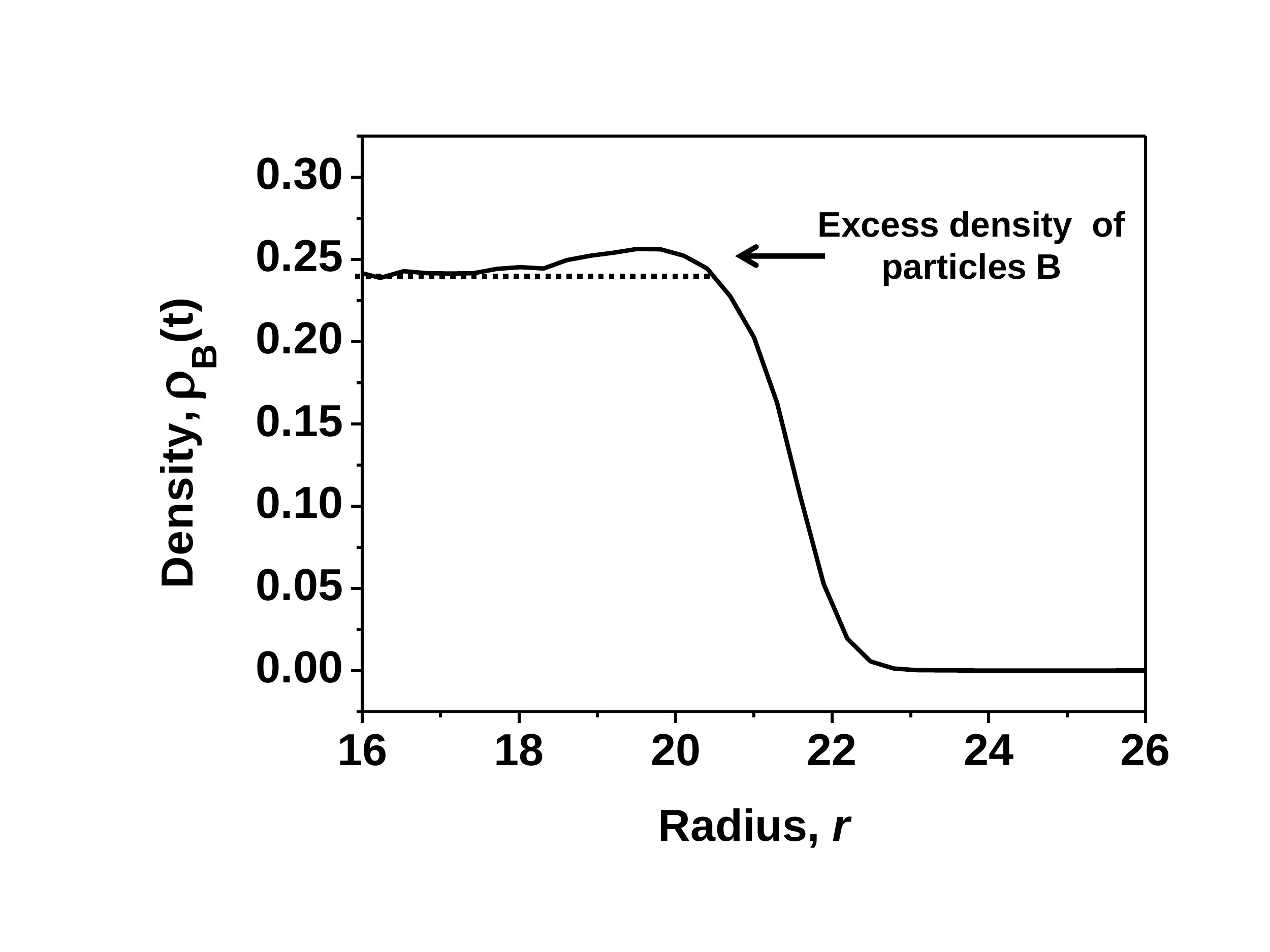}
\end{center}
\caption{Density profile of the $B$ component at $t=50$ in a binary liquid drop at $T=0.5$  with $\approx 40000$ particles after the proportional cut off.} \label{Fig18new}
\end{figure}

Consider first numerical results obtained by means of the proportional cut off. Evolution of the surface tension $\gamma(t)$ and the width of the density profile of particles $A$, $\Delta_A(t)$, is shown in Figs. \ref{Fig16}--\ref{Fig17} at different temperatures. One can see that initially, the dynamic behaviour of surface tension is reminiscent of that found in monatomic drops. After the initial microscopic stage of relaxation, surface tension grows to a value, which is close to that at equilibrium. At lower temperatures, for example at $T=0.45$, one can observe a long tail of overshoot relaxation, although the value of surface tension is only slightly above the equilibrium value. As we will see a bit later, this is connected with the diffusion of particles $B$ into the bulk of the droplet. Quantitatively, this dynamics is also illustrated through the parameters of the Gaussian fit (\ref{GaussFit}) listed  in Table. \ref{Table7}.

If we compare the density profiles $\Delta_{A,B}(t)$, Fig. \ref{Fig18}, one can see that their evolution is almost synchronised and both density profiles of particles $A$ and $B$ are moving together. They start to diverge only during the late stage of relaxation, since $\Delta_A^0<\Delta_B^0$. The exponential fit (\ref{ExpF}) applied to $\Delta_{A}(t)$ and  $\Delta_{B}(t)$ reveals three characteristic time scales for both components, which are presented in Table \ref{Table6}. One can see that the second relaxation time, which has been associated previously with the macroscopic relaxation time of the surface tension, is of the same magnitude for both profiles, though $t_{A2}^{\rho} \le  t_{B2}^{\rho}$ as expected.  At the same time, while the width of both profiles at the end of this stage, $t\approx t_{A2}^{\rho}, t_{B2}^{\rho}$, is approximately the same $\Delta_A(t_{A2}^{\rho})\simeq \Delta_B(t_{B2}^{\rho})$,  the density profile of $A$ component plays the leading role here. This is because  $\Delta_A(t_{A2}^{\rho})\simeq \Delta_B(t_{B2}^{\rho})  \simeq \Delta_A^0$, but $\Delta_B(t)$ at $t=t_{B2}^{\rho}$ is still far away from its equilibrium value $\Delta_B^{0}$. This kind of behaviour is directly related to the fact that while we have tried to reproduce a structure of the interfacial layer after the proportional cut off, which would be congruent to the equilibrium structure, we still obtained a profile which needs for relaxation some $B$ particles to diffuse into the bulk area over much larger distance than $\Delta_B^{0}$. This may be evidenced from the density profile of particles $B$ at $t=50>t_{B2}^{\rho}$ after the cut off, where one can see a small bulge, Fig. \ref{Fig18new}. The effect is not very pronounced and is only related to the fact that equilibrium profile of particles $B$ is wider than that of particles $A$.  As a consequence, the surface tension is larger than the equilibrium value and the long tail of overshoot relaxation of the surface tension is observed, Fig.~\ref{Fig16}. 

If we now compare the second relaxation time $t_{A2}^{\rho}$ and its temperature scaling with that of  $\tau_{\Delta_A^0}$  or $\tau_{\Delta_B^0}$, Table \ref{Table6}, we see that there is a strong correlation between all of them indicating that as in the previous case of monatomic liquids, the second relaxation time scale of the density profile is related with relaxation on the length scale of its intrinsic width.  If we now define the surface tension relaxation time by eq. (\ref{FitST}), as we did in the case of monatomic liquid drops, then one can find that $1\le \tau_{\gamma}\le 2$ at $T=0.65$, $\tau_{\gamma}\approx 7.2\pm1.6$ at $T=0.5$, $\tau_{\gamma}\approx 6.5\pm 1.2$ at $T=0.45$. As it was the case in monatomic liquid drops, the fit is applied after the first microscopic stage of relaxation is completed,  see Fig. \ref{ST-relaxation-time-bin} for illustration. From that one can see that the numerical values of two time scales $t_{A2}^{\rho}$ or $\tau_{\Delta_A^0}$ and $\tau_{\gamma}$ and their temperature dependence are in good agreement. It is interesting to note that the relaxation time scaling, for example $\tau_{\Delta_A^0}$, with temperature has opposite trend than that found in the case of monatomic liquids, namely the relaxation time grows when the temperature is getting smaller. This means that the coefficient of self-diffusion decreases with decreasing temperature faster than the intrinsic width of the interface decreases.

Our results imply that as in the previous case of monatomic liquids, one can define the surface tension relaxation time through the characteristic time $\tau_{\Delta_A^0}$ of the dominant component
$$
\tau_{\gamma} = \tau_{\Delta_A^0}.
$$

The third relaxation time of the density profile width $t_{A3}^{\rho}$, is correlated well with the capillary wave longest relaxation time, calculated by means of eq. (\ref{overD}) with $l=2$, see Table \ref{Table6}. On the other hand, similar time scale for particles $B$, $t_{B3}^{\rho}$, is difficult to identify exactly as it seems to be related to both processes, effective diffusion on the length scale $\Delta_{DL}>1$ into the bulk area and the excitation of capillary waves.  

\subsection{Bulk cut off in binary liquid drops}

\begin{figure}
\begin{center}
\includegraphics[trim=1cm 1cm 1cm 1cm,width=0.4\columnwidth]{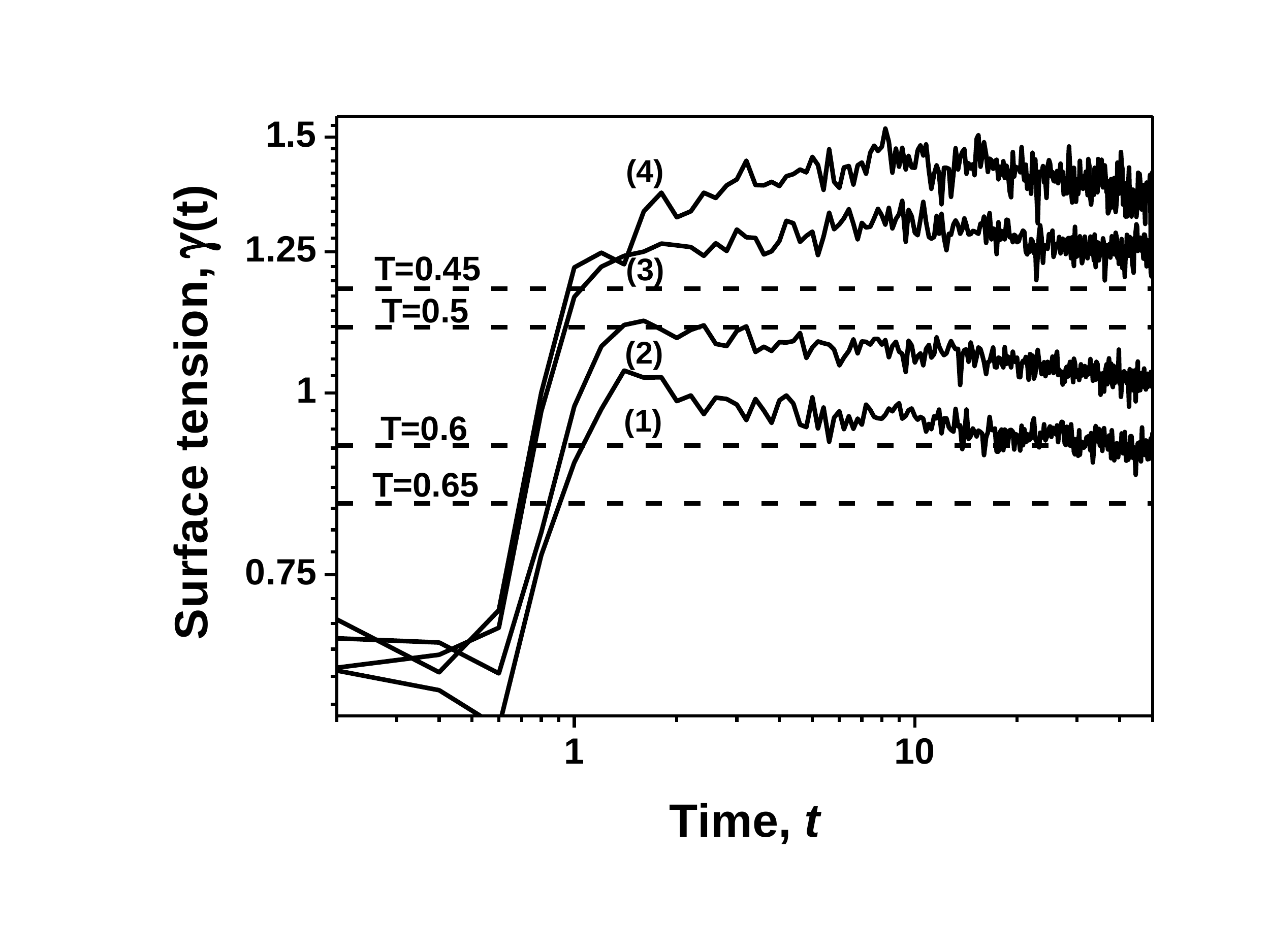}
\end{center}
\caption{Evolution of surface tension $\gamma(t)$ in binary LJ liquid drops at different temperatures after the bulk cut off: (1) $T=0.65$, (2) $T=0.6$, (3) $T=0.5$, (4) $T=0.45$. The dashed lines correspond to equilibrium values of surface tension.} \label{Fig19}
\end{figure}

\begin{figure}
\begin{center}
\includegraphics[trim=1cm 1cm 1cm 1cm,width=0.4\columnwidth]{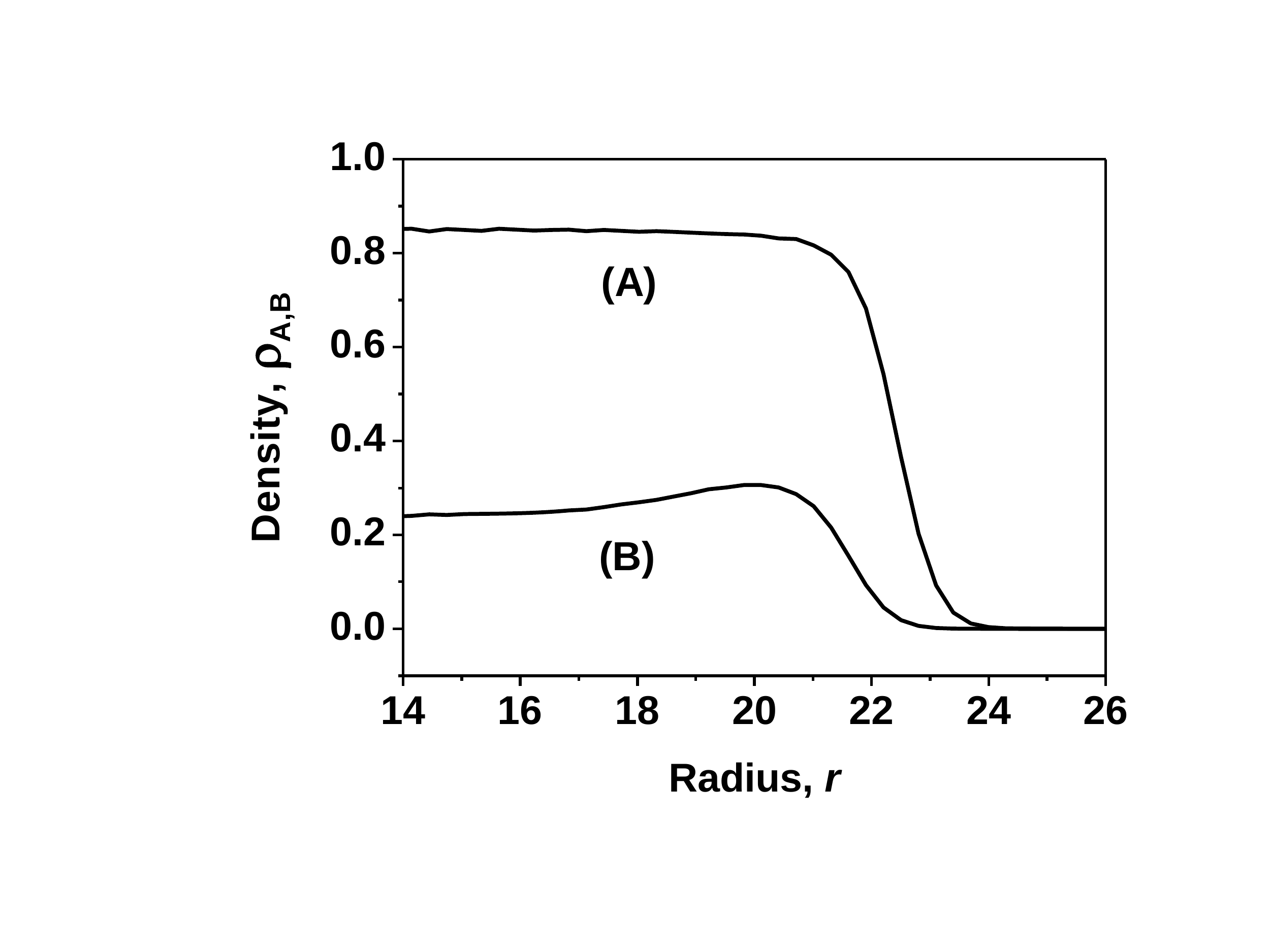}
\end{center}
\caption{Density profiles of $A$ and $B$ components at $t=50$ in a binary LJ liquid drop at $T=0.5$  with $\approx 40000$ particles after the bulk cut off. } \label{Fig20}
\end{figure}

Evolution of interfacial profiles after the bulk cut off at the initial macroscopic stage of relaxation is similar to the evolution after the proportional cut off, Fig.  \ref{Fig19}.  The surface tension in the newly formed sharp interface is at its minimum, as before, which is slightly larger than the half of the equilibrium value. Then the surface tension grows to some value, which is larger than the equilibrium value and is defined by the concentrations of particles $A$ and $B$ at the interface. After the fast macroscopic initial stage of relaxation, defined by the intrinsic width of the interface dominant component and characterised by $\tau_{\gamma 1}=\tau_{\Delta_A^0}$,  the surface tension is moving slowly to the equilibrium  on the time scale controlled by diffusion of particles $B$ into the bulk area, Fig. \ref{Fig20}, to restore the equilibrium structure of the interface. This second slow macroscopic stage of the surface tension relaxation is reminiscent of relaxation in interfaces with surfactants, [10, 11]. We fit evolution curves of $\gamma(t)$ at $t>10$, when the initial fast stage is completed, with a single exponential function
$$
\gamma(t)=\gamma_0+A e^{-\frac{t}{\tau_{\gamma}^{s}}}.
$$
We can find the characteristic time scales of this slow relaxation process, that is $\tau_{\gamma}^{s}\simeq 320\pm 20$ at $T=0.45$, $\tau_{\gamma}^{s}\simeq 143\pm 23$ at $T=0.5$, $\tau_{\gamma}^{s}\simeq 84\pm 10$ at $T=0.6$, $\tau_{\gamma}^{s}\simeq 75\pm 10$ at $T=0.65$. The slow relaxation time scale is apparently related to the diffusion of particles $B$ into the bulk area, which is evidenced from the fact that their product, $\tau_{\gamma}^s D_B$, is almost constant with temperature $2.1 \le \tau_{\gamma}^s D_B\le 2.9$, while both $D_B$ and $\tau_{\gamma}^s$ undergo an order of magnitude changes. The equivalent length scale $\Delta_{DL}^2=2\tau_{\gamma}^{s} D_B$ is then estimated as $2.1<\Delta_{DL}<2.4$, which is close to $2$ corresponding to the double-layer structure of the binary liquid interface.

\section{Conclusions}

In conclusion, we have studied the process of recreation of a liquid-gas interface from an initially sharp interfacial density profile with essentially bulk local structure using MD simulations. We have investigated three different scenarios: recreation of a liquid-gas interface in simple monatomic LJ liquids, in the binary LJ liquids after the proportional and the bulk cut off procedures. 

We have established that the surface tension relaxation time in a liquid-gas interfacial layer of monatomic LJ liquids is directly related to the diffusion-like process on the length scale of intrinsic width of the interface $\Delta_0^{(eq)2}$ and can be calculated by means 
\begin{equation}
\tau_{\gamma}= \frac{\Delta_0^{(eq)2}}{2D}
\label{tau}
\end{equation} 
if the coefficient of self-diffusion $D$ in the bulk is given. We have shown that this process of the surface tension relaxation is separated from another dynamic process associated with liquid-gas interfaces, the excitation of capillary waves. During this relaxation process, the surface tension grows in time from a minimal value $\gamma_{min}$ achieved at a sharp interfacial layer. The minimal value of the surface tension $\gamma_{min}$ constitutes about $60\%$ of the equilibrium value $\gamma_0$. 

As is seen from eq. (\ref{tau}), the surface tension relaxation time is inversely proportional to the coefficient of self-diffusion, which in turn should be inversely proportional to the liquid viscosity $\mu$. This trend found in our MD simulations is consistent with what was anticipated in [12] on the basis of macroscopic arguments, that is $\tau_{\gamma}\propto \mu$. However, the absolute value of surface tension relaxation time found in our MD experiments is three orders of magnitude smaller than that estimated in macroscopic analysis [12, 16, 17]. For example for water, $3\times 10^{-9}\, \mbox{s} <\tau_{\gamma}^{m}< 10^{-8}\, \mbox{s}$ from [12] at room temperature, while from eq. (\ref{tau}) using $\Delta_0^{(eq)}=8\times 10^{-11}\,\mbox{m}$ and $D=2\times 10^{-9}\,\mbox{m}^2/\mbox{s}$ [8, 9], we have $\tau_{\gamma}=1.6\times 10^{-12}\,\mbox{s}$. This discrepancy is to the large extent due to much smaller interfacial width, the intrinsic width $\Delta_0^{(eq)}=8\times 10^{-11}\,\mbox{m}$ found in [8] (though this value may be essentially underestimated [61]) than that originally anticipated in the macroscopic studies $\Delta_0^{(eq)}\sim  10^{-9}\,\mbox{m}$, [12]. The coefficient of proportionality $\tau^{\dag}$ between surface tension relaxation time and viscosity $\tau_{\gamma}^m=\tau^{\dag} \Delta_0^{(eq)2} \mu$ from [12] is $\tau^{\dag}=3\times 10^{12}\,\mbox{m}^{-2}\mbox{Pa}^{-1}$ for water, if we take $\mu=1\,\mbox{mPa s}$, $\tau_{\gamma}^m=3\times 10^{-9}\,\mbox{s}$ and $\Delta_0^{(eq)}=10^{-9}\,\mbox{m}$. Then for $\Delta_0^{(eq)}=8\times 10^{-11}\,\mbox{m}$ obtained in [8], we would have, using $\tau^{\dag}$, $\tau_{\gamma}^m\approx 2\times 10^{-11}\,\mbox{s}$. This value is only one order of magnitude away from $\tau_{\gamma}=1.6\times 10^{-12}\,\mbox{s}$.

In case of binary LJ liquids using two different ways of preparation of sharp interfaces (the bulk and proportional cut off), we have found that in general there are two different macroscopic relaxation time scales. The first, shortest relaxation time is associated with the diffusion-like process on the length scale of the intrinsic width of the first interfacial layer (consisting of $A$ particles in our case), similar to the process in monatomic liquids. Surface tension grows in time during this relaxation process from a minimum value $\gamma_{min}$ with the characteristic relaxation time, which can be calculated as
$$
\tau_{\gamma 1}= \frac{\Delta_{A}^{02}}{2D_A}.
$$

A comparison of the characteristic time scales has shown that this fast relaxation process is also separated from the excitation of capillary waves. The second, longer relaxation time is found to be related with the restitution of the double-layer interfacial structure of binary liquids. This characteristic time can be calculated by means of
$$
\tau_{\gamma 2}=\frac{\Delta_{DL}^2}{2D_B},
$$
which is $\displaystyle \tau_{\gamma 2} \approx \frac{2}{D_B}$ at $\Delta_{DL}\approx 2$. It is not difficult to see that $\tau_{\gamma 2}$ is on the time scale of the excitation of capillary waves. We have not studied any connection between the capillary wave dynamics and the slow surface tension relaxation in the present work and left this for future analysis. We note here that a coupling between capillary waves and an "intrinsic" density profile has been studied using the density-functional approach for binary mixtures in [50].

In summary, we have shown that although the surface tension can be about factor of
$2$ lower than the equilibrium value after fresh interface creation, the
equilibrium value is restored rather quickly. This relaxation time can be
accurately predicted as the characteristic diffusion time on the scale of intrinsic
width of the density profile, which is extremely small (less than the atom
size). It is however larger than the bulk stress relaxation time, so in
principle it is possible to create experimental conditions when liquid is still
Newtonian but the interfacial tension is out-of-equilibrium. The fast
relaxation can overshoot, resulting in surface tension above equilibrium value.
This scenario, observed in binary mixtures, is expected to be quite generic if
the system possesses some order parameters different from the overall density
(concentration of $A/B$ molecules in our case). In such case we expect a fast
increase of the surface tension followed by a slower decrease towards equilibrium, as
illustrated in Fig. \ref{Fig19}.

\appendix
\appendixpage 
\section{\bf Microscopic stress tensor and calculation of surface tension in the drops}
The value of the surface tension generated in the interfacial layer is calculated from the microscopic stress tensor defined by [29, 37] 
$$
T_{\alpha\beta}({\bf r})=\frac{1}{2}\sum _{i} \sum _{j\neq i} \left\langle  \frac{r_{ij}^{\alpha}}{r_{ij}} \frac{dU(r_{ij}) }{dr_{r_{ij}}} \int_{C_{ij}} dl^{\beta}\delta(\bf r-l) \right\rangle - \left\langle \sum _{i}  v_i^{\alpha}   v_i^{\beta}  \delta(\vect{r}-\vect{r}_i) \right\rangle 
$$
where ${\bf v}_i$ is the velocity of particle $i$, $C_{ij}$ is a straight line connecting particles $i$ and $j$, which corresponds to the Irving-Kirkwood choice of the contour connecting the interacting particles [56], 
$$
{\bf l}=\frac{1}{2}\left\{ {\bf r}_i +{\bf r}_j +\lambda\, {\bf r}_{ij}   \right\}, \quad {\bf r}_{ij} ={\bf r}_j-{\bf r}_i, \qquad -1\le \lambda\le 1
$$
and 
$\left\langle ... \right\rangle$ is ensemble average. In the case of a spherically symmetric system, after the averaging over angular coordinates, the normal and tangential components of the microscopic stress tensor can be represented as

$$
\begin{array}{l}
\displaystyle T_N(r)=\frac{1}{4\pi r^2}\sum_{k}c_{ij} \frac{dU(r_{ij}) }{dr_{r_{ij}}} -T \rho({\bf r}), \\
\displaystyle T_T(r)=\frac{1}{8\pi r^2}\sum_{k} c_{ij} \frac{dU(r_{ij}) }{dr_{r_{ij}}}\left\{  \left(c_{ij}\right)^{-2}-1 \right\} - T \rho({\bf r}),
\end{array}
$$
where
$$
c_{ij}=\frac{1}{2}\frac{r_{ij}}{r} \left\{   \left( \frac{r_i^2-r_j^2}{r_{ij}} \right)^2+1-2\left( \frac{r_i^2+r_j^2}{r_{ij}} \right)+\frac{4r^2}{r_{ij}^2}  \right\},
$$
$r_{ij}=|{\bf r}_{ij}|$ and the summation over $k$ goes over all the intersections of the line connecting all particles $i$ and $j$ and the surface of observation $r=const$ [37]. The surface tension is then given by an integral, which is in the macroscopic limit $\Delta/R_0\to 0$ (the only limit we are interested in the current study of sufficiently large drops) 
$$
\gamma=\int_0^{\infty} \frac{R_s}{r}(T_T-T_N) \, dr \approx \int_0^{\infty} (T_T-T_N) \, dr,
$$
where $R_s\approx R_0$ (with the accuracy $\delta_{Tl}/R_0$) is the surface of tension and the actual integration takes place in the interval $r\in [R_{ST}^{(1)}, R_{ST}^{(2)},]$, that is between the points where the integrand distribution vanishes, see Fig. \ref{Fig4}. 

\begin{figure}
\begin{center}
\includegraphics[trim=1cm 1cm 1cm 1cm,width=0.6\columnwidth]{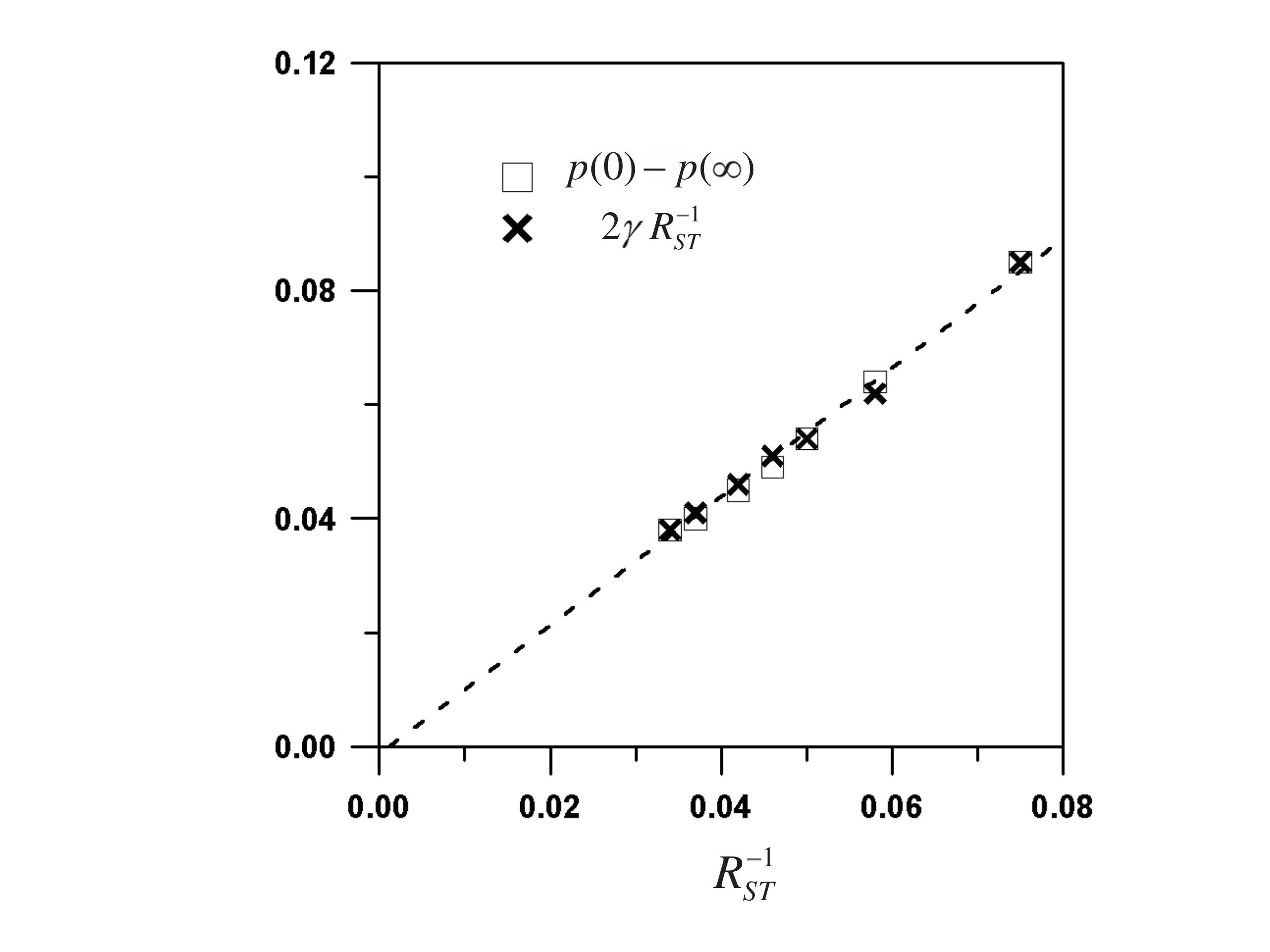}
\end{center}
\caption{Pressure difference $p(0)-p(\infty)$ as a function of the inverse drop radius $R_{ST}^{-1}$ at $T=0.7$ - empty boxes, $\frac{2\gamma}{R_{ST}}$ at $\gamma=0.55$ - crosses and the straight line is a fit  $\frac{A}{R_{ST}}$.} \label{Fig5}
\end{figure}

To validate our numerical procedure, we have calculated independently distributions of $T_T({\bf r})$, $T_N({\bf r})$ and pressure $p=-T_{\alpha\alpha}/3$, averaged over $\Delta t_a=1000$, for several drops of different dimensions and compared the result with the macroscopic "capillary" Young--Laplace equation (neglecting corrections due to Tolman length $\delta_{Tl}\sim 0.1\div0.8$, which are important when $R_0<5$, see [32, 33] for details) 
$$
p(0) - p(\infty )=\frac{2\gamma}{R_{st}},
$$
where $R_{st}$ is the point of maximum of the distribution $T_T({\bf r})-T_N({\bf r})$, Fig \ref{Fig4}. One can observe excellent agreement, Fig \ref{Fig5}, with average $\gamma=0.55\pm 0.02$ at $T=0.7$, which is similar to the value obtained in [40].

\vskip0.8truecm
{\bf Acknowledgement.} This work is supported by the EPSRC grant EP/H009558/1. The authors are grateful to Marcus M\"{u}ller and Yulii Shikhmurzaev for useful discussions. 
\vskip0.8truecm

\vskip0.2truecm
\centerline{\rule{5truecm}{0.5pt}}
\vskip0.2truecm

\RefAIP{1} \by{Kowalewski, T.A.} \yr{1996} \paper{On the separation of droplets from a liquid jet} \jour{Fluid Dyn. Res.} \vol{17} \pages{121--145}

\RefAIP{2} \by{Eggers, J.} \yr{1993} \paper{Universal pinching of 3D axisymmetric fre-surface flow} \jour{Phys. Rev. Lett.} \vol{71} \pages{3458--3460} 

\RefAIP{3} \by{Eggers, J.} \yr{1997} \paper{Nonlinear dynamics and breakup of free-surface flows} \jour{Rev. Mod. Phys.} \vol{69} \pages{865--929}

\RefAIP{4} \by{Joseph, D.D. et al.} \yr{1990} \paper{Two-dimensional cusped interfaces} \jour{\JFM} \vol{223} \pages{383--409}

\RefAIP{5} \by{Brocchini, M., Peregrine, D.H.} \yr{2001} \paper{The dynamics of strong turbulence at free surfaces. Part 1. Description} \jour{\JFM} \vol{449} \pages{225--254}

\RefAIP{6} \by{Thoroddsen, S.T., Tan, Y.K.} \yr{2004} \paper{Free-surface entrainment into a rimming flow containing surfactants} \jour{Phys. Fluids} \vol{16} \pages{13--16}

\RefAIP{7} \by{Basaran, O.A.} \yr{2002} \paper{Small-Scale Free Surface Flows with Breakup: Drop Formation and Emerging Applications} \jour{AIChE Journal} \vol{48} \pages{1842--1848}

\RefAIP{8} \by{Ismail, A.E., Grest, G.S., Stevens, M.J.} \yr{2006} \paper{Capillary waves at the liquid-vapor interface and the surface tension of water} \jour{J. Chem. Phys.} \vol{125} \pages{014702}

\RefAIP{9} \by{Ertl, H., Dullien, F.A.L.} \yr{1973} \paper{Self-Diffusion and Viscosity of Some Liquids as a Function of Temperature} \jour{AICHE Journal} \vol{19} \pages{1215--1223}

\RefAIP{10} \by{Miller, R., Joos, P., Fainerman, V.B.} \yr{1994} \paper{Dynamic Surface and Interfacial Tensions of
Surfactants and Polymer Solutions} \jour{Adv. Colloid Interface Sci.} \vol{49} \pages{249--302}

\RefAIP{11} \by{Saint Vincent, M.R.D. et al.} \yr{2012} \paper{Dynamic interfacial tension effects in the rupture of liquid necks} \jour{J. Fluid Mech.} \vol{692} \pages{499--510}

\RefAIP{12} \by {Blake, T. D. \and Shikhmurzaev, Y. D.} \yr{2002} \paper {Dynamic wetting by liquids of different viscosity} \jour{\JCIS} \vol{253} \pages{196--202} 

\RefAIP{13} \by{Eggers, J., Evans, R.} \yr{2004} \paper{Comment on "Dynamic wetting by liquids of different viscosity," by T.D. Blake and Y.D. Shikhmurzaev} \jour{J. Colloid. Interf. Sci.} \vol{280} \pages{537--538}

\RefAIP{14} \by{Shikhmurzaev, Y.D., Blake, T.D.} \yr{2004} \paper{Response to the Comment} \jour{J. Colloid. Interf. Sci.} \vol{280} \pages{539--541}

\RefAIP{15} \by {Pismen L.M.}  \yr{2004}   \paper{Diffuse-interface effects near a cusp singularity on a free surface} \jour{Phys. Rev. E}  \vol{70}  \pages{051604}

\RefAIP{16} \by {Shikhmurzaev, Y. D.}  \yr{2005}  \paper {Capillary breakup of liquid threads: A singularity-free solution} \jour{IMA J. Appl.\ Maths} \vol{70}   \pages{880--907}

\RefAIP{17} \by{Shikhmurzaev, Y.D.} \yr{2007} {\it Capillary Flows with Forming Interfaces.} Taylor \& Francis.

\RefAIP{18} \by{Castrej\'{o}n-Pita, J.R., Castrej\'{on}-Pita, A.A., Hinch, E.J. et al.} \yr{2012} \paper{Self-similar breakup of near-inviscid liquids} \jour{Phys. Rev. E} \vol{86} \pages{015301}

\RefAIP{19} \by{Eggers, J.} \yr{2008} \paper{Physics of liquid jets} \jour{Rep. Prog. Phys.} \vol{71} \pages{ 036601}

\RefAIP{20} \by{Shikhmurzaev, Y.D.} \yr{2011} \paper{Some dry facts about dynamic wetting} \jour{Eur. Phys. J-Spec. Top.} \vol{197} \pages{47--60 }

\RefAIP{21} \by{Blake, T.D.} \yr{2011} \paper{Discussion Notes: A more collaborative approach to the moving contact-line problem?} \jour{Eur. Phys. J-Spec. Top.} \vol{197} \pages{343--345 }

\RefAIP{22} \by{Pismen, L. M.} \yr{2011} \paper{Discussion Notes on "Some dry facts about dynamic wetting", by YD Shikhmurzaev Diffuse interface between mathematical and physical thinking} \jour{Eur. Phys. J-Spec. Top.} \vol{197} \pages{63--65 }

\RefAIP{23} \by{Thiele, U.} \yr{2011} \paper{Discussion Notes: Thoughts on mesoscopic continuum models} \jour{Eur. Phys. J-Spec. Top.} \vol{197} \pages{67--71 }

\RefAIP{24} \by{Henderson, J. R} \yr{2011} \paper{Discussion Notes on "Some dry facts about dynamic wetting", by YD Shikhmurzaev} \jour{Eur. Phys. J-Spec. Top.} \vol{197} \pages{61--62 }

\RefAIP{25} \by{Kawano, S.} \yr{1998} \paper{Molecular dynamics of rupture phenomena in a liquid thread} \jour{Phys. Rev. E} \vol{58} \pages{4468--4472}

\RefAIP{26} \by{Moseler, M., Landman, U.} \yr{2000} \paper{Formation, stability and breakup of nanojets} \jour{Science} \vol{289} \pages{1165--1169}

\RefAIP{27} \by{White, J.A.} \yr{1999} \paper{Lennard-Jones as a model for argon and test of extended renormalization group calculations} \jour{J. Chem. Phys.} \vol{111} \pages{9352--9356}

\RefAIP{28} \by{Tolman, R.C.} \yr{1949} \paper{The Effect of Droplet Size on Surface Tension} \jour{J. Chem. Phys.} \vol{17} \pages{333-337}

\RefAIP{29} \by{Schofield, P., Henderson, J.R. } \yr{1982} \paper{Statistical Mechanics of Inhomogeneous Fluids} \jour{P. Roy. Soc. A-Math. Phy.} \vol{379} \pages{231--246}

\RefAIP{30} \by{Rowlinson, J.S., Widom, B.} \yr{1989} {\it Molecular Theory of Capillarity.} Clarendon, Oxford.

\RefAIP{31} \by{Blokhius, E.M., Bedeaux, D.} \yr{1992} \paper{Derivation of microscopic expressions for the rigidity constants of a simple liquid-vapor
interface} \jour{Physica A} \vol{184} \pages{42-70}

\RefAIP{32} \by{Block, B.J., Das, S.K., Oettel, M., Virnau, P. and Binder, K.} \yr{2010} \paper{Curvature dependence of surface free energy of liquid drops and bubbles:
A simulation study} \jour{J. Chem. Phys.} \vol{133} \pages{154702} 

\RefAIP{33} \by{Tr\"{o}ster, A., Oettel, M., Block, B., Virnau, P. and Binder, K.} \yr{2012} \paper{Numerical approaches to determine the interface tension of curved
interfaces from free energy calculations} \jour{J. Chem. Phys.} \vol{136} \pages{064709} 

\RefAIP{34} \by{Chapela, G.A., Saville, G., Rowlinson, J.S.} \yr{1975} \paper{Computer simulations of gas-liquid surface} \jour{Faraday discuss.} \vol{59} \pages{22--28}

\RefAIP{35} \by{Chapela, G.A., Saville, G., Thompson, S.M. et al.} \yr{1977} \paper{Computer simulation of a gas-liquid surface 1.} \jour{J. Chem. Soc. Faraday Trans. II} \vol{73} \pages{1133--1144}

\RefAIP{36} \by{Powles, J.G., Fowler, R.F., Evans, W.A.B.} \yr{1983} \paper{The surface thickness of simulated microscopic liquid drops} \jour{Phys. Lett.} \vol{98A} \pages{421--425}

\RefAIP{37} \by{Thompson, S.M. et al.} \yr{1984} \paper{A molecular dynamics study of liquid drops} \jour{J. Chem. Phys.} \vol{81} \pages{530--542}

\RefAIP{38} \by{Nijmeijer, M.J.P., Bakker, A.F., Bruin, C. et al.} \yr{1988} \paper{A molecular dynamics simulation of the Lennard-Jones liquid–vapor interface} \jour{J. Chem. Phys.} \vol{89} \pages{3789--3792}

\RefAIP{39} \by{Matsumoto, M., Kataoka, Y.} \yr{1988} \paper{Study on liquid vapor interface of water 1. Simulational results of thermodynamic properties and orientational structure} \jour{J. Chem. Phys.} \vol{88} \pages{3233--3245}

\RefAIP{40} \by{Holcomb, C.D., Clancy, P., Zollweg, J.A.} \yr{1993} \paper{A critical study of the simulation of the liquid-vapour interface of a Lennard-Jones fluid} \jour{Mol. Phys.} \vol{78} \pages{437--459}

\RefAIP{41} \by{Mecke, M., Winkelmann, J., Fischer, J.} \yr{1997} \paper{Molecular dynamics simulation of the liquid-vapor interface: The Lennard-Jones fluid} \jour{J. Chem. Phys.} \vol{107} \pages{9264--9270}

\RefAIP{42} \by{Goujon, F., Malfreyt, P.,   Simon, J.M. et al.} \yr{2004} \paper{Monte Carlo versus molecular dynamics simulations in heterogeneous
systems: An application to the n-pentane liquid-vapor interface} \jour{J. Chem. Phys.} \vol{121} \pages{12559} 

\RefAIP{43} \by{Hansen, J.P., Verlet, L.} \yr{1969} \paper{Phase Transitions of the Lennard-Jones System} \jour{Phys. Rev.} \vol{184} \pages{151--161}

\RefAIP{44} \by{Buff, F.P., Lovett, R.A. and Stillinger, P.H.} \yr{1965} \paper{Interfacial density profile for fluids in the critial region} \jour{Phys. Rev. Lett.} \vol{15} \pages{621--623}

\RefAIP{45} \by{Weeks, J.D.} \yr{1977} \paper{Structure and thermodynamics of liquid-vapor interface} \jour{J. Chem. Phys.} \vol{67} \pages{3106-3121}

\RefAIP{46} \by{Evans, R.} \yr{1979} \paper{The nature of the liquid-vapour interface and other topics in the statistical mechanics of non-uniform, classical fluids} \jour{Adv. Phys.} \vol{28} \pages{143--200}

\RefAIP{47} \by{Evans, R.} \yr{1981} \paper{The role of capillary wave fluctuations in determining the liquid-vapour interface} \jour{Molecular Physics} \vol{42} \pages{1169--1196} 

\RefAIP{48} \by{Lacasse, M.D., Grest, G.S., Levine, A.J.} \yr{1998} \paper{Capillary-wave and chain-length effects at polymer/polymer interfaces} \jour{Phys. Rev. Lett.} \vol{80} \pages{309--312}

\RefAIP{49} \by{Binder, K., Muller, M., Schmid, F. et al.} \yr{2001} \paper{Intrinsic profiles and capillary waves at interfaces between coexisting phases in polymer blends} \jour{Adv. Colloid Interface Sci.} \vol{94} \pages{237--248}

\RefAIP{50} \by{Hiester, T., Dietrich, S., Mecke, K.} \yr{2006} \paper{Microscopic theory for interface fluctuations in binary liquid mixtures.} \jour{J. Chem. Phys.} \vol{125} \pages{184701}

\RefAIP{51} \by{Arcidiacono, S., Poulikakos, D., Ventikos, Y.} \yr{2004} \paper{Oscillatory behavior of nanodroplets.} \jour{Phys. Rev. E} \vol{70} \pages{011505}

\RefAIP{52} \by{Beaglehole, R.} \yr{1979} \paper{Thickness of the surface of liquid Argon near the triple point} \jour{Phys. Rev. Lett.} \vol{43} \pages{2016--2018}; \yr{1980} \paper{Ellipsometric study of the surface of simple liquids} \jour{ Physica B \& C} \vol{100} \pages{163--174}

\RefAIP{53} \by{Beaglehole, R.} \yr{1987} \paper{Capillary-wave and intrinsic thicknesses of the surface of a simple liquid} \jour{Phys. Rev. Lett.} \vol{58} \pages{1434--1436}

\RefAIP{54} \by{Beysens, D., Robert, M.} \yr{1987} \paper{Thickness of fluid interfaces near the critical point from optical reflectivity measurements} \jour{J. Chem. Phys.} \vol{87} \pages{3056}

\RefAIP{55} \by{Fradin, C., Braslau, A., Luzet, D. et al.} \yr{2000} \paper{Reduction in the surface energy of liquid interfaces at short length scales} \jour{Nature} \vol{403} \pages{871--874 }

\RefAIP{56} \by{Irving, J.H., Kirkwood, J.G.} \yr{1950} \paper{The Statistical Mechanical Theory of Transport Processes. IV. The Equations of Hydrodynamics} \jour{J. Chem. Phys.} \vol{18} \pages{817--829}

\RefAIP{57} \by{Likhtman, A.E., Sukumaran, S.K., Ramirez, J.} \yr{2007} \paper{Linear Viscoelasticity from Molecular Dynamics Simulation of
Entangled Polymers} \jour{Macromolecules}  \vol{40} \pages{ 6748-6757}

\RefAIP{58} \by{Landau, L.D., Lifshitz, E.M.} \yr{1987} {\it Fluid Mechanics. Vol. 6.}  Butterworth-Heinemann.

\RefAIP{59} \by{ Fernandez, E.M., Chacon, E., Tarazona, P., Parry, A.O., Rascon, C.} \yr{2013} \paper{Intrinsic Fluid Interfaces and Nonlocality} \jour{Phys. Rev. Lett.}  \vol{111} \pages{096104}

\RefAIP{60} \by{Kob, W., Anderson, H.C.} \yr{1995} \paper{Testing mode-coupling theory for a supercooled binary Lennard-Jones mixture: the van Hove correlation function} \jour{Phys. Rev. E}  \vol{51} \pages{ 4626}

\RefAIP{61} \by{Penfold, J.} \yr{2001} \paper{The structure of the surface of pure liquids} \jour{Rep. Prog. Phys.} \vol{64} \pages{777–-814}

\clearpage

\end{document}